# TOWARDS AN
# ATOMIC AGENCY FOR QUANTUM-AI

A COMPARATIVE ANALYSIS OF AI & QUANTUM TECHNOLOGY REGULATION AND INNOVATION MODELS IN THE U.S., EU AND CHINA

Mauritz Kop*[1]

## ABSTRACT

This essay analyzes emerging AI and quantum technology (including their increasing complementarity and interdependence embodied in quantum-AI hybrids) regulation, export controls, and technical standards in the U.S., EU, and China, comparing legislative efforts anno 2025 to strategically balance the benefits and risks of these transformative technologies through the lens of their distinct innovation systems. Analytically, it finds convergence on the need for principled, responsible technology governance, despite conflicting market-driven, values-based, and state-driven innovation philosophies. The paper posits that the U.S.'s security-centric policy lead creates a *de facto* 'Washington effect' setting global rules, but regulating under profound uncertainty about quantum's future trajectory presents significant challenges. This U.S. first-mover position carries the inherent risk of premature regulation based on incomplete future insights, which could stifle innovation or cause suboptimal technological lock-in. Conversely, China's push for state-aligned standardization via initiatives like the Digital Silk Road signals a 'Beijing effect,' potentially exporting autocratic norms and fragmenting global interoperability, trends exacerbated by unilateral export controls and decoupling pressures.

Faced with planetary challenges like inequality and climate change, aligning on Responsible Quantum Technology norms and standards is a critical global opportunity. The paper explores pathways toward a harmonized Quantum Acquis Planétaire, anchored in universal ethical values ('what connects us') translated into foundational standards and agile legal guardrails. Critically, smart regulation must move beyond mere restrictions to actively incentivize desired responsible behaviors from nations and industry, for instance through frameworks promoting 'Responsible Quantum Technology (RQT) by design'. This necessitates frameworks for equitable benefit/risk redistribution, calling for nations to cooperatively steward quantum resources.

[1]* Mauritz Kop is the Founder of the Stanford Center for Responsible Quantum Technology at Stanford University, mkop@stanford.edu. This work was supported by the Centre for Quantum Networks (CQN) Engineering Research Centers Program of the National Science Foundation under NSF Cooperative Agreement No. 1941583. Any opinions, findings and conclusions or recommendations expressed in this material are those of the author and do not necessarily reflect those of the National Science Foundation. The author thanks Mark Lemley for helpful comments on an earlier version of this article.

The Quantum Acquis Planétaire is envisioned as a global body of Quantum Law, complemented by sector-specific practices. Achieving this requires inter-continental policy making and strategic Sino-American "recoupling" based on shared values and restorative diplomacy. This could be coupled with collaborative research platforms for quantum and AI (which increasingly rely on each other) akin to CERN or ITER - emulating successful international resource and research pooling to foster coordinated responsible quantum innovation. Realizing ambitious goals like fault tolerant quantum-centric supercomputing alongside algorithmic development and use case discovery toward quantum benefit requires such collective global expertise, challenging protectionist measures that stifle necessary collaboration and supply chains.

Codification of a global body of Quantum Law could proceed via a UN Quantum Treaty inspired by precedents like the 2024 UN AI Resolution, the Council of Europe Framework Convention on AI, and the 1968 Nuclear Non-Proliferation Treaty (NPT), designed explicitly to align quantum advancements with global imperatives such as the UN Sustainable Development Goals (SDG).

To enforce such an acquis, disincentivize a quantum arms race, and ensure non-proliferation, establishing an *'Atomic Agency for quantum-AI'* modelled after the International Atomic Energy Agency (IAEA) leveraging its expertise in safeguards implementation for dual-use quantum/AI technologies, potentially overseeing a global treaty, warrants serious examination.

\* \* \*

TABLE OF CONTENTS

## I. INTRODUCTION

The intense competition for quantum leadership, particularly between the state-driven advancements of China and the market driven and values-based innovation ecosystems of the United States and European Union, necessitates robust global governance structures. China's demonstrated lead in quantum communication infrastructure, like the Beijing-Shanghai quantum key distribution (QKD) network, its rapid progress in computing and sensing, alongside policies towards coercing ASEAN nations to adopt Chinese quantum standards (SAC/TC578) to access Belt & Road initiative (BRI) funding, signals China's clear ambition to set international technical standards, potentially embedded with autocratic norms.[2] Sidelining NIST/ISO/IEEE frameworks, the U.S. and EU face bifurcation risks as China develops parallel quantum standards regimes, resulting in incompatible quantum ecosystems with incompatible security architectures.

This strategic rivalry makes a globally harmonized body of Quantum Law a *pragmatic imperative* – not merely an ideal – essential for establishing predictable rules for ethical development, ensuring technological interoperability, managing liability, and fostering equitable access across nations. Concurrently, achieving effective global technical standards is crucial to counteract fragmentation and ensure that quantum technologies develop within a framework of shared values. Unilateral export controls may inadvertently strengthen China's self-sufficiency in quantum and weaken U.S. and EU quantum technology development caused by Chinese retaliatory critical mineral export controls. Dual use ambiguity refers to the difficulty of distinguishing between military and civilian applications (e.g., quantum sensors for navigation vs. surveillance). The inherent dual-use nature of quantum technologies, with significant military and security implications, demands effectively harmonized export controls to prevent weapons proliferation and mitigate risks to supply chain security, moving beyond current, often uncoordinated national restrictions.[3] Therefore, establishing a dedicated non-proliferation framework for quantum technologies inspired by the 1968 Nuclear Non-Proliferation Treaty and the 2024 United Nations Declaration on AI -overseen by an IAEA-like, Atomic Agency for Quantum & AI- represents a necessary mechanism to manage these strategic risks, prevent unchecked militarization, and provide a stable foundation for international scientific collaboration focused on beneficial applications.

While the establishment of such comprehensive international frameworks remains a complex diplomatic endeavor, the perceived urgency of managing quantum risks is prompting significant national-level policy interventions. The actions taken by the United States exemplify this trend towards proactive domestic regulation in response to the strategic challenges posed by quantum technologies.

---

[2] *See e.g.* Antonia Hmaidi & Jeroen Groenewegen-Lau, *China's Long View: Quantum Tech Has US and EU Playing Catch-Up*, MERICS (2023), https://merics.org/en/report/chinas-long-view-quantum-tech-has-us-and-eu-playing-catch and Philip Luck, *The Hidden Risk in Rising U.S.-PRC Tensions: Export Control Symbiosis*, CSIS (Apr. 2025), https://www.csis.org/analysis/hidden-risk-rising-us-prc-tensions-export-control-symbiosis.

[3] *See e.g.* Robert D. Atkinson et al., *China Is Rapidly Becoming a Leading Innovator in Advanced Industries*, ITIF (Sept. 16, 2024), https://itif.org/publications/2024/09/16/china-is-rapidly-becoming-a-leading-innovator-in-advanced-industries/

One significant moment occurred on August 9, 2023, when President Biden issued an Executive Order, proclaiming a national emergency to safeguard U.S. investments in quantum information technologies within countries of concern.[4] This mandate instructs the Treasury Secretary, alongside various agency leaders, to enact regulations that require certain transactions involving quantum information technologies to be reported or prohibited. These regulations aim to control the transfer of advanced quantum technologies, products and services, materials and devices, especially those with expected impacts on encryption, cybersecurity, and defense communications.

Additionally, the House Science Committee has pushed forward the National Quantum Initiative Reauthorization Act, which is an extension of the original 2018 National Quantum Initiative Act.[5] This legislation seeks to catalyze breakthrough research, foster the development of new quantum applications, enhance partnerships within the industry, and invigorate the quantum technology ecosystem. It mandates the National Institute of Standards and Technology (NIST) to collaborate with other agencies to implement quantum-resistant cryptography.

These regulatory efforts demonstrate the U.S. government's acute awareness of quantum technology's geostrategic value, highlighting an imperative to oversee its progression and integration, particularly in areas affecting national security, economic stability, and global competitiveness.

The governance initiatives discussed in this essay represent concrete manifestations of national approaches to harnessing and controlling quantum potential, reflecting distinct policy priorities and risk assessments. To systematically evaluate the similarities, differences, and possible impacts of these diverse national strategies requires the application of a dedicated analytical framework. It is crucial to recognize that the scope of 'quantum technologies' considered herein extends beyond computing, explicitly encompassing significant advancements and implications within quantum sensing, quantum simulation, quantum networking & communications, quantum cryptography, and the emerging field of quantum-AI hybrid technologies. To this end, the analysis utilizes a comparative methodology, examining the distinct approaches towards regulating quantum through the lens of safeguarding, engaging, and advancing (SEA) the suite of quantum technologies for societal benefit.

---

[4] "Executive Order 14105 on Addressing United States Investments in Certain National Security Technologies," Federal Register, vol. 88, no. 170, August 2023, https://www.federalregister.gov/documents/2023/08/11/2023-17449/addressing-united-states-investments-in-certain-national-security-technologies-and-products-in

[5] Committee on Science, Space & Technology, H.R. 6213 - The National Quantum Initiative Reauthorization Act (November 23, 2023), https://science.house.gov/2023/11/the-national-quantum-initiative-reauthorization-act

**II. QUANTUM TECHNOLOGY REGULATION IN THE U.S., EU & CHINA**

**A. U.S.**

Since 2018, the United States has been the first to pass legislation pertaining to quantum governance. As of 2025, the United States has enacted several regulatory interventions concerning quantum technology, indicating the nation's commitment to not only advance but also steer the societal implications of this burgeoning domain. Notable quantum specific legislation includes:

1. **The Quantum Initiative Act (NQIA):** The NQIA was signed into law on December 21, 2018, and aims to ensure U.S. leadership in quantum information science and technology applications.[6] It focuses on accelerating quantum research and development to enhance the country's economic and national security.[7] The NQIA calls for a coordinated federal program to support QIS research, development, and demonstration, and to improve interagency planning and coordination.

2. **Executive Order 13885 on Establishing the National Quantum Initiative Advisory Committee**: Issued in August 2019, this order established the National Quantum Initiative Advisory Committee. The Committee's role was to advise the Secretary of Energy and the Subcommittee on Quantum Information Science on the National Quantum Initiative Program, which focuses on quantum information science and technology applications.[8]

3. **Quantum User Expansion for Science and Technology (QUEST) Act of 2020:** The Act – issued in Sept. 2020 and signed into code in August 2022 - amends the NQIA to establish the Quantum User Expansion for Science and Technology (QUEST) Program, authorized by the CHIPS & Science Act.[9] The Program aims to improve access to quantum computing resources for U.S.-based researchers and laboratories. The Department of Energy (DOE) is tasked with implementing the QUEST Program through a merit-reviewed application process.[10]

4. **Executive Order 14073 on Enhancing the National Quantum Initiative Advisory Committee**: Signed in May 2022, this order enhances the National Quantum Initiative

---

[6] H.R.6227 - National Quantum Initiative Act, 115th Congress (2017-2018), June 26, 2018, https://www.congress.gov/bill/115th-congress/house-bill/6227
[7] See: https://www.quantum.gov/
[8] "Executive Order 13885—Establishing the National Quantum Initiative Advisory Committee," Federal Register, vol. 84, no. 159, August 2019, https://www.federalregister.gov/documents/2019/09/05/2019-19367/establishing-the-national-quantum-initiative-advisory-committee
[9] H.R.8303 - QUEST Act of 2020, 116th Congress (2019-2020), Sept 17 2020, https://www.congress.gov/bill/116th-congress/house-bill/8303/text and 15 U.S. Code § 8854 - Department of Energy Quantum User Expansion for Science and Technology program, Aug 19, 2022, https://www.law.cornell.edu/uscode/text/15/8854
[10] *See* for the quantum aspects of CHIPS and Science Act: Gregory Arcuri & Hideki Tomoshige, A Look at the Quantum-related Portions of CHIPS+, CSIS, Aug 29, 2022, https://www.csis.org/blogs/perspectives-innovation/look-quantum-related-portions-chips and Santanu Basu & Jacqueline A. Basu, Breaking down the 2022 CHIPS and Science Act, QED-C, Dec 2022, https://quantumconsortium.org/blog/breaking-down-the-2022-chips-and-science-act/

Advisory Committee to ensure continued American leadership in quantum information science and technology. The Committee, consisting of various members from industry, universities, and Federal laboratories, is tasked with advising the President on quantum information science and technology research, development, demonstrations, standards, education, technology transfer, commercial application, and national security and economic stability concerns.[11]

5. **National Security Memorandum on Promoting United States Leadership in Quantum Computing While Mitigating Risks to Vulnerable Cryptographic Systems**: This memorandum, also from May 2022, focuses on the security risks posed by quantum computers. It outlines measures to transition to quantum-resistant cryptography, including establishing a "Migration to Post-Quantum Cryptography Project" at the National Cybersecurity Center of Excellence. The memorandum mandates the Secretary of Commerce, through the Director of NIST, and other federal agencies to engage in activities to secure cryptographic systems against the threats posed by quantum computing.[12]

6. **National Security Presidential Memorandum-33 (NSPM-33):** This memorandum, issued January 2021 by the Trump Presidency and implemented January 2022, represents a significant step by the U.S. government in safeguarding and advancing national security interests in the context of quantum information science and its potential impacts. It directs federal agencies to strengthen protections for U.S. government-funded research and development (R&D) against foreign interference and exploitation, with particular concern for activities by countries such as China. The directive requires federal research funding agencies to enhance and standardize disclosure requirements for universities and researchers who receive federal awards. It mandates the creation of research security programs -including certification- at institutions obtaining federal funds, aiming to safeguard intellectual property, prevent the misappropriation of research, and ensure the responsible use of U.S. taxpayer money. The directive promotes U.S. national and economic security by supporting the secure advancement of quantum information science, fostering collaboration and investment in quantum research, emphasizing workforce development in the quantum sector, and shaping policy and regulatory frameworks for quantum technologies.[13]

---

[11] "Enhancing the National Quantum Initiative Advisory Committee," Federal Register, vol. 87, no. 110, May 2022, https://www.federalregister.gov/documents/2022/05/09/2022-10076/enhancing-the-national-quantum-initiative-advisory-committee. *See also* National Quantum Initiative, Why Quantum Matters to You, https://www.quantum.gov/why-quantum-matters-to-you/; and Potomac Quantum Innovation Center, Enhancing the National Quantum Initiative, https://www.pqic.org/enhancing-the-national-quantum-initiative

[12] "National Security Memorandum on Promoting United States Leadership in Quantum Computing," White House, May 2022, https://bidenwhitehouse.archives.gov/briefing-room/statements-releases/2022/05/04/national-security-memorandum-on-promoting-united-states-leadership-in-quantum-computing-while-mitigating-risks-to-vulnerable-cryptographic-systems/

[13] Presidential Memorandum on United States Government-Supported Research and Development National Security Policy, NSPM-33, https://trumpwhitehouse.archives.gov/presidential-actions/presidential-memorandum-united-states-government-supported-research-development-national-security-policy and https://www.whitehouse.gov/wp-content/uploads/2022/01/010422-NSPM-33-Implementation-Guidance.pdf

7. **Executive Order 14105 on Addressing United States Investments in Certain National Security Technologies and Products in Countries of Concern**: Signed in August 2023, this order is focused on regulating U.S. investments in sensitive technologies, including quantum information technologies, in countries of concern. It aims to prevent the development and production of quantum technologies that could compromise encryption, cybersecurity controls, and military communications. The Department of the Treasury is responsible for developing regulations under this order.[14]

8. **Executive Order 14117 on Preventing Access to U.S. Sensitive Personal Data and Government-Related Data by Countries of Concern or Covered Persons:** Signed on February 28, 2024, this order seeks to address the growing exploitation of Americans' sensitive personal data which threatens the development of an international technology ecosystem that protects U.S. citizens' security, privacy, and human rights.[15] On January 8 2025 the Department of Justice issued a final rule to implement Executive Order 14117 by prohibiting and restricting certain data transactions with certain countries or persons. The rule is designed to encourage the adoption of sufficiently effective methods of encryption, aggregation, and/or other privacy-preserving technologies, including any United States Government-approved standards for quantum-resistant public-key cryptographic algorithms.[16]

9. **Treasury Department's Advanced Notice of Proposed Rulemaking (ANPRM) on Quantum Investments:** Issued on Nov. 15, 2024, the ANPRM proposes regulations targeting investments in sensitive quantum technologies such as quantum sensors and networking systems.[17] The Final Rule, to implement Executive Order 14105 of August 9, 2023, "Addressing United States Investments in Certain National Security Technologies and Products in Countries of Concern" (the Outbound Order), prioritizes restrictions on transactions with foreign entities linked to military applications or government intelligence.[18]

10. **U.S.-EU Quantum Task Force:** On April 5 2024, the United States and the European Union issued a joint statement regarding the outcomes of the Trade and Technology Council (TTC).[19] The document highlights key areas of collaboration and future priorities, emphasizing the importance of transatlantic cooperation in technology and trade policies.

---

[14] "Executive Order 14105 on Addressing United States Investments in Certain National Security Technologies," Federal Register, vol. 88, no. 170, August 2023, https://www.federalregister.gov/documents/2023/08/11/2023-17449/addressing-united-states-investments-in-certain-national-security-technologies-and-products-in

[15] Preventing Access to United States Sensitive Personal Data and Government-Related Data by Countries of Concern, 90 Fed. Reg. 1226 (Jan. 8, 2025), https://www.federalregister.gov/documents/2025/01/08/2024-31486/preventing-access-to-us-sensitive-personal-data-and-government-related-data-by-countries-of-concern

[16] Preventing Access to United States Sensitive Personal Data and Government-Related Data by Countries of Concern, 90 Fed. Reg. 1226, 1323 (Jan. 8, 2025), https://www.federalregister.gov/d/2024-31486/p-323

[17] Federal Register, Provisions Pertaining to U.S. Investments in Certain National Security Technologies and Products in Countries of Concern, Nov 15, 2024, https://www.federalregister.gov/documents/2024/11/15/2024-25422/provisions-pertaining-to-us-investments-in-certain-national-security-technologies-and-products-in

[18] Treasury Department, Treasury Issues Regulations to Implement Executive Order Addressing U.S. Investments in Certain National Security Technologies and Products in Countries of Concern, Oct 28, 2024, https://home.treasury.gov/news/press-releases/jy2687

The TTC concludes that the rapid pace of digital transformation opens vast possibilities for expansion and innovation but also introduces many risks and obstacles that necessitate enhanced collaboration and strong ongoing coordination in joint strategies for developing guidelines for new technologies like artificial intelligence (AI), quantum technologies, and 6G wireless communications. The TTC's goal is to promote interoperability and uphold shared democratic principles and human rights protection, alongside encouraging innovation. Additionally, the TTC is committed to continually updating U.S.-EU workforce's skills to adapt to the demands of swiftly evolving technologies, including AI.

During the TTC, the United States and the European Union have created a Quantum Task Force focused on enhancing science and technology cooperation in quantum technologies. The task force's main goal is to align research and development efforts between these regions by establishing a common understanding of technology readiness levels, setting unified benchmarks, identifying key components of quantum technology, and advancing international values-based, rights respecting technical standards. Beyond merely aligning research, the Task Force is developing common definitions for technology readiness levels and benchmarks, and crucially addresses reciprocity in research openness and intellectual property rights, aiming for a unified Western governance approach balancing innovation and security.

11. **National Quantum Initiative Advisory Committee's Quantum Networking Report:** Published in September 2024, the National Quantum Initiative Advisory Committee (NQIAC) report, "Quantum Networking: Findings and Recommendations for Growing American Leadership," assesses the state of U.S. quantum networking research and development. It emphasizes the strategic importance of this field for national security and economic prosperity, calling for sustained, coordinated R&D involving government, academia, and industry. The report highlights the need for appropriately scaled testbeds, standardized metrics to measure progress, enhanced international collaboration, and dedicated efforts to build a diverse quantum workforce.[20]

12. **Executive 14144 Order on Strengthening and Promoting Innovation in the Nation's Cybersecurity**: Issued by President Biden on January 16, 2025, the order serves as a capstone directive consolidating and advancing the U.S. federal government's cybersecurity strategy in response to evolving digital threats and lessons learned from recent major cyber incidents.[21]

---

[19] The White House. (2024, April 5). *U.S.-EU joint statement of the Trade and Technology Council*. https://www.whitehouse.gov/briefing-room/statements-releases/2024/04/05/u-s-eu-joint-statement-of-the-trade-and-technology-council-3/
[20] Nat'l Quantum Initiative Advisory Comm. [NQIAC], *Quantum Networking: Findings and Recommendations for Growing American Leadership* (Sept. 6, 2024), https://www.quantum.gov/nqiac-report-on-quantum-networking/
[21] Executive Order 14144 on Strengthening and Promoting Innovation in the Nation's Cybersecurity, Biden White House (Jan. 16, 2025), https://bidenwhitehouse.archives.gov/briefing-room/presidential-actions/2025/01/16/executive-order-on-strengthening-and-promoting-innovation-in-the-nations-cybersecurity/.

The order mandates federal agencies to implement next-generation cybersecurity practices such as zero trust architecture, secure software development frameworks, and the adoption of post-quantum and hybrid encryption standards. It also requires software vendors to provide verifiable attestations of secure development practices, enhances the use of advanced technologies like artificial intelligence for cyber defense, and expands public-private partnerships to protect critical infrastructure sectors-especially those at high risk from state-sponsored and criminal cyber actors. The Executive Order further institutionalizes programs like the U.S. Cyber Trust Mark for consumer IoT security and directs agencies to prioritize research, development, and rapid adoption of innovative cybersecurity solutions, ensuring the resilience and competitiveness of the nation's digital ecosystem in the face of quantum and AI-enabled threats.

13. **National Quantum Initiative Reauthorization Act**: This legislative act, advanced by the House Science Committee, reauthorizes the National Quantum Initiative Act (NQIA) of 2018 (that expired in 2023) to extend funding and programs through 2028. The NQIA focuses on breakthrough quantum research hubs, driving new quantum applications, strengthening industry partnerships, investing in the U.S. quantum workforce prioritizing quantum information science & technology (QIST) and STEM education, and invigorating the quantum ecosystem. The act includes provisions for adopting post-quantum cryptography (PQC), instructs collaboration with the National Institute of Standards and Technology (NIST), and expands the National Quantum Initiative Program to include quantum networking and supply chain resilience.[22] Status: pending reauthorization as of April 2025.

14. **The Department of Energy Quantum Leadership Act of 2025**: Introduced in February 2025 by Senators Durbin and Daines, the Quantum Leadership Act 2025 focuses on enhancing the U.S. Department of Energy's (DOE) quantum infrastructure, research and development, by proposing over $2.5 billion in funding for the fiscal years 2026-2030.[23] The Act is pending legislative action as of April 2025. The DOR Quantum Leadership Act complements the NQIA Reauthorization by addressing gaps in DOE's quantum infrastructure, while the latter provides overarching policy continuity. Both bills aim to address China's rapid advancements in quantum hardware, strengthen U.S. competitiveness in quantum-classical hybrid systems, and align with the CHIPS and Science Act's supply chain resilience goals. The figure below describes key differences between the two Acts:

| Aspect | NQIA Reauthorization | Quantum Leadership Act 2025 |
|--------|---------------------|----------------------------|

---

[22] "National Quantum Initiative Reauthorization Act," Congress.gov, 2023, https://www.congress.gov/bill/118th-congress/house-bill/6213
[23] S.579 - Department of Energy Quantum Leadership Act of 2025 118th Cong. (2025) (proposing over $2.5 billion in funding for quantum research and development under the Department of Energy), https://www.congress.gov/bill/119th-congress/senate-bill/579/text and https://www.durbin.senate.gov/newsroom/press-releases/durbin-daines-introduce-bipartisan-legislation-to-support-the-future-of-quantum-research-at-energy-department

| Scope | Broad reauthorization of federal quantum programs | Focused on DOE infrastructure and materials science |
|---|---|---|
| **Funding Targets** | Cross-agency (NSF, NIST, DOE) | DOE-specific facility upgrades |
| **Technical Focus** | General quantum R&D and PQC | Quantum materials and networking |
| **Workforce Development** | STEM education and industry partnerships | Targeted fellowships and lab-academia collaboration |

Figure 1: Key differences between the NQIA Reauthorization and the Quantum Leadership Act.

The U.S. approach, evident in the NQIA, executive orders, and task forces, strongly emphasizes 'Advancing' technological leadership and 'Safeguarding' national security through investment screening and research security protocols, and stronger cybersecurity measures against quantum computing threats. The 'Engaging' aspect appears primarily through strategic expert consultations. The focus on post-quantum cryptography (PQC) showcases proactive safeguarding. However, this operates under the uncertainty of the 'Quantum Event Horizon,' making optimal regulatory design challenging

**B. EU**
By 2025, the European Union has engaged in exploring diverse policies and regulations within the realm of quantum technologies, acknowledging their transformative potential and societal impact across numerous sectors.[24] The current EU's approach to quantum technology is largely coordinated through and characterized by funding and research frameworks, Member State level national quantum strategies and roadmaps, and translating research into applications and concrete use cases, rather than through direct quantum legislation akin to laws and regulations for AI and data privacy.[25] As quantum specific rules are still evolving, these key initiatives play a pivotal role in shaping the quantum technology industry, policy, and regulatory landscape in the block:

1. **The Quantum Technologies Flagship** was launched in May 2016 as part of the European Commission's Horizon 2020 program. This €1 billion FET (Future and Emerging Technologies) Flagship initiative is designed to develop quantum technology over a 10-year period. Its goal is to consolidate and expand European scientific leadership and excellence in this field, ensuring that the EU remains at the forefront of the second quantum revolution.[26]

---

[24] European Commission Joint Research Centre. (n.d.). Quantum technologies. European Commission. Retrieved February 2, 2024, from https://joint-research-centre.ec.europa.eu/scientific-activities-z/quantum-technologies_en
[25] Quantum Flagship. (2023, October 17). Quantum Technologies Public Policies Report 2023. https://qt.eu/news/2023/2023-10-17_quantum-technologies-public-policies-report-2023
[26] European Commission. (2016). Quantum Technologies Flagship, https://ec.europa.eu/digital-single-market/en/quantum-technologies-flagship

2. **The European Quantum Communications Infrastructure (EuroQCI)** from June 2018 was announced by the European Commission and put into effect after all 27 Member States signed the EuroQCI Declaration of June 2019.[27]. The EuroQCI aims to build a secure quantum communication infrastructure that spans the entire EU, including its overseas territories. This infrastructure is intended to protect European critical infrastructures and sensitive communications while enhancing privacy for European citizens.[28]

3. Significant funding for high-performance computing and quantum computing is included in the **Digital Europe Programme** from September 2020. It aims to deploy a quantum communication infrastructure across Europe and support the development of quantum technologies.[29]

4. The European Commission's vision for Europe's digital transformation by 2030, known as **Europe's Digital Decade** published in February 2021, identifies quantum computing as a key area for achieving digital sovereignty. The initiative sets out digital targets for the EU to be digitally sovereign in an open and interconnected world.[30]

5. **The Joint Undertaking on European High-Performance Computing (EuroHPC Joint Undertaking)** from Sept 2018 was established to support the development and integration of high-performance computing and quantum computing infrastructures across Europe. Though initially launched earlier, it has been continuously updated to reflect the growing importance of quantum computing.[31] As of January 2024, AI Factories make a new pillar within EuroHPC.[32] These AI factories will focus on both software and hardware: producing large general purpose AI models, and manufacturing GPUs for AI, quantum, and quantum-AI hybrid computational paradigms.

6. **EU-U.S. Trade and Technology Council**: The EU has established an ad hoc task force on quantum as part of the EU-U.S. Trade and Technology Council, focusing on transparency, information exchange, and transatlantic cooperation in the field of quantum technologies.[33] The task force will elaborate on the applicable intellectual property rights framework, identifying essential components, setting technical and interoperability standards, determining benchmarks for quantum computing performance, and addressing issues related to export controls. This collaboration is part of a broader strategy to align with international partners on quantum technology development and regulation.

---

[27] European Commission, The Future of Quantum: EU Countries Plan Ultra-Secure Communication Network, DIGITAL STRATEGY (June 6, 2023), https://digital-strategy.ec.europa.eu/en/news/future-quantum-eu-countries-plan-ultra-secure-communication-network.
[28] European Commission. (2018). European Quantum Communications Infrastructure (EuroQCI), https://ec.europa.eu/digital-single-market/en/news/european-quantum-communication-infrastructure-euroqci
[29] European Parliament. (2020). Digital Europe Programme. https://www.europarl.europa.eu/legislative-train/theme-a-new-push-for-european-democracy/file-digital-europe-programme
[30] European Commission. (2021). Europe's Digital Decade, https://ec.europa.eu/digital-single-market/en/digital-decade
[31] EuroHPC Joint Undertaking (2018). https://eurohpc-ju.europa.eu/ and European Commission. (n.d.). High performance computing joint undertaking. Retrieved March 16, 2024, from https://digital-strategy.ec.europa.eu/en/policies/high-performance-computing-joint-undertaking
[32] European Commission. (January 2024). AI factories. Retrieved from https://digital-strategy.ec.europa.eu/en/policies/ai-factories

7. **European Declaration on Quantum Technologies:** On December 6, 2023, a group of EU member states signed a pact that recognizes the strategic importance of quantum technologies for the scientific and industrial competitiveness of the EU.[34] The declaration serves as a comprehensive framework designed to synchronize quantum-related endeavors at EU, national, and regional levels, with the aim of constructing a cohesive, values-based quantum technology ecosystem.[35] The European Declaration on Quantum Technologies expands on prior initiatives mentioned in this section, aimed at establishing a quantum technology infrastructure within the EU. These include the Quantum Technologies Flagship, which unites research bodies, the industrial sector, and public financiers to foster the commercialization of quantum research, and the European High-Performance Computing Joint Undertaking (EuroHPC JU) initiative, which seeks to create cutting-edge experimental quantum computers. For instance, Spain inaugurated its first quantum computer based on European technology in early 2025, integrating it with the MareNostrum5 supercomputer as part of the Quantum Spain project and the EuroHPC JU network.[36] With the Flagship and the HPC ventures inaugurated in 2018, the 2023 EU Quantum Declaration marks a significant step toward Europe's quantum technological progression.

8. **European Critical Raw Materials Act:** The ECRM of March 16, 2023, is a legislative initiative designed to secure the European Union's access to critical raw materials essential for green technologies, digital innovation, and strategic sectors such as quantum computing and renewable energy.[37] Adopted as part of the European Green Deal Industrial Plan, the Act sets targets for extraction, processing, and recycling within the EU, aiming to reduce dependence on third-country suppliers—particularly those with geopolitical sensitivities like China and Russia. It establishes a framework for monitoring supply chains, streamlining permitting procedures, and supporting strategic projects to ensure a resilient, sustainable, and competitive

---

[33] The White House. (2023, May 31). U.S.-EU Joint Statement of the Trade and Technology Council, https://www.whitehouse.gov/briefing-room/statements-releases/2023/05/31/u-s-eu-joint-statement-of-the-trade-and-technology-council-2/
[34] European Commission. (2023, Dec 6)). European declaration on quantum technologies. Retrieved from https://digital-strategy.ec.europa.eu/en/library/european-declaration-quantum-technologies and European Declaration on Quantum Technologies
[35] Quantum Flagship. (2024, March 22). EU leaders showcase quantum technology ambitions after signing landmark pact. Retrieved from https://qt.eu/news/2024/2024-03-22_eu-leaders-showcase-quantum-technology-ambitions-after-signing-landmark-pact
[36] Quantum Spain Project, *Quantum Spain presents the first quantum computer in Spain developed with 100% European technology* (Feb. 6, 2025), https://quantumspain-project.es/en/quantum-spain-presents-the-first-quantum-computer-in-spain-developed-with-100-european-technology/
[37] European Commission, European Critical Raw Materials Act, COMMISSION, https://commission.europa.eu/strategy-and-policy/priorities-2019-2024/european-green-deal/green-deal-industrial-plan/european-critical-raw-materials-act_en

European industry.[38] The Act is central to the EU's strategy for technological sovereignty and economic security in the face of global supply chain disruptions and rising demand for rare earth elements and other critical materials.

8. **Strategic Research and Industry Agenda (SRIA) 2030:** Published in February 2024 by the European Quantum Flagship under the supervision of the Strategic Advisory Board, the SRIA serves as a reference blueprint for advancing quantum technologies within the European Union. It is structured into two main parts: the first examines the four core pillars of quantum technology—computing, simulation, communication, and sensing and metrology—detailing their impact, challenges, and strategic objectives. The second part addresses transversal topics critical for the field's development, including basic science, enabling technologies, education, standardization, funding, intellectual property, international collaboration, and ethical considerations. The SRIA 2030 aims to harmonize the objectives of both research and industry, fostering an independent and world-leading European quantum ecosystem that can drive innovation and address societal challenges.

9. **QuIC Strategic industry Roadmap (SIR) 2025:** The SIR of April 2025, written by the European Quantum Industry Consortium (QuIC), provides a comprehensive overview of the current state and future trajectory of quantum technologies in Europe.[39] It details five key pillars: quantum computing, quantum simulation, quantum communication, quantum sensing and metrology, and enabling technologies such as photonics and cryogenics. For each area, the roadmap outlines technological foundations, existing European capabilities, ongoing research, application potential, standardization needs, intellectual property considerations, and workforce development requirements, projecting developments towards 2035. The report's primary purpose is to inform policymakers and decision-makers across Europe, guiding them in understanding the ambitions of the European quantum industry and identifying strategic areas for support to accelerate its growth and ensure Europe's competitive position in the global quantum landscape.

In March 2025, QuIC also published a whitepaper with recommendations for the upcoming EU Quantum Strategy.[40] It outlines the strategic importance of quantum computing, communication, and sensing for Europe's future economic and societal well-being. The report is structured around key areas such as fostering a 'Made in Europe' quantum computer, advancing quantum chips, ensuring secure quantum communication, developing quantum sensing, strengthening the European supply chain, attracting investment, establishing standards and protecting intellectual property, building a skilled workforce, ensuring European sovereignty, and leveraging quantum for societal benefit. Ultimately, the document aims to position Europe as a global leader in the quantum revolution through targeted investments, strategic collaborations, and the creation of a robust ecosystem. In January 2025, QuIC also published a global patent landscape study of quantum technologies anno 2025.[41]

---

[38] European Commission, Critical Raw Materials Act, SINGLE MARKET ECONOMY, https://single-market-economy.ec.europa.eu/sectors/raw-materials/areas-specific-interest/critical-raw-materials/critical-raw-materials-act_en
[39] European Quantum Industry Consortium, Strategic Industry Roadmap 2025, EUROQUIC (2025), https://www.euroquic.org/strategic-industry-roadmap-2025/
[40] European Quantum Industry Consortium, Recommendations from the European Quantum Industry Consortium (QuIC) for the EU Quantum Strategy, EUROQUIC (2025), https://www.euroquic.org/recommendations-from-the-european-quantum-industry-consortium-quic-for-the-eu-quantum-strategy/

Oftentimes, the initiatives listed above involve significant funding allocations, collaboration between Member States, and public-private partnerships to advance quantum computing, sensing, communication, and technology research and infrastructure. Although these initiatives and frameworks are not sweeping horizontal legislative acts in the traditional sense -such as unifying regulations or harmonizing directives as part of the imminent European Quantum Strategy-, they play a crucial role in guiding the development and application of quantum technology within the EU while ensuring these align with the block's policies and strategic interests.[42]

The EU's strategy -combining collaborative research funding (Quantum Flagship, Horizon Europe), infrastructure development (EuroQCI, EuroHPC JU), and strategic declarations- strongly aligns with the 'Engaging' and 'Advancing' aspects of responsible innovation. Its underlying regulatory philosophy suggests future 'Safeguarding' will prioritize values, ethics, and fundamental rights, expectedly via a dedicated EU Quantum Act. The focus on strategic autonomy and economic security also informs its safeguarding posture, offering a contrast to the U.S. market-first approach.

**C. China**

As of 2025, China's approach to regulating quantum technology is part of its broader strategy for technological advancement, with special emphasis on state investment in key system technologies including artificial intelligence (AI) and quantum information science. Detailed central planning documents regarding quantum technology policy strategies are sparingly publicly accessible. Notable government directives, policy frameworks, and legislative actions designed to establish China as a global leader in quantum and related technologies include:

1. **863 Program: National High-Technology Research and Development Program on Quantum Information Technology** (2016) - Part of China's broader effort to advance high-tech research, the 863 Program has increasingly emphasized quantum computing and quantum communications.[43]

2. **13th Five-Year Plan (2016-2020)** (March 2016) - Under its 13th five-year plan, China has launched a megaproject for quantum communications and computing. This project aims for major breakthroughs by 2030, including the development of a general quantum computer prototype and a practical quantum simulator. These initiatives demonstrate China's commitment to becoming a global leader in quantum technology. It marked the official inclusion of quantum technology in China's national strategic goals.[44]

---

[41] European Quantum Industry Consortium, A Portrait of the Global Patent Landscape in Quantum Technologies 2025, EUROQUIC (2025), https://www.euroquic.org/a-portrait-of-the-global-patent-landscape-in-quantum-technologies-2025/ Compare to: Aboy, M., Minssen, T. & Kop, M. Mapping the Patent Landscape of Quantum Technologies: Patenting Trends, Innovation and Policy Implications. *IIC* **53**, 853–882 (2022). https://doi.org/10.1007/s40319-022-01209-3
[42] European Commission, Quantum Technologies, DIGITAL STRATEGY, https://digital-strategy.ec.europa.eu/en/policies/quantum
[43] Ministry of Science and Technology of the People's Republic of China, "National High-Technology Research and Development Program (863 Program)" (2016), http://www.most.gov.cn/eng/programmes1/200610/t20061009_36223.htm

3. **Establishment of the National Laboratory for Quantum Information Sciences** (September 2017) - Part of China's significant investment in quantum research infrastructure, focusing on quantum computing and communications research.[45]

4. **China's Standards for Quantum Communication Technology** (2018) - These standards cover various aspects of quantum communication, including quantum key distribution (QKD) to ensure secure communications.[46]

5. **Quantum Information Science Development Strategy** (2018) Although specific details and the full text may not be publicly available, reports indicate that China announced a strategy to guide the development of quantum information sciences, signaling strong state support for quantum research and applications. It includes measures to secure communications through quantum key distribution (QKD) networks like the Beijing-Shanghai link. This strategy underscores China's ambition to lead in this area by 2030.[47]

6. **Regulations on Scientific Data Management** (2018) - While not exclusively focused on quantum technology, these regulations are relevant for quantum research as they govern the management, sharing, and protection of scientific data in China. This framework impacts how quantum research data is handled, emphasizing the security and strategic value of scientific and technological data.[48]

7. **14th Five-Year Plan (2021-2025) for National Economic and Social Development and the Long-Range Objectives Through the Year 2035** (October 2020) - President Xi Jinping's ambition for China to be a global leader in innovation by 2035 includes significant investments in quantum science, aiming for major breakthroughs. In this economic blueprint, quantum technology was once more emphasized as critical area of innovation, with plans for significant investment in quantum computing, sensing and communications research and development.

---

[44] National Development and Reform Commission (NDRC), "13th Five-Year Plan for Economic and Social Development of the People's Republic of China (2016-2020)" (March 2016), http://en.ndrc.gov.cn/newsrelease_8232/201603/t20160318_796923.html

[45] Chinese Academy of Sciences, "Establishment of the National Laboratory for Quantum Information Sciences" (September 2017), http://www.cas.cn/syky/201709/t20170912_4645769.shtml

[46] State Administration for Market Regulation and the Standardization Administration, "China's Standards for Quantum Communication Technology" (2018), http://www.samr.gov.cn/hd/zjdc/201812/t20181229_282951.html

[47] Ministry of Science and Technology, "Quantum Information Science Development Strategy" (2018), http://www.most.gov.cn/kjbgz/201812/t20181230_143959.htm

[48] General Office of the State Council, "Regulations on Scientific Data Management" (2018), http://www.gov.cn/zhengce/content/2018-04-02/content_5279276.htm

The national strategy for "innovation-driven" development is expected to influence regulatory frameworks and policies in quantum technology.[49]

8. **National Laboratory for Quantum Information Sciences** (2021) - The launch of this national laboratory, although not a legislative action per se, represents a significant government-backed initiative aimed at consolidating China's quantum research efforts. The laboratory is part of China's broader strategy to develop quantum computing and communication technologies.[50]

9. **Comprehensive AI Law** (2023): China is working on a comprehensive law that addresses AI as a whole, rather than focusing on specific subsets such as deepfakes. This law is expected to be broad in scope akin to the sweeping, cross-sectoral EU AI Act, yet more problem based, agile, and surgical. It reflects China's aim to exert state control over how AI impacts various economic and industrial sectors, including aspects relevant to quantum technology. China's approach to AI regulation, including the creation of a "negative list" of areas and products AI companies should avoid unless they have government approval, and a whitelist of approved algorithms, provides insights into its regulatory philosophy. This approach might extend to quantum technologies including quantum-AI hybrids, while focusing on areas that align with China's national objectives and strategic interests.[51]

10. **Draft Guidelines for AI Industry Standardization**: China's industry ministry issued draft guidelines to standardize the AI industry, proposing to form national and industry-wide standards for AI by 2026. This initiative includes participation in forming international standards, indicating a holistic approach that likely encompasses quantum-AI hybrid technologies as well.[52]

11. **Quantum Computing Application Roadmap 2024-2030**: This January 2024 roadmap marks a shift from fundamental research toward application-focused quantum development, outlines prioritized quantum application domains for government support and introduces a certification framework for quantum algorithms.[53] The roadmap focuses on achieving fault tolerant quantum computing and establishing a strong domestic quantum ecosystem.[54] Key areas include developing full-stack solutions, strengthening standards leadership, and securing technological independence through initiatives like Hefei's Quantum Avenue, the 1,200-mile Beijing-Shanghai QKD network and the Micius satellite. The goal is to move from niche demonstrations to practical applications in areas like quantum chemistry and drug discovery

---

[49] Central Committee of the Communist Party of China and the State Council, "14th Five-Year Plan for National Economic and Social Development and the Long-Range Objectives Through the Year 2035" (October 2020), http://www.gov.cn/xinwen/2021-03/13/content_5592681.htm
[50] University of Science and Technology of China, "Launch of the National Laboratory for Quantum Information Sciences" (2021), http://en.ustc.edu.cn/
[51] Carnegie Endowment for International Peace, 'China's AI Regulations and How They Get Made,' last modified July 10, 2023, https://carnegieendowment.org/2023/07/10/china-s-ai-regulations-and-how-they-get-made-pub-90117
[52] Ministry of Industry and Information Technology (China), "Draft Guidelines for AI Industry Standardization" (Year not specified), http://www.miit.gov.cn/
[53] Ministry of Science & Technology (China), "Quantum Computing Application Roadmap 2024–2030" (Jan. 2024), see summary at Sci. & Tech. Daily, Jan. 18, 2024.
[54] *See e.g.* The State Council Information Office (SCIO) of the People's Republic of China, How China's quantum leap is set to redefine future of computing, March 5, 2025, http://english.scio.gov.cn/m/chinavoices/2025-03/05/content_117746794.html

within five years, eventually aiming for universal fault tolerant quantum computing in about 15 years.[55]

12. **China's Innovation Roadmap 2025**: China's 2025 innovation roadmap, part of the larger "Made in China 2025" initiative, aims to reduce reliance on foreign technology, enhance industrial capabilities, and achieve global competitiveness in key industries.[56] The government workforce report dated March 5 2025 focuses on strengthening innovation, improving product quality, and promoting efficient and integrated manufacturing across sectors like advanced information technology, AI, quantum, robotics, aerospace, and new materials. The ultimate goal is to transition China into a technology superpower by 2049.

China's nascent regulatory landscape for quantum technology and AI reflects its broader objective to achieve global leadership in high-tech industries.[57] This approach combines government-led national development strategies, state investments in quantum research and applications, and efforts to institute standards and regulations for AI and quantum information science, in particular for quantum networking & communications, quantum sensing, quantum computing, and quantum cryptography.

China's state-driven approach clearly prioritizes 'Advancing' national capabilities towards global leadership (Five-Year Plans, QIS Strategy) and 'Safeguarding' state control and security interests (Data/Cybersecurity laws, standards strategy). Significant state investment fuels progress in targeted areas like quantum communication. 'Engagement' primarily involves mobilizing national resources towards state goals, rather than broad public deliberation. While ethical guidelines exist, their application is subordinate to state objectives, raising alignment concerns with global democratic norms.

**D. International Organizations & Industry Self-Governance**

International organizations and industry players are actively shaping the normative landscape for quantum technologies through distinct, yet sometimes overlapping, efforts:

**NATO:** Identifying quantum technologies as a critical Emerging and Disruptive Technology (EDT) with possible paradigm-shifting impacts on defense and security,[58] NATO has adopted its first comprehensive Quantum Technologies Strategy.[3] Its core aim is to help the Alliance prepare for the quantum era by fostering and protecting quantum technologies for Allied security, ensuring NATO is "quantum-ready".[59] This involves nurturing a vibrant transatlantic quantum ecosystem through coordinated investment and close cooperation among Allies, partners, industry, and academia; leveraging quantum technologies (sensing, computing, communications) for defense applications; driving the crucial transition to quantum-safe cryptography; and actively shaping global norms and standards.[60] Guiding these efforts are principles of responsible development and use, aligned with Allied values, norms, and international law, reflecting NATO's broader approach to emerging technologies like AI. Initiatives supporting this strategy include the Science for Peace and Security (SPS) Programme, the Defence Innovation Accelerator for the North Atlantic (DIANA), and the NATO Innovation Fund.[61]

---

[55] *See also* Hodan Omaar and Martin Makaryan, How Innovative Is China in Quantum? ITIF, Sept. 9 2024, https://itif.org/publications/2024/09/09/how-innovative-is-china-in-quantum/ and Matt Swayne, *Report: China Is Challenging U.S. Leadership in Quantum*, THE QUANTUM INSIDER (Sept. 9, 2024), https://thequantuminsider.com/2024/09/09/report-china-is-challenging-u-s-leadership-in-quantum/

[56] *See e.g.* Imran Khalid, China's Ambitious Economic Roadmap For 2025: Innovation, Consumption, And Stability – Analysis, March 9, 2025, Eurasia Review, https://www.eurasiareview.com/09032025-chinas-ambitious-economic-roadmap-for-2025-innovation-consumption-and-stability-analysis/ and Hasan Muhammad, China's Innovation Roadmap 2025 - AI, Quantum Tech, 6G, China Economic Net, March 13, 2025, http://en.ce.cn/Insight/202503/13/t20250313_39318977.shtml

[57] *See e.g.* Ciel Qi, China's Quantum Ambitions: A Multi-Decade Focus on Quantum Communications, Yale Journal of International Affairs, Feb.2, 2024, https://www.yalejournal.org/publications/chinas-quantum-ambitions

[58] NATO Sci. & Tech. Org. [STO], Science & Technology Trends 2020-2040: Exploring the S&T Edge (Mar. 2020).

**OECD:** The Organisation for Economic Co-operation and Development (OECD) provides policy analysis and fosters strategic dialogue on quantum technologies through initiatives like its Global Forum on Technology. Its work examines the economic potential, societal contributions (e.g., towards SDGs), and associated risks (especially digital security) of quantum.[62] The OECD focuses on policy opportunities and challenges, including government support for quantum ecosystems, skills development, supply chain constraints, and the crucial role of international collaboration and anticipatory governance based on shared, human-centric values.

**UNESCO:** Through bodies like the World Commission on the Ethics of Scientific Knowledge and Technology (COMEST), UNESCO influences global discourse on the ethics of quantum computing.[63] It advocates for ethical guardrails based on human dignity, human rights, democratic accountability, international solidarity, and intergenerational equity. UNESCO emphasizes addressing environmental impacts, privacy and security risks (including the need for post-quantum cryptography), dual-use concerns, and ensuring equitable global access, positioning quantum development as a global public good aligned with the Sustainable Development Goals.

**World Economic Forum (WEF):** Acting as an international organization for public-private cooperation, the WEF convenes diverse global stakeholders through initiatives like its Global Future Council on Quantum Computing.[64] It facilitates dialogue on the responsible development and use of quantum technology, aiming to build trust and pre-empt risks. A key output is the co-designed "Quantum Computing Governance Principles" (published January 2022), which outline core values (e.g., common good, accountability, accessibility, non-maleficence) and thematic guidance on managing transformative capabilities, ensuring hardware access, fostering open innovation, creating awareness, developing the workforce, addressing cybersecurity and privacy, promoting standardization, and ensuring sustainability.[65]

---

[59] NATO Sci. for Peace & Sec., Quantum Technologies and the Science for Peace and Security Programme (Nov. 2023)
[60] NATO, Summary of NATO's Quantum Technologies Strategy (Jan. 16, 2024), https://www.nato.int/cps/en/natohq/official_texts_221777.htm.
[61] NATO, NATO'S Digital Transformation Implementation Strategy (Oct. 17, 2024), https://www.nato.int/cps/en/natohq/official_texts_229801.htm.
[62] OECD, *A Quantum Technologies Policy Primer*, OECD Digital Economy Papers No. 371 (Jan. 2025).
[63] UNESCO World Comm'n on the Ethics of Sci. Knowledge & Tech. (COMEST), https://www.unesco.org/en/ethics-science-technology/comest
[64] World Econ. F., Glob. Future Council on Quantum Computing,

**Industry Self-Governance:** Beyond formal state regulation, the tech industry itself increasingly functions as a co-regulator, particularly evident in AI governance and potentially extending to quantum. Driven by economic interests and the need to navigate fragmented regulatory landscapes ("techno-federalism"), tech firms shape the "rules of the road" through terms of use, technical standards embedded in hardware and infrastructure, and the design of software algorithms.[66] This includes developing voluntary responsible quantum principles or quantum impact assessments, exploiting jurisdictional differences and forum shopping, thereby acting as a third regulatory force alongside central and local authorities.[67]

## III. ARTIFICIAL INTELLIGENCE (AI) REGULATION IN THE U.S., EU & CHINA

This section lists emergent AI regulations in the U.S., EU, and China.

### A. U.S.

As of 2025, in the United States, there are several developments and considerations regarding the regulation of Artificial Intelligence (AI) at the federal and state level. These regulatory efforts reflect a growing recognition of AI's impact across various sectors and the need for oversight to ensure responsible development and deployment of AI technologies, while fostering innovation. Under the Trump Administration, regulatory efforts have shifted focus from oversight to removing barriers to AI innovation, emphasizing the elimination of restrictive policies in order to accelerate private sector development, safeguard US intellectual property, and strengthen America's global leadership in AI technologies.

1. **Securities and Exchange Commission (SEC) Proposed AI Rules**: The SEC proposed rules in 2023 to address conflicts of interest related to the use of AI by broker-dealers and investment advisers. These proposed rules are still pending.[68] In its 2025 examination priorities, the SEC announced it will focus on risks related to artificial intelligence, cybersecurity, and crypto assets in its oversight of regulated entities.[69]

---

[65] World Econ. F., Quantum Computing Governance Principles (Jan. 2022).
[66] Wu, Jason Jia-Xi, Techno-Federalism: How Regulatory Fragmentation Shapes the U.S.-China AI Race (February 14, 2025). 17 *Harv. Nat'l Sec. J.* __ *(forthcoming 2026), Available at SSRN:* https://ssrn.com/abstract=5138815 *or* http://dx.doi.org/10.2139/ssrn.5138815
[67] *See also* Mauritz Kop, *Quantum Technology Impact Assessment,* Futurium, European Commission, (Apr. 20, 2024), https://futurium.ec.europa.eu/en/european-ai-alliance/best-practices/quantum-technology-impact-assessment.
[68] U.S. Securities and Exchange Commission. (2023, July 26). SEC proposes new requirements to address risks to investors from conflicts of interest associated with the use of predictive data analytics by broker-dealers and investment advisers, see https://www.sec.gov/news/press-release/2023-140 and https://www.sec.gov/files/sec-ai-compliance-plan.pdf
[69] U.S. Sec. & Exch. Comm'n, 2025 Examination Priorities (Mar. 2025), https://www.sec.gov/file/2025-exam-priorities

2. **American Data Privacy and Protection Act (ADPPA)**: This proposed piece of U.S. federal legislation was aimed at establishing comprehensive national standards for data privacy and protection, but failed to pass in 2022. It embodied a significant effort to create a unified regulatory framework for the protection of personal information in the U.S., responding to growing concerns about privacy breaches, data misuse, and the need for more robust consumer protection in the digital age. The ADPPA sought to address the fragmented patchwork of state-level privacy laws with a federal standard that would provide clear rules for businesses and stronger rights for consumers.[70]

3. **Federal Artificial Intelligence Risk Management Act of 2023**: Introduced in the U.S. Congress, this bill aims to manage the risks associated with AI use by federal agencies. It includes guidance for compliance with the NIST AI Risk Management Framework, cybersecurity strategies for AI systems, and contract language requirements for AI procurement.[71]

4. **Federal Trade Commission (FTC) Rulemaking and Enforcement**: The FTC is using its existing authority under various consumer protection laws, such as the Fair Credit Reporting Act and the FTC Act, to expand AI enforcement. The commission authorized compulsory processes to address deceptive or fraudulent business practices involving AI, including Algorithmic Disclosure Rules.[72] The FTC has also issued an advance notice of proposed rulemaking to explore regulations for "automated decision-making systems," which may lead to a more holistic federal regulatory framework for AI.[73]

5a. **FDA AI medical Device Regulation**: The U.S. Food and Drug Administration (FDA) regulates AI-driven medical devices under its existing framework for Software as a Medical Device (SaMD). In 2021, the FDA introduced guidelines for AI/ML-based medical devices, requiring manufacturers to demonstrate safety, efficacy, and risk mitigation through premarket submissions. A key innovation is the Predetermined Change Control Plan, which allows iterative updates to AI models while maintaining regulatory compliance.[74]

5b. **FDA Quality System Regulation** (21 CFR Part 820) sets the minimum requirements for medical device manufacturers to establish and maintain a Quality Management System (QMS). This regulation applies to all medical device manufacturers who intend to market their products in the U.S. and includes requirements for the design, manufacture, packaging, labeling, storage, installation, and servicing of medical devices. For medical devices with AI, 21 CFR Part 820 requirements must be adapted to address the unique aspects of these devices, including cybersecurity and post-market monitoring. The FDA has recently amended 21 CFR Part 820 to align more closely with ISO 13485.[75]

---

[70] United States Congress. (2022). H.R.8152 - American Data Privacy and Protection Act. Congress.gov. https://www.congress.gov/bill/117th-congress/house-bill/8152.
[71] U.S. Congress, "Federal Artificial Intelligence Risk Management Act of 2023," https://www.congress.gov/bill/118th-congress/senate-bill/3205
[72] Federal Trade Commission, AI Companies: Uphold Your Privacy and Confidentiality Commitments, FTC (Jan. 9, 2024), https://www.ftc.gov/policy/advocacy-research/tech-at-ftc/2024/01/ai-companies-uphold-your-privacy-confidentiality-commitments.
[73] Federal Trade Commission. (2023, November 21). FTC authorizes compulsory process for AI-related products and services. https://www.ftc.gov/news-events/news/press-releases/2023/11/ftc-authorizes-compulsory-process-ai-related-products-services
[74] FDA, *Artificial Intelligence/Machine Learning (AI/ML)-Based Software as a Medical Device (SaMD) Action Plan* (2021), https://www.fda.gov/news-events/press-announcements/fda-releases-artificial-intelligencemachine-learning-action-plan

5c. Relatedly, **HIPAA (Health Insurance Portability and Accountability Act of 1996)** is a critical U.S. law governing the privacy and security of protected health information (PHI). While not AI-specific, HIPAA's Privacy and Security Rules directly impact AI systems used in healthcare, requiring safeguards for electronic PHI (ePHI) through encryption, access controls, and audit trails. For example, AI tools analyzing patient data must ensure compliance with HIPAA's Breach Notification Rule, mandating disclosure of unauthorized PHI exposures.[76]

6. **State-Level Initiatives**: Various states have created advisory councils or ordered state agencies to study and monitor AI use and develop policies accordingly, especially in the context of data privacy, employee rights and consumer protection. It can even be argued that states lead in U.S. AI regulation.[77] For example, Virginia's Artificial Intelligence Developer Act created operating standards for AI developers and deployers (this Act was passed by the House on February 20, 2025 but vetoed by Virginia Governor Glenn Youngkin[78]).[79] The Illinois' H.B. 3773 Biometric Data Act and NYC's AEDT (automated employment decision tools) Law aim to restrict the use of generative AI in HR to target workplace algorithmic discrimination.[80] California's 2024 legislative session saw the introduction of a comprehensive package of artificial intelligence bills, with one major bill passed and 17 others advancing through the legislature. These bills focus on regulating AI systems to ensure transparency, accountability, and the mitigation of algorithmic bias, particularly in sensitive areas such as employment, housing, and public services. Requirements include conducting impact assessments, providing disclosures about automated decision-making, and implementing safeguards to prevent discriminatory outcomes. The legislative approach aims to balance innovation with robust oversight, reflecting California's role as a policy leader in responsible AI governance.[81] The

---

[75] U.S. Food & Drug Admin., Quality Management System Regulation Final Rule: Amending the Quality System Regulation—Frequently Asked Questions, FDA (Feb. 2, 2024), https://www.fda.gov/medical-devices/quality-system-qs-regulationmedical-device-current-good-manufacturing-practices-cgmp/quality-management-system-regulation-final-rule-amending-quality-system-regulation-frequently-asked.

[76] HIPAA, Pub. L. No. 104-191, 110 Stat. 1936 (1996), https://www.hhs.gov/hipaa/for-professionals/privacy/laws-regulations/index.html

[77] Wu, Jason Jia-Xi, *supra* note 66.

[78] *See* the veto letter here: Virginia General Assembly, H.B. 2094, 2025 Gen. Assemb., Reg. Sess. (Va. 2025), https://lis.virginia.gov/bill-details/20251/HB2094/text/HB2094VG. See also: Governor of Virginia, Exec. Order No. 30 (Mar. 21, 2024), https://www.governor.virginia.gov/media/governorvirginiagov/governor-of-virginia/pdf/eo/EO-30.pdf.

[79] Virginia's General Assembly, "Artificial Intelligence Developer Act," Jan 2024, https://lis.virginia.gov/cgi-bin/legp604.exe?241+ful+SJ14+hil

[80] *See e.g.* Adam Aft et al., *Illinois Joins Colorado and NYC in Restricting Generative AI in HR*, The Employer Report (August 2024), https://www.theemployerreport.com/2024/08/illinois-joins-colorado-and-nyc-in-restricting-generative-ai-in-hr-a-comprehensive-look-at-us-and-global-laws-on-algorithmic-bias-in-the-workplace/

[81] *See* Myriah V. Jaworski & Ali Bloom, A View from California: One Important Artificial Intelligence Bill Down, 17 Others Good to Go!, CLARK HILL (Apr. 15, 2024), https://www.clarkhill.com/news-events/news/a-view-from-california-one-important-artificial-intelligence-bill-down-17-others-good-to-go/.

California Privacy Protection Agency (CPPA) recently proposed legislation to protect consumers' rights from artificial intelligence infused automated decision systems.[82]

7. **NIST AI Risk Management Framework (RMA)**: This framework, developed by the National Institute of Standards and Technology (NIST), is designed to manage risks associated with the design, development, use, and evaluation of AI products, services, and systems. It provides guidelines to ensure that AI is developed and deployed responsibly, ethically, and securely, aligning with broader national and economic security interests.[83]

8. **Executive Order 14110 on Safe, Secure, and Trustworthy AI (October 2023)**: This Executive Order directs federal agencies to develop and implement policies for AI systems that are safe, secure, reliable, and trustworthy. It focuses on promoting the development of AI in a manner that upholds civil liberties, privacy, and civil rights, and also encourages trust in AI technologies.[84] The EO directed the U.S. Patent and Trademark Office and Copyright Office to address AI-related intellectual property issues, such as inventorship and eligibility, encouraged the adoption of the NIST AI Risk Management Framework across government and industry, and outlined principles for an AI Bill of Rights to protect Americans from algorithmic bias and discrimination. This executive order represented the most comprehensive federal action on AI governance to date, aiming to balance innovation with safeguards for national security and individual rights. On January 20, 2025, the Trump administration removed these federal requirements, prioritizing deregulation and industry flexibility.[85]

9. **Intellectual Property Concerns**: The Patent & Trademark Office and the Copyright Office are currently developing guidance on AI-related intellectual property issues, such as AI Inventorship, patent eligibility, geostrategic IP risks, public domain and copyright implications of AI-generated content.[86] The Patent Office has been instructed to investigate these issues by (the now revoked) President Biden's October 2023 Executive Order on Safe, Secure, and Trustworthy AI.[87]

10. **Blueprint for an AI Bill of Rights**: Released by the White House, this document outlines a set of principles to protect the rights of Americans in the AI era. It aims to ensure that AI

---

[82] California Privacy Protection Agency. (2023, November 27). A new landmark for consumer control over their personal information: CPPA proposes regulatory framework for automated decision-making technology. https://cppa.ca.gov/announcements/2023/20231127.html
[83] National Institute of Standards and Technology. (2023). AI Risk Management Framework. https://www.nist.gov/itl/ai-risk-management-framework
[84] The White House. (2023, October 30). Executive Order on the Safe, Secure, and Trustworthy Development and Use of Artificial Intelligence. https://www.whitehouse.gov/briefing-room/presidential-actions/2023/10/30/executive-order-on-the-safe-secure-and-trustworthy-development-and-use-of-artificial-intelligence/
[85] *See also* Andrew Wilcox and Matt Dumiak, Security and Privacy - Takeaways From the AI Executive Order, CompliancePoint, November 7, 2023, https://www.compliancepoint.com/cyber-security/security-and-privacy-takeaways-from-the-ai-executive-order/
[86] U.S. Patent and Trademark Office. (2023, September 5). Copyright Office issues notice of inquiry on copyright and artificial intelligence. https://www.uspto.gov/subscription-center/2023/copyright-office-issues-notice-inquiry-copyright-and-artificial
[87] The White House. (2023, October 30). FACT SHEET: President Biden issues executive order on safe, secure, and trustworthy artificial intelligence. https://www.whitehouse.gov/briefing-room/statements-releases/2023/10/30/fact-sheet-president-biden-issues-executive-order-on-safe-secure-and-trustworthy-artificial-intelligence/

systems are used in a way that is equitable, accountable, and free from bias, and that they respect privacy, safety, and democratic values.[88]

11. **Artificial Intelligence Research, Innovation, and Accountability Act of 2023 (AIRIA)**: This act is focused on promoting research and innovation in AI while ensuring accountability in its development and deployment. It includes provisions for increased funding, ethical guidelines, and oversight mechanisms for AI technologies.[89] As of April 2025, the AIRIA act is pending.

12. **U.S.-UK Partnership on Science of AI Safety:** On April 1, 2024, the U.S. Department of Commerce announced a Memorandum of Understanding (MOU) with the UK, focusing on the science of AI safety. Signed by U.S. Commerce Secretary and the UK Technology Secretary, this partnership aims to develop and iterate evaluations for AI models, systems, and agents, sharing capabilities to ensure effective risk management. This collaboration follows commitments made at the 2023 AI Safety Summit and intends to foster a shared approach to advanced AI safety testing and research.[90]

13. **Executive Order 14179, Removing Barriers to American Leadership in Artificial Intelligence**: President Trump's January 2025 executive order, "Removing Barriers to American Leadership in Artificial Intelligence," revokes prior federal directives that imposed regulatory requirements on the development and deployment of AI in the United States.[91] The order instructs agencies to develop an Artificial Intelligence Action Plan to achieve the policy of the United States to sustain and enhance America's global AI dominance in order to promote human flourishing, economic competitiveness, and national security. The order emphasizes deregulation, instructing federal agencies to eliminate or revise existing rules that may hinder AI innovation and competitiveness. It directs agencies to prioritize American leadership by reducing compliance burdens, streamlining approval processes, and promoting private-sector investment in AI. Notably, the order directs federal agencies to review all actions they have taken under President Biden's 2023 AI Executive Order and to revoke, revise, or suspend any such actions that conflict with the policies set forth in President Trump's new executive order on artificial intelligence. It also calls for a reassessment of risk management and transparency mandates, shifting the federal approach from precautionary oversight to enabling rapid technological advancement and market-driven growth in AI. On January 24, 2025, the White House released a Fact Sheet that explains how 'President Trump Takes Action to Enhance America's AI Leadership'.[92]

---

[88] The White House Office of Science and Technology Policy. (October 2022). Blueprint for an AI Bill of Rights. The White House. https://www.whitehouse.gov/ostp/ai-bill-of-rights/
[89] U.S. Senate. (2023). S.3312 - 118th Congress (2023-2024): Artificial Intelligence Research, Innovation, and Accountability Act of 2023. Congress.gov. Retrieved February 2, 2024, from https://www.congress.gov/bill/118th-congress/senate-bill/3312?s=1&r=7
[90] U.S. Department of Commerce. (2024, April 1). U.S. and UK announce partnership on science of AI safety. U.S. Department of Commerce. https://www.commerce.gov/news/press-releases/2024/04/us-and-uk-announce-partnership-science-ai-safety
[91] See Exec. Order No. 14,179, Removing Barriers to American Leadership in Artificial Intelligence (Jan. 2025), https://www.whitehouse.gov/presidential-actions/2025/01/removing-barriers-to-american-leadership-in-artificial-intelligence/.
[92] White House, Fact Sheet: President Donald J. Trump Takes Action to Enhance America's AI Leadership (Jan. 24, 2025), https://www.whitehouse.gov/fact-sheets/2025/01/fact-sheet-president-donald-j-trump-takes-action-to-enhance-americas-ai-leadership/.

14. **M-25-21 and M-25-22 Federal Use and Procurement of AI Memoranda**: On April 7, 2025, the White House issued two revised AI policy memoranda, aligning federal AI governance with Executive Order 14179 (*Removing Barriers to American Leadership in Artificial Intelligence*). M-25-21 prioritizes accelerating AI adoption through innovation and risk management, mandating agencies to appoint Chief AI Officers, develop compliance plans, and implement safeguards for high-impact AI systems.[93] M-25-22 streamlines AI procurement by promoting competition, preventing vendor lock-in, and ensuring transparency in data rights and performance monitoring.[94] Both policies, revised in coordination with the Office of Science and Technology Policy (OSTP) and the President's Science Advisor, emphasize deregulation to enhance U.S. competitiveness while maintaining safeguards for privacy, civil liberties, and national security.

This continually evolving legal landscape exemplifies the broad and multifaceted approach of the U.S. federal government in regulating and guiding the development and use of AI technologies, with incremental steps being taken towards more comprehensive, broadly scoped regulation.

**B. EU**

Since 2016, the EU has been taking the lead in AI governance. As of 2025, the European Union is actively working on drafting and implementing various regulations concerning AI and related technologies. This also taps into the EU Strategy for Data, which aims to strengthen data protection, facilitate data sharing and re-use, and foster trust and innovation in the EU's digital economy, and the EU Digital Services package. The main initiatives are:

1. **AI Act**: Proposed by the European Commission and effective in 2024, this regulation aims to ensure the safety and fundamental rights of people and businesses while strengthening AI uptake, literacy, investment, and innovation across the EU. It categorizes AI systems based on their risk levels and sets out corresponding requirements. The EU AI Act manifests a significant step towards creating a regulatory environment that harnesses the benefits of AI while mitigating its risks.[95] The EU AI Act includes provisions for assessing risks posed by quantum-enhanced AI systems, particularly in cybersecurity applications.[96] The EU AI Act is cross-sectoral legislation which means requirements with existing QMS (quality management systems) may overlap, such as in healthcare. This implies, for example, that AI applications used in healthcare qualified as 'medical devices' and are also regulated under the EU Medical Devices Regulation (MDR) and the In Vitro Diagnostic medical devices Regulation (IVDR), with EU AI Act conformity assessments being embedded in MDR/IVDR assessment.[97] Following its August 2024 entry into force, rules on prohibited practices -including guidelines on AI system definition[98]- applied from February 2025, with governance rules for the AI Office and obligations for general-purpose AI models applying from August 2025, and high-risk system requirements phasing in through 2026-2027. The AI Office began issuing clarifying guidelines in early 2025.[99]

---

[93] *See* White House, White House Releases New Policies on Federal Agency AI Use and Procurement (Apr. 7, 2025), https://www.whitehouse.gov/articles/2025/04/white-house-releases-new-policies-on-federal-agency-ai-use-and-procurement/; White House, Fact Sheet: Eliminating Barriers for Federal Artificial Intelligence Use and Procurement (Apr. 7, 2025), https://www.whitehouse.gov/fact-sheets/2025/04/fact-sheet-eliminating-barriers-for-federal-artificial-intelligence-use-and-procurement/. and Off. of Mgmt. & Budget, Exec. Office of the President, Memorandum M-25-21, Accelerating Federal Use of AI through Innovation, Governance, and Public Trust (Apr. 3, 2025).
[94] Off. of Mgmt. & Budget, Exec. Office of the President, Memorandum M-25-22, Driving Efficient Acquisition of Artificial Intelligence in Government (Apr. 3, 2025)
[95] European Commission, "Proposal for a Regulation laying down harmonized rules on artificial intelligence (Artificial Intelligence Act) and amending certain Union legislative acts," COM(2021) 206 final, April 21, 2021. https://eur-lex.europa.eu/legal-content/EN/TXT/?uri=CELEX%3A52021PC0206
[96] For further reading on emerging legislation for quantum-AI Hybrids, see section III.D below.

2. **AI Liability Directive**: This September 2022 directive proposal focuses on harmonizing product liability rules for AI, making it easier for victims of AI-related damage to obtain compensation, while providing legal certainty to both developers, providers, and users of AI. It complements the EU's broader strategy on artificial intelligence, including the AI Act, by providing a framework to ensure that victims of harm caused by AI systems can obtain compensation.[100] Status: withdrawn by the EC on February 12, 2025.[101]

3. **Data Act**: While not exclusively about AI, this act is significant for AI as it governs data access, sharing, and use, particularly data generated by Internet of Things (IoT) devices, which are crucial for AI applications. The act aims to ensure fairness in the digital environment, stimulate a competitive data market, and boost data sharing across sectors and EU member states. It entered into force in January 2024.[102]

4. **General Data Protection Regulation (GDPR)**: Came into effect in the European Union (EU) on May 25, 2018. It establishes unified guidelines for the collection and processing of personal data of individuals within the EU. The GDPR has had a significant impact (extraterritorial effect) not only within the EU but globally, as it applies to any organization, regardless of location, that processes the personal data of individuals residing in the EU.[103]

---

[97] *See e.g.* European Commission, New Regulations for the Medical Devices Sector, HEALTH, https://health.ec.europa.eu/medical-devices-sector/new-regulations_en

[98] European Comm'n, Commission Publishes Guidelines on AI System Definition to Facilitate First AI Act's Rules Application (last updated Mar. 7, 2025), https://digital-strategy.ec.europa.eu/en/library/commission-publishes-guidelines-ai-system-definition-facilitate-first-ai-acts-rules-application

[99] European Comm'n, *European AI Office* (last updated Feb. 18, 2025), https://digital-strategy.ec.europa.eu/en/policies/ai-office

[100] European Commission, "Proposal for a Directive of the European Parliament and of the Council on adapting non-contractual civil liability rules to artificial intelligence," COM(2022) 495 final, September 28, 2022. https://eur-lex.europa.eu/legal-content/EN/TXT/?uri=CELEX%3A52022PC0495

[101] European Commission, Liability Rules for Artificial Intelligence, COMMISSION, https://commission.europa.eu/business-economy-euro/doing-business-eu/contract-rules/digital-contracts/liability-rules-artificial-intelligence_en

[102] European Commission, "Proposal for a Regulation of the European Parliament and of the Council on harmonized rules on fair access to and use of data (Data Act)," COM(2022) 68 final, February 23, 2022. URL: https://eur-lex.europa.eu/legal-content/EN/TXT/?uri=CELEX%3A52022PC0068

[103] European Parliament and Council of the European Union, "Regulation (EU) 2016/679 of the European Parliament and of the Council of 27 April 2016 on the protection of natural persons with

5. **Data Governance Act (DGA)**: Entered into force in September 2023, and focuses on data sharing across the EU, promoting data availability and supporting data sharing mechanisms. The DGA sets out rules and measures to encourage the reuse of certain categories of protected public sector data, improve data sharing mechanisms across the EU, and establish a trustworthy environment for data sharing.[104]

6. **Open Data Directive (ODD)**: Adopted in June 2019, the ODD is concerned with the re-use of public sector information, opening up access to publicly funded data. It required EU Member States to transpose the directive into national law by July 2021. The ODD aims to make public sector data more accessible and usable, thereby stimulating innovation, growth, and transparency for all stakeholders. [105]

7. **ePrivacy Regulation**: Complements GDPR, focusing on the confidentiality of electronic communications and personal data protection in the digital communications sector. The ePrivacy Regulation is designed to update and replace the ePrivacy Directive (Directive 2002/58/EC), and aims to modernize the rules for electronic communications to ensure higher levels of privacy protection for individuals and to harmonize data privacy laws across all EU Member states for electronic communications.[106]

8. **The Regulation on the Free Flow of Non-Personal Data**: As a complement to the GDPR, this regulation aims to facilitate the cross-border movement of non-personal data, enhancing data availability and fluidity across the European Union. It focuses on removing restrictions on data localization and improving the availability of data for businesses and public administrations, thereby contributing to the EU's single market for data.[107]

9. **Machinery Regulation**. The EU Machinery Regulation was adopted in May 2023 -replacing the Machinery Directive (2006/42/EC)- and is a major piece of legislation aimed at harmonizing the safety requirements for machinery within the European Union. This regulation is designed to ensure a high level of safety for machinery products being placed or made available on the EU market through a CE-marking, including market surveillance to ensure compliance, while also facilitating the free movement of these products within the single market. The Machinery

---

regard to the processing of personal data and on the free movement of such data (General Data Protection Regulation)," OJ L 119, May 4, 2016, p. 1–88. https://eur-lex.europa.eu/legal-content/EN/TXT/?uri=CELEX%3A32016R0679

[104] European Parliament and Council of the European Union, "Regulation (EU) 2022/868 of the European Parliament and of the Council of 30 May 2022 on European data governance (Data Governance Act)," OJ L 153, June 8, 2022, p. 1–28. https://eur-lex.europa.eu/legal-content/EN/TXT/?uri=CELEX%3A32022R0868

[105] European Parliament and Council of the European Union, "Directive (EU) 2019/1024 of the European Parliament and of the Council of 20 June 2019 on open data and the re-use of public sector information," OJ L 172, June 26, 2019, p. 56–83. https://eur-lex.europa.eu/legal-content/EN/TXT/?uri=CELEX%3A32019L1024

[106] European Commission. (2017). Proposal for an ePrivacy Regulation (Regulation of the European Parliament and of the Council concerning the respect for private life and the protection of personal data in electronic communications and repealing Directive 2002/58/EC). Digital Strategy. https://digital-strategy.ec.europa.eu/en/policies/eprivacy-regulation and https://digital-strategy.ec.europa.eu/en/library/proposal-regulation-privacy-and-electronic-communications

[107] European Parliament and Council of the European Union, "Regaplation (EU) 2018/1807 of the European Parliament and of the Council of 14 November 2018 on a framework for the free flow of non-personal data in the European Union," OJ L 303, November 28, 2018, p. 59–68. https://eur-lex.europa.eu/legal-content/EN/TXT/?uri=CELEX%3A32018R1807

Regulation also intends to better cover new technologies such as autonomous mobile machinery (robots), internet of things with connected equipment, or artificial intelligence (AI), where specific modules of AI using learning techniques ensure safety functions.[108]

10. **EU Digital Services Package: Digital Services Act (DSA) and Digital Markets Act (DMA)**: The EU Digital Services Package, enacted in July 2022, is a comprehensive set of reforms introduced by the European Commission to update and reshape the digital landscape within the European Union, particularly focused on online platforms owned by the large AI companies. The package consists of two main components: the Digital Services Act (DSA) and the Digital Markets Act (DMA).[109] These legislative proposals aim to create a safer digital space where the fundamental rights of users are protected and to establish a level playing field for businesses. The DMA took action in May 2023 and targets large online platforms acting as "gatekeepers" in the digital market and introduces prohibited practices. It aims to ensure that these platforms behave fairly and can be challenged by new entrants and existing competitors, ensuring consumer choice and innovation.[110] The DSA took action in August 2023 and intends to modernize the legal framework for digital services across the EU, focusing on online platforms and intermediary services. It introduces new obligations for digital services that act as intermediaries in their role of connecting consumers with goods, services, and content. The DSA's provisions for algorithmic transparency and accountability enhance and complement the other EU initiatives listed above such as the EU AI Act.[111]

11. **EU Code of Practice for General-Purpose AI (GPAI)**: The European Commission is in the final stages of developing comprehensive guidelines for General-Purpose AI (GPAI) through its forthcoming Code of Practice, which will operationalize the AI Act's rules for providers of GPAI models, including those with systemic risks.[112] The Code, facilitated by the European AI Office, is being drafted in a highly inclusive and transparent process involving nearly 1,000 stakeholders from industry, academia, civil society, and Member States, as well as international observers. Its core objectives are to ensure safe, trustworthy, and transparent AI by detailing obligations such as risk assessment, mitigation, transparency, and copyright compliance for GPAI providers. The process is iterative, with multiple drafting rounds, workshops, and consultations, and is expected to culminate in May 2025, nine months after the AI Act's entry into force.[113] The AI Office is also developing a standardized template for the summary of training data, directly linked to transparency and copyright requirements under Article 53(1)(d) of the AI Act. These guidelines are intended to set state-of-the-art practices for compliance, foster innovation without undue regulatory burden, and serve as a global benchmark, much like the GDPR has for data protection. The initiative reflects the EU's commitment to balancing innovation with robust safeguards, ensuring that GPAI development aligns with ethical, legal, and societal values, and that it remains beneficial for all stakeholders, including academia, industry, policymakers, and end-users.[114]

---

[108] European Agency for Safety and Health at Work. (2023). Regulation 2023/1230/EU - machinery. https://osha.europa.eu/en/legislation/directive/regulation-20231230eu-machinery and https://ec.europa.eu/commission/presscorner/detail/en/ip_22_7741
[109] European Commission. (2022). Digital Services Act Package. European Commission - Digital Strategy. https://digital-strategy.ec.europa.eu/en/policies/digital-services-act-package
[110] European Commission. (2022). The Digital Markets Act: ensuring fair and open digital markets. European Commission. https://commission.europa.eu/strategy-and-policy/priorities-2019-2024/europe-fit-digital-age/digital-markets-act-ensuring-fair-and-open-digital-markets_en
[111] European Commission. (2022). The EU's Digital Services Act. European Commission. https://commission.europa.eu/strategy-and-policy/priorities-2019-2024/europe-fit-digital-age/digital-services-act_en
[112] European Commission, General-Purpose AI Code of Practice, EUR. COMM'N (Dec. 10, 2024), https://digital-strategy.ec.europa.eu/en/policies/ai-code-practice
[113] Philipp Hacker, Responsible Generative AI, Saïd Business School, University of Oxford (Mar. 5, 2025), https://www.youtube.com/watch?v=l9cxkObineE

Together these initiatives represent a pioneering legislative effort to regulate artificial intelligence systems within the EU and beyond. To incentivize the desired behavior, fines for non-compliance can amount to up to 6% of annual turnover. Introduced to establish a legal framework for the development, deployment, and use of AI, the rules jointly aim to safeguard fundamental rights, respect the rule of law and ensure safety while fostering innovation and widespread adoption of AI technologies.

**C. China**

From 2017 onwards, China has been making significant strides in regulating advanced AI technologies, including generative AI models akin to ChatGPT. The general direction of Chinese AI regulation has consistently emphasized information control, data security, and the ethical use of AI technologies. This approach reflects China's proactive stance on managing the societal impacts of AI and ensuring that advancements in AI technologies align with national security and public interest objectives, including improving the Party and state supervision system. As of 2025, China has implemented several AI regulations, guidelines and initiatives, focusing on ethical development, responsible use, security, and information control:

1. **Next Generation Artificial Intelligence Development Plan** (July 2017) - This strategic plan outlines China's ambition to become a world leader in AI and quantum machine learning by 2030, focusing on ethical norms and policies for AI development.[115]

2. **Governance Principles for a New Generation of Artificial Intelligence: Develop Responsible Artificial Intelligence** (February 2019) - Issued by the National New Generation Artificial Intelligence Governance Specialist Committee, this document sets forth eight principles for the responsible development of AI in China, focusing on fairness, accountability, and transparency.[116]

3. **Ethics Guidelines for New Generation Artificial Intelligence** (December 2020) - Released by the Ministry of Science and Technology, these guidelines emphasize fairness, justice, and respect for privacy in AI development.[117]

---

[114] For in depth reading on this topic, see: Hacker, Philipp, and others (eds), *The Oxford Handbook of the Foundations and Regulation of Generative AI* (online edn, Oxford Academic, 22 Apr. 2025), https://doi.org/10.1093/oxfordhb/9780198940272.001.0001,

[115] Next Generation Artificial Intelligence Development Plan (July 2017), State Council of the People's Republic of China, http://www.gov.cn/zhengce/content/2017-07/20/content_5211996.htm

[116] Library of Congress. (2019, September 9). China: AI Governance Principles Released. https://www.loc.gov/item/global-legal-monitor/2019-09-09/china-ai-governance-principles-released/

[117] Center for Security and Emerging Technology. (2021). Ethical norms for new generation artificial intelligence released. CSET Georgetown University. https://cset.georgetown.edu/publication/ethical-norms-for-new-generation-artificial-intelligence-released/.

4. **Regulation on Recommendation Algorithms** (2020): Introduced by the Cyberspace Administration of China (CAC), requiring algorithms to be filed with a government registry and granting users the right to opt out of personalization.[118]

5. **Regulations on the Management of Automobile Data Security** (Trial) (April 2021) - This regulation, while specific to the automotive sector, indicates the broader concerns about data security and privacy in the era of AI and autonomous vehicles.[119]

6. **Cybersecurity Review Measures** (July 2021) - Updated cybersecurity review measures that include scrutiny of AI companies holding data that could pose a security risk if it falls under foreign control.[120]

7. **Personal Information Protection Law (PIPL)** (August 2021) - China's comprehensive privacy law that affects how AI companies can collect, use, and store personal data, ensuring ensure data security and privacy.[121]

8. **Data Security Law (DSL)** (August 2021) - This law provides a framework for data security and management, outlining requirements for AI companies regarding data collection, storage, and cross-border data transfer.[122]

9. **Provisions on the Management of Algorithmic Recommendations in Internet Information Services** (December 2021)[123]

---

[118] Mat Sheehan and Sharon Du, Carnegie Endowment for International Peace. (2022, December 9). What China's algorithm registry reveals about AI governance. https://carnegieendowment.org/2022/12/09/what-china-s-algorithm-registry-reveals-about-ai-governance-pub-88606.

[119] Mark Schaub, Atticus Zhao, Mark Fu, China Strengthens Regulating on Data Security in Auto Industry (May 26, 2021), https://www.chinalawinsight.com/2021/05/articles/corporate-ma/ma/china-strengthens-regulating-on-data-security-in-auto-industry/#page=1

[120] Stanford University DigiChina Project. (2022). Translation: Cybersecurity Review Measures (Revised) - Effective Feb 15, 2022. *DigiChina Stanford University*. https://digichina.stanford.edu/work/translation-cybersecurity-review-measures-revised-effective-feb-15-2022/.

[121] Stanford University DigiChina Project. (2021). Translation: Personal Information Protection Law of the People's Republic of China - Effective Nov 1, 2021. *DigiChina Stanford University*. https://digichina.stanford.edu/work/translation-personal-information-protection-law-of-the-peoples-republic-of-china-effective-nov-1-2021/

[122] Christian Perez (2022, January 28). China's Data Governance and Security Law Privacy. *Foreign Policy*. Retrieved from https://foreignpolicy.com/2022/01/28/china-data-governance-security-law-privacy/

[123] Internet Information Service Algorithmic Recommendation Management Provisions (promulgated by the Cyberspace Admin. of China, Ministry of Indus. & Info. Tech., Ministry of Pub. Sec., & State Admin. for Mkt. Regul., Dec. 31, 2021, effective Mar. 1, 2022) (China), *available at* China Law Translate, https://www.chinalawtranslate.com/en/algorithms/ and https://www.cac.gov.cn/2022-01/04/c_1642894606364259.htm

10. **Measures for the Security Assessment of Data Export** (January 2022) - These measures regulate the export of data, including data used in AI, requiring a security assessment for cross-border data transfers.[124]

11. **Guidelines for the Construction of a Standard System for the National Integrated Data Center** (March 2022) - Although not exclusively about AI, this document is relevant for AI development as it addresses data center standards, which are critical infrastructure for AI operations.[125]

12. **Guidelines on Strengthening the Ethical Governance of Artificial Intelligence** (July 2022) - Issued by the Cyberspace Administration of China, these guidelines aim to strengthen ethical governance over AI technologies, emphasizing the importance of AI serving people and promoting fairness and justice.[126]

13. **Rules for Deep Synthesis** (2022) Governs synthetically generated content, including images, video, audio, and text, to address the challenges posed by deepfake technologies.[127]

14. **Rules on Generative AI** (2023) Effective August 15, 2023, China's "Interim Measures for the Administration of Generative Artificial Intelligence Services" target developers and deployers of generative AI, reflecting China's proactive stance on new AI advancements like ChatGPT. These measures were published on July 13, 2023, by the Cyberspace Administration of China (CAC) in collaboration with six other governmental bodies, subsequent to draft measures circulated in April 2023.[128]

15. **AI Content Labeling Measures** (Mar 2025): Effective September 2025, these CAC measures, supported by mandatory national standard GB 45438-2025, command clear labeling of AI-generated content to ensure traceability, control, and combat misinformation, reflecting the states' priorities of ongoing 'Qinglang' campaign enforcement.[129]

---

[124] Anas Baig, "China: New Measures for Security Assessment of Data Exports," SECURITI Blog, available at https://securiti.ai/blog/china-new-measures-for-security-assessment-of-data-exports/

[125] Jian Xu, "China's National Data Bureau and Global Data Governance," Internet Policy Review, available at https://policyreview.info/articles/news/chinas-national-data-bureau-and-global-data-governance

[126] *See* Ministry of Foreign Affairs of the People's Republic of China, Position Paper of the People's Republic of China on Strengthening Ethical Governance of Artificial Intelligence (AI) (2022-11-17)**,** available at
https://www.mfa.gov.cn/eng/wjdt_665385/wjzcs/202211/t20221117_10976730.html and http://www.most.gov.cn/tztg/202107/t20210728_176136.html

[127] Provisions on the Administration of Deep Synthesis Internet Information Services, 2022-11-25, http://www.cac.gov.cn/2022-12/11/c_1672221949354811.htm. See also: *Administrative Provisions on Deep Synthesis Internet Information Services* (promulgated by the Cyberspace Admin. of China, Ministry of Indus. & Info. Tech., & Ministry of Pub. Sec., Nov. 25, 2022, effective Jan. 10, 2023) (China), *available at* China Law Translate, https://www.chinalawtranslate.com/en/deep-synthesis/

[128] *See* Carnegie Endowment for International Peace, *supra* note 51 and Cyberspace Administration of China, 生成式人工智能服务管理暂行办法 [Interim Measures for the Management of Generative Artificial Intelligence Services] (July 13, 2023), http://www.cac.gov.cn/2023-07/13/c_1690898327029107.htm and Cyberspace Administration of China, 国家互联网信息办公室关于《生成式人工智能服务管理办法（征求意见稿）》公开征求意见的通知 [Measures for the Management of Generative Artificial Intelligence Services] (11 April 2023)**.**

The initiatives listed herein signal China's rapid adaption to harnessing AI's potential including addressing privacy, data security, and ethical challenges, as well as ensuring that generative AI technologies do not produce or disseminate prohibited content, including misinformation, content that threatens national security, or violates public morality. China's three most concrete and impactful regulations on algorithms and AI are its 2021 regulation on recommendation algorithms, the 2022 rules for deep synthesis (synthetically generated content), and the 2023 draft rules on generative AI.[130] Information control is a central goal of all three measures, which can be illustrated by Beijing's swift response to ChatGPT through implementing new rules for developers and deployers of generative AI. Given China's market power, its AI regulations may have an extraterritorial effect comparable to the EU AI Act and the GDPR. Key actors proposing AI regulatory measures are the Chinese government and regulatory bodies such as the Cyberspace Administration of China (CAC), the Ministry of Industry and Information Technology (MIIT), the Ministry of Science and Technology (MOST), and the State Administration for Market Regulation (SAMR). Given the dynamic nature of AI policy and regulation, especially concerning cutting-edge technologies like generative and multimodal AI, one should expect these rules to be updated on regular basis, responding to newly arising challenges.[131]

**D. Quantum-AI Hybrids: Novel Legal-Technical Taxonomies**

Quantum-classical hybrid approach is the industry preferred term for combinations of quantum computational paradigms, and classical computing, be it software or hardware. Examples are quantum-inspired computing, hybrid quantum annealing, quantum algorithms, quantum data, and quantum-AI hybrids such as quantum machine learning (QML) and quantum artificial intelligence (QAI).[132]

Governing quantum-AI hybrid systems presents a complex challenge for existing legal regimes, prompting fundamental questions about regulatory applicability and scope.[133] It remains unclear whether current AI laws can adequately address the AI component while distinct quantum regulations govern the quantum part, or if a holistic framework encompassing the entire hybrid system, product, or service is required. While the principle of technology-neutral regulation aims for enduring, outcome-based rules, the unique synergistic effects and emergent properties of quantum-AI hybrids may challenge this approach, likely necessitating the development of novel legal and technical taxonomies to accurately classify these systems and delineate regulatory boundaries.[134] A further crucial question is whether regulations should primarily target the specific functionality, practical use case and application risks of the hybrid system – aligning with risk-based approaches common in AI governance – or if distinct rules are needed for the quantum base layer technology itself, especially as quantum mechanical effects become intrinsic to foundational technologies like beyond-nano scale semiconductors or quantum-centric computing architectures. Regardless of any horizontal quantum or AI laws, sector-specific regulations will invariably apply *per se*; for example, a quantum-AI hybrid used for medical diagnostics will be subject to rigorous healthcare and medical device laws focused on patient safety and efficacy, adding critical layers of domain-specific governance based on the application's context and impact.

---

[129] Yan Luo & Ken Dai, *China Releases New Labeling Requirements for AI-Generated Content*, Inside Privacy (Jan. 11, 2023), https://www.insideprivacy.com/international/china/china-releases-new-labeling-requirements-for-ai-generated-content/ and https://www.cac.gov.cn/2025-03/14/c_1743654684782215.htm (the 2025 AI Labeling Measures) and https://www.tc260.org.cn/front/postDetail.html?id=20250315113048 (AI standard GB 45438-2025).

[130] Hew Roberts and Emmie Hine, "The future of AI policy in China," East Asia Forum, September 27, 2023, available at https://eastasiaforum.org/2023/09/27/the-future-of-ai-policy-in-china/

[131] Mark MacCarthy, "The U.S. and its allies should engage with China on AI law and policy," Brookings (Brookings Institution), available at https://www.brookings.edu/articles/the-us-and-its-allies-should-engage-with-china-on-ai-law-and-policy/

[132] *See e.g.* Irie, H., Liang, H., Doi, T. *et al.* Hybrid quantum annealing via molecular dynamics. *Sci Rep* **11**, 8426 (2021). https://doi.org/10.1038/s41598-021-87676-z

[133] *See* Kop, Mauritz, *Establishing a Legal-Ethical Framework for Quantum Technology*, Yale Journal of Law & Technology, The Record, (Mar. 30, 2021), https://yjolt.org/blog/establishing-legal-ethical-framework-quantum-technology.

In addition to the AI and Quantum legislation listed above, the EU Chips Act -which entered into force on 21 September 2023- and the U.S. CHIPS Act of 2022 are relevant for quantum technologies -indirectly one might say- as they are aimed at bolstering the semiconductor industry, particularly when it comes to semiconductors used in quantum computing and hybrid quantum-classical computing. Both regulations have different scopes and implications:

**1. EU Chips Act**: The EU's initiative is primarily focused on strengthening the European semiconductor ecosystem, addressing supply chain resilience, and promoting research and development in advanced semiconductor technologies. This includes support for cutting-edge semiconductor technologies, which are crucial for the development of quantum computing. While the EU Chips Act is not specifically tailored towards quantum computing, the advancements and investments in semiconductor technologies under this act are likely to indirectly benefit the quantum computing sector, including hybrid quantum-classical systems.[135] 'Indicative of this link, the EU's Chips Joint Undertaking earmarked €500 million in March 2024 for integrated photonic chips for quantum-classical interfaces, while the US CHIPS Program Office guidance from February 2024 prioritized manufacturing capabilities for related hardware like superconducting control chips.[136]

**2. U.S. CHIPS & Science Act**: Similarly, the U.S. CHIPS & Science Act (or CHIPS for America Act) primarily focuses on revitalizing and securing United States' semiconductor supply chains.[137] It includes provisions for research and development, manufacturing incentives, and workforce development in the semiconductor industry. The U.S. CHIPS Act could indirectly benefit the quantum computing sector by fostering a robust semiconductor ecosystem -either domestically or through friend-shoring-, which is foundational for the development and production of quantum computing technologies. Investments in semiconductor research and manufacturing will inevitably lead to innovations that are applicable across quantum technology domains, including hybridized systems that combine quantum, AI and classical high-performance computing (HPC) elements.[138]

---

[134] Kop, Mauritz, *Regulating Quantum Technology*, Stanford Center for Responsible Quantum Technology, Stanford Law School (2022), https://law.stanford.edu/stanford-center-for-responsible-quantum-technology/projects/regulating-quantum-technology-2/.

[135] "European Chips Act," European Commission, available at https://commission.europa.eu/strategy-and-policy/priorities-2019-2024/europe-fit-digital-age/european-chips-act_en

[136] *See e.g.* Chips Joint Undertaking, *Chips JU entered negotiations to fund a Pilot Line for Advanced Photonic Integrated Circuits* (Nov. 7, 2024), https://www.chips-ju.europa.eu/News-detail/?id=fc89ef1a-109d-ef11-8a6a-7c1e52377187

[137] *See* "Fact Sheet: CHIPS and Science Act Will Lower Costs, Create Jobs, Strengthen Supply Chains, and Counter China," The White House, August 9, 2022, available at https://www.whitehouse.gov/briefing-room/statements-releases/2022/08/09/fact-sheet-chips-and-science-act-will-lower-costs-create-jobs-strengthen-supply-chains-and-counter-china/ and https://www.nist.gov/chips

In both cases, the acts are more broadly directed at advancing the semiconductor industry as a whole rather than specifically targeting quantum computing, sensing, simulation, or networking. Moreover, both semiconductors and quantum technologies can be thought of as base layer technologies. Given that cutting-edge semiconductor technology is a critical component of quantum computers and hybrid quantum-classical systems, it is however relevant to include them in our comparative analysis. The impact of these legislative acts on the various quantum technology domains will become more apparent as the policies and investments outlined within them are executed and evolve over time.

**E. International Frameworks for AI Governance**

The rapid proliferation of AI necessitates robust international governance. While a unified global regime is still evolving, a diverse landscape of influential frameworks has emerged, ranging from intergovernmental principles and recommendations to technical standards and legally binding treaties. This section provides an overview of key international initiatives shaping the global approach to AI governance, highlighting their objectives and core tenets in promoting responsible AI development and deployment worldwide.

**1. OECD AI Principles and Risk Classification Framework**

The OECD Recommendation on Artificial Intelligence, first adopted in 2019 and revised in 2024, is the first intergovernmental standard promoting trustworthy AI aligned with human rights and democratic values. These principles guide the responsible stewardship of AI systems, emphasizing inclusive growth, human-centred values, fairness, transparency, robustness, security, and accountability, while also providing recommendations for national policies and international co-operation.[139]

Published in 2022, the OECD Framework for the Classification of AI Systems provides a tool for policymakers, regulators, and others to characterize AI systems based on dimensions like context, data, model, and task/output. It facilitates a nuanced understanding of different AI applications and their potential impacts, thereby supporting risk assessment and the tailoring of governance approaches consistent with the OECD AI Principles.[140]

**2. UNESCO AI Ethics Recommendation**

---

[138] For the quantum specific parts of the U.S. CHIPS & Science Act, *see* Arcuri & Tomoshige, and & Basu supra note 10.
[139] Org. for Econ. Coop. & Dev. [OECD], *Recommendation of the Council on Artificial Intelligence*, OECD/LEGAL/0449 (adopted May 22, 2019; revised May 3, 2024), https://oecd.ai/en/ai-principles
[140] Org. for Econ. Coop. & Dev. [OECD], *OECD Framework for the Classification of AI systems*, OECD Digital Econ. Paper No. 323 (Feb. 22, 2022), https://oecd.ai/en/classification

Adopted by UNESCO's Member States in November 2021, the Recommendation on the Ethics of Artificial Intelligence serves as the first global standard-setting instrument specifically for AI ethics. It establishes an internationally agreed-upon framework based on universal values like human rights, environmental flourishing, diversity, and peace, along with core principles such as proportionality, fairness, transparency, and accountability. The Recommendation provides concrete policy action areas to guide Member States in translating these ethical principles into legislation and practice, ensuring AI development benefits humanity responsibly.[141]

**3. NIST AI RMF**

The NIST AI Risk Management Framework (AI RMF 1.0) of January 26, 2023, is a voluntary guidance document developed by the U.S. National Institute of Standards and Technology to help organizations identify, assess, manage, and govern risks associated with artificial intelligence systems throughout their lifecycle. It provides a structured approach organized around four core functions—Govern, Map, Measure, and Manage—aiming to cultivate trust in AI technologies and promote responsible AI innovation by improving the understanding and mitigation of potential negative impacts while maximizing benefits. The framework is intended to be adaptable across different sectors and organizational contexts.[142]

**4. UN AI Resolution**

In March 2024, the United Nations General Assembly unanimously adopted its first standalone resolution addressing artificial intelligence, titled "Seizing the opportunities of safe, secure and trustworthy artificial intelligence systems for sustainable development." This non-binding resolution represents a significant global consensus, promoting international cooperation to ensure AI is developed and governed in a way that respects human rights, protects personal data, bridges digital divides, and aligns with the Sustainable Development Goals. It encourages member states and stakeholders to foster safe, secure, and trustworthy AI systems for the benefit of all nations.[143]

**5. Council of Europe Framework Convention on AI**

The Council of Europe's Framework Convention on Artificial Intelligence, Human Rights, Democracy and the Rule of Law stands as the first international legally binding treaty specifically addressing AI. Adopted in May 2024 and opened for signature in September 2024, it establishes a legal framework requiring signatory states (including non-Council of Europe members) to ensure activities within the AI lifecycle comply with human rights, democracy, and the rule of law. While primarily focused on public sector activities, it encourages application to the private sector and sets key principles such as transparency, oversight, non-discrimination, and accountability, obliging Parties to implement effective risk management and remedies.[144]

---

[141] U.N. Educ., Sci. & Cultural Org. [UNESCO], *Recommendation on the Ethics of Artificial Intelligence* (adopted Nov. 23, 2021), https://unesdoc.unesco.org/ark:/48223/pf0000381137
[142] Nat'l Inst. of Standards & Tech., U.S. Dep't of Com., NIST AI 100-1, Artificial Intelligence Risk Management Framework (AI RMF 1.0) (Jan. 2023), https://nvlpubs.nist.gov/nistpubs/ai/NIST.AI.100-1.pdf.
[143] G.A. Res. 78/265, *Seizing the opportunities of safe, secure and trustworthy artificial intelligence systems for sustainable development*, U.N. Doc. A/RES/78/265 (Mar. 21, 2024), https://news.un.org/en/story/2024/03/1147831

Collectively, these international frameworks represent a concerted global effort to govern AI through the lens of Safeguarding, Engaging, and Advancing. They aim to Safeguard fundamental human rights, democratic processes, and the rule of law by establishing ethical principles, risk management guidelines (like NIST's RMF and OECD's Classification), accountability structures, and, in the case of the Council of Europe, legally binding obligations. Simultaneously, they foster Engagement by creating platforms for international cooperation (UN Resolution, OECD Principles), multi-stakeholder dialogue (UNESCO Recommendation), and the development of shared norms and standards, crucial for interoperability and trust. The initiatives seek to Advance responsible AI innovation by ensuring that the development and deployment of these powerful technologies align with societal values, contribute positively to sustainable development goals, and distribute benefits equitably across the globe.

## IV. EXPORT CONTROLS

Alongside direct regulation of AI systems, export controls serve as a critical, albeit distinct, governance tool for managing the risks associated with the international proliferation of sensitive technologies, including quantum and AI. After explaining the concepts of import, export, and trade controls, this section lists technology and rare minerals & earths related export controls implemented by the U.S., EU, and China. These controls primarily represent a 'Safeguarding' function aimed at protecting national security and economic interests, though their implementation carries significant trade-offs and risks that may exert counterproductive effects on global innovation and collaboration.

**Import controls** encapsulate the myriad governmental measures and policies implemented to oversee the influx of goods and services across a nation's frontiers from abroad. The purpose behind these regulatory mechanisms is multifaceted, aiming not only to shield local industries from overseas competition but also to safeguard national security, uphold public health and safety, and execute trade restrictions or embargoes. The nature of import controls is diverse, encompassing mechanisms like tariffs (taxes levied on imported goods), quotas (restrictions on the amount of certain goods that can enter), licensing conditions for importation, as well as stringent standards and regulations concerning the quality and safety of products.[145]

**Export controls** capture the range of limitations set by governments on the exportation of goods, technology, software and hardware, and information, anchored in the objectives of safeguarding national security, steering foreign policy, and protecting trade interests. The essence of these restrictions lies in their aim to prevent the transfer of critical technologies and commodities to parties that could threaten a nation's security or misuse these assets in ways that are at odds with the exporter's national interests, such as in the proliferation of weapons of mass destruction, acts of terrorism, or military confrontations. Key mechanisms within export controls include the necessity for licensing, mandating exporters to secure authorization prior to the shipment of regulated items, alongside embargoes or sanctions directed at particular nations, entities, or individuals. Most countries have export control licensing systems in place that include processes to support audits and acquire the licenses.[146] The extent and enforcement of export controls are subject to variation across different countries, mirroring their unique security priorities and obligations on the international stage. Quantum's dual use ambiguity, exemplified by China's integration of civilian and military sectors- creates additional challenges for the U.S. and EU in implementing effective export controls on dual-use quantum technology, as it blurs the distinction between civilian and military end-users and complicates risk assessments and enforcement. The main trade-offs of export controls are disruption of global supply lines, interoperability, market dynamism, and innovation.[147]

---

[144] Council of Europe Framework Convention on Artificial Intelligence, Human Rights, Democracy and the Rule of Law, *opened for signature* Sept. 5, 2024, C.E.T.S. No. 225, https://www.coe.int/en/web/artificial-intelligence/the-framework-convention-on-artificial-intelligence

[145] *See e.g.* Hauge, J. (2020). Industrial policy in the era of global value chains: Towards a developmentalist framework drawing on the industrialisation experiences of South Korea and Taiwan. *The World Economy*, *43*(8), 2070-2092, https://doi.org/10.1111/twec.12922

**Trade controls** refer to an expansive array of regulations and policies established to oversee and manage cross-border trade activities, covering the importation and exportation of goods and services. These regulatory measures in international commerce are instituted by governments with the intention of fulfilling various objectives related to economic stability, national security, diplomatic relations, and environmental conservation. The spectrum of trade controls spans tariffs, as well as non-tariff barriers like quotas, embargoes, sanctions, and licensing stipulations, all designed to curb the dissemination of weapons, safeguard intellectual property, and uphold environmental protocols. Such controls are typically formulated and implemented through negotiations within international trade agreements and are governed by the standards set forth by global trade bodies, including the World Trade Organization (WTO).[148]

**A. U.S. Export Controls**

As of 2025, the United States has enacted targeted export controls and regulatory measures concerning quantum technology, with an emphasis on safeguarding national security and preventing the dissemination of critical technologies to certain countries. As the UK, France, and Spain have recently introduced new export controls on quantum computers and QKD, the U.S., Canada, Japan, and South-Korea are expected to initiate new controls soon, on any quantum computing related export. This includes the technology to produce quantum computers, the software, hardware, cryogenics, physical qubits, and logical, error corrected qubits of any topology.[149] Central elements of these controls and regulations include:

1. **Executive Order on Investments in Certain National Security Technologies**: In August 2023, an Executive Order was signed by President Biden, empowering the Secretary of the Treasury with the authority to oversee U.S. investments in entities that are active in fields related to sensitive technologies, including quantum information technologies. This order specifically pinpointed the People's Republic of China as a country of concern. Following this order, the Department of the Treasury issued an Advanced Notice of Proposed Rulemaking (ANPRM) to further detail the scope of the program slated to regulate investments, introducing proposed definitions for key terms. The ANPRM document is set to undergo a public review and feedback process prior to its finalization and execution.[150]

---

[146] Such as SPIRE in the UK, see: https://www.spire.trade.gov.uk/spire/fox/espire/LOGIN/login
[147] *See e.g.* Khan, S. M. (2020). US Semiconductor Exports to China: Current Policies and Trends. *Washington, DC: Center for Security and Emerging Technology*, https://cset.georgetown.edu/publication/u-s-semiconductor-exports-to-china-current-policies-and-trends/
[148] *See e.g.* Fuller, D. B. (2021). China's counterstrategy to American export controls in integrated circuits. *China Leadership Monitor*, (67), https://papers.ssrn.com/sol3/papers.cfm?abstract_id=3798291
[149] At present, the annealers are not covered by these export controls.

Perhaps counterintuitively, studies show that Chinese overcapacity of legacy semiconductor chips poses a greater threat to the industry than export controls.[151] In light of geopolitical and economic implications of China's semiconductor strategy, the United States and its allies should use informed, data-driven trade policies to address future overcapacity and ensure supply chain resilience.

2. **Advanced Notice of Proposed Rulemaking (ANPRM):** The Treasury Department is considering regulations targeting highly specialized and cutting-edge quantum information technologies and products. These contemplated regulations envisage restrictions on transactions by U.S. persons with designated foreign entities involved in the development or use of quantum computers and their components, quantum sensors, as well as systems for quantum networking and communication. These systems are particularly scrutinized when designed for specific applications, including military operations, government intelligence activities, or secure communication channels. The ANPRM seeks public comment on these topics, signaling a phase of ongoing refinement and implementation of these investment controls.[152]

3. **Traffic in Arms Regulations (ITAR):** The Directorate of Defense Trade Controls (DDTC), under the U.S. Department of State, oversees the export and import of defense-related articles and services. Its responsibilities include ensuring that commercial exports of defense articles and defense services comply with national security and foreign policy objectives. This includes regulating arms sales, brokering services, and defense training, through the International Traffic in Arms Regulations (ITAR).[153] ITAR results in the US not being able to share its most advanced dual use cybersecurity, AI and quantum technology with its closest allies (not even with Tier 1 countries), such as the UK and Australia.

4. **BIS 2024-2025 Emerging Technology Controls:** The "Emerging Technology Controls Strategy: 2024 Framework Proposal" published by the U.S. Department of Commerce, Bureau of Industry and Security (BIS), outlines a new approach to regulating exports and reexports of certain emerging technologies. BIS in January 2025 acknowledged the limits of hardware-focused export controls as well as challenges in controlling quantum software/algorithms and proposed refining controls based on quantum application domains (e.g., cryptanalysis, military sensing) rather than just hardware specifications. The BIS framework aims to protect U.S. national security and foreign policy interests by controlling the movement of advanced technologies and updating the Entity List.[154] In January 2025 BIS introduced new Export Control Classification Numbers (ECCNs) to specifically identify and control certain emerging technologies, particularly those with dual-use capabilities (both commercial and military applications). BIS issued interim final rules significantly tightening controls on advanced computing ICs (expanding license requirements globally for some ECCN 3A090 items, with exceptions for allies/validated data centers) and introducing controls on 'frontier' AI model weights above a computational threshold (new ECCN 4E091) requiring licenses worldwide for closed-source models. The BIS framework strengthens existing controls and restrictions on end-users and end-uses of controlled technologies, particularly those related to military and intelligence applications and advanced technologies prone to diversion to the PRC, and expands controls on activities of U.S. persons.[155] BIS establishes a new framework within the Export Administration Regulations (EAR) to identify items for which controls are harmonized with the Implemented Export Controls (IEC) of international partners.[156] In addition, the strategy includes a specific focus on artificial intelligence (AI) technologies, particularly advanced computing integrated circuits (ICs) and AI model weights, aiming to cultivate secure ecosystems for responsible diffusion and use of AI.[157] The framework also updates the Data Center Validated End User authorization to facilitate the export, reexport, and transfer of advanced computing ICs to end users in destinations that do not raise national security or foreign policy concerns. The 2025 BIS export control framework represents a significant shift in how the U.S. government approaches export controls on emerging technologies, aiming to balance national security demands with the need to foster innovation and international cooperation.

---

[150] Office of Investment Security. (2023, August 14). Provisions pertaining to U.S. investments in certain national security technologies and products in specific foreign countries. *Federal Register*. https://www.federalregister.gov/documents/2023/08/14/2023-17164/provisions-pertaining-to-us-investments-in-certain-national-security-technologies-and-products-in

[151] Benson, E., Mouradian, C., & Alvarez-Aragones, P. (2024, April 5). *Evaluating chip overcapacity and the transatlantic trade tool kit*. Center for Strategic and International Studies. https://www.csis.org/analysis/evaluating-chip-overcapacity-and-transatlantic-trade-tool-kit

[152] *ibid.*

[153] Directorate of Defense Trade Controls. (n.d.). [Title of the Article]. Retrieved from https://www.pmddtc.state.gov/ddtc_public/ddtc_public?id=ddtc_kb_article_page&sys_id=24d528fddbfc930044f9ff621f961987

5. **AI Diffusion Framework and Foundry Due Diligence Rule:** Introduced by the Biden-Harris Administration in January 2025, the framework establishes guidelines for ethical AI deployment, emphasizing transparency, security, and accountability across sectors like healthcare and defense[158]. This U.S.-led regulatory approach aims to balance AI innovation with risk management. The Foundry Due Diligence Rule mandates rigorous compliance checks for AI developers, requiring audits of training data, algorithmic bias, and potential misuse before deployment[159]. These measures aim to foster public-private collaboration, aligning with ANSI's standards for interoperable AI systems while addressing risks like disinformation and privacy breaches[160]. Together, they prioritize responsible scaling of AI, ensuring competitiveness without compromising safety.

---

[154] Press Release, Bureau of Industry & Security, Commerce Makes Revisions to the Entity List to Strengthen U.S. National Security (Apr. 2025), https://www.bis.gov/press-release/commerce-makes-revisions-entity-list-strengthen-u.s.-national-security
[155] Press Release, Bureau of Industry & Security, Commerce Strengthens Restrictions on Advanced Computing Semiconductors, Enhances Foundry Due Diligence to Prevent Diversion (Mar. 2025), https://www.bis.gov/press-release/commerce-strengthens-restrictions-advanced-computing-semiconductors-enhance-foundry-due-diligence-prevent
[156] Bureau of Industry & Security, Export Administration Regulations (EAR), https://www.bis.gov/regulations/ear
[157] Press Release, Bureau of Industry & Security, Biden-Harris Administration Announces Regulatory Framework for Responsible Diffusion of Advanced Artificial Intelligence (Jan. 2025), https://www.bis.gov/press-release/biden-harris-administration-announces-regulatory-framework-responsible-diffusion-advanced-artificial

6. **China Technology Transfer Control Act**. This Act was proposed in February 2025 to further restrict exports of 'national interest technology' and related IP to China.[161] This proposed legislation seeks to block China from obtaining technology suitable for military activities, human rights violations, or cyber warfare by empowering the President to regulate exports of designated "covered national interest technology or intellectual property.

7. **Considerations of Effectiveness and Multilateral Support:** The effectiveness of the U.S. export controls on quantum technology could depend on garnering support and collaboration from multiple countries. Due to the competitive nature of quantum technologies globally, unilateral U.S. controls could inadvertently provide commercial incentives to foreign, non-U.S. firms to supply restricted technology to export controlled entities and regions. In addition, export controls could have counterproductive effects and (1 impede on fragile global quantum devices, rare earths, and critical mineral supply chains, and 2) incentivize systemic rivals to advance quantum technologies to become global leaders themselves, unintentionally making them self-sufficient instead of dependent on the U.S..[162] Policymakers might target specific applications or use cases of quantum technologies, such as quantum key distribution networks. However, distinguishing peaceful from military applications remains challenging as quantum technologies are inherently dual use. While optimizing quantum technology risk/benefit curves, the trade-offs of implementing export controls should be carefully studied, to prevent export controls from doing more harm than good.[163] From this it follows, that their scope should be constantly updated in accordance with changing geopolitical circumstances.[164]

---

[158] Gregory C. Allen & Will Hunt, The AI Diffusion Framework and the Foundry Due Diligence Rule: A Compliance Perspective, CSIS (Apr. 2025), https://www.csis.org/analysis/ai-diffusion-framework-and-foundry-due-diligence-rule-compliance-perspective.

[159] Bolstering AI Innovation with a Regulatory Framework for AI, ANSI (Jan. 21, 2025), https://www.ansi.org/standards-news/all-news/2025/01/1-21-25-bolstering-ai-innovation-regulatory-framework-ai.

[160] Press Release, Bureau of Industry & Security, Biden-Harris Administration Announces Regulatory Framework for Responsible Diffusion of Advanced AI (Jan. 2025), https://www.bis.gov/press-release/biden-harris-administration-announces-regulatory-framework-responsible-diffusion-advanced-ar.

[161] China Technology Transfer Control Act of 2025, H.R. 1122, 119th Cong. (2025), https://www.congress.gov/bill/119th-congress/house-bill/1122/text

[162] *See e.g.* Sujai Shivakumar, Charles Wessner, and Thomas Howell, *The Limits of Chip Export Controls in Meeting the China Challenge*, CSIS (Mar. 2025), https://www.csis.org/analysis/limits-chip-export-controls-meeting-china-challenge.

8. **Challenges in Regulating Quantum Technology**: The regulation of the suite of 2G quantum technologies presents intricate challenges, notably due to the uncertainties surrounding the eventual form that viable second quantum computers, sensors, and simulators will take and the lack of clear chokepoints for regulatory intervention. The situation is further complicated by considerations related to quantum-AI hybrids, technology readiness levels, and thorny issues surrounding intellectual property, national security, and fair competition.[165] Moreover, the United States encounters additional hurdles in synchronizing its regulatory framework with that of other countries deeply invested in quantum technology advancements. These nations might not share the same enthusiasm for imposing stringent controls that resonate an America First isolationist policy stance that emphasizes U.S. exceptionalism, thereby posing a risk of the U.S. ultimately falling behind in the global quantum technology race.

In sum, the U.S. strategy to export controls on quantum technology reflects a delicate balancing act between safeguarding national security and economic stability, and advancing technological innovation.[166] The complexity of regulating quantum technology pillars like computing, sensing, and networking arises from uncertainties about their future development, challenges in software and hardware regulation, especially open-source quantum software and critical materials and devices supply chains, and the need for international alignment on import, export, and trade control measures. As quantum technology continues to evolve, so too will the regulatory landscape, necessitating constant monitoring and adjustments of these controls to remain effective and relevant.

The U.S. strategy, heavily reliant on export controls and investment screening (E.O. 14105), clearly prioritizes 'Safeguarding' national interests. However, this approach risks disrupting global supply chains, hindering necessary international scientific 'Engagement', potentially incentivizing rivals toward self-sufficiency, and thus may prove counterproductive to long-term

---

[163] Gallagher, N., Rand, L., Entrikin, D., & Aoki, N. (2023, June 22). *The desirability and feasibility of strategic trade controls on emerging technologies*. Center for International and Security Studies at Maryland. https://cissm.umd.edu/research-impact/publications/desirability-and-feasibility-strategic-trade-controls-emerging

[164] Reinsch, W. A., Denamiel, T., & Schleich, M. (2024, February 14). *Optimizing U.S. export controls for critical and emerging technologies: Working with partners*. Center for Strategic & International Studies. https://www.csis.org/analysis/optimizing-us-export-controls-critical-and-emerging-technologies-working-partners.

[165] Kop, M. *et al. Towards Responsible Quantum Technology* (Harvard Berkman Klein Center for Internet & Society, 2023), https://cyber.harvard.edu/publication/2023/towards-responsible-quantum-technology. See also: Rand, L. E. (2023). *Schrödinger's technology is here and not: A socio-technical evaluation of quantum sensing implications for nuclear deterrence* [Doctoral dissertation, University of Maryland]. Digital Repository at the University of Maryland. https://doi.org/10.13016/dspace/ac9d-umv1

[166] Kop *et al.*, 10 Principles for Responsible Quantum Innovation, *Quantum Sci. Technol.* **9** 035013 (2024), https://iopscience.iop.org/article/10.1088/2058-9565/ad3776

'Advancing'. The focus on unilateral controls requires careful calibration to target high-risk applications effectively while minimizing collateral damage to responsible innovation.

**B. EU Export Controls**

By 2025, the European Union has taken significant steps to evaluate and implement export controls on critical technologies, including quantum computing. These measures are integral to a broader strategy aimed at safeguarding critical economic assets and essential technologies against geopolitical rivals, particularly in the context of evolving EU-China relations. Key components of this strategy comprise:

1. **Focus on Sensitive Technologies**: The EU has recognized quantum computing, sensing, and communication as crucial technologies that require protection from external geopolitical influences. In this light, the EC will further support EU technological sovereignty and resilience of EU value chains, by developing key technologies through Strategic Technologies for Europe Platform (STEP).[167] This set of measures is aimed at more rigorous screening of investments from abroad and stricter export controls, particularly concerning dual use technologies.[168]

2. **Risk Assessments and Proposed Rules**: The European Commission, in collaboration with EU member states, is undertaking collective, in depth risk assessments across several technology domains including quantum technologies.[169] The recommendation to carry out these assessments stem from a Joint Communication on a European Economic Security Strategy.[170] The results of these assessments will guide future regulatory initiatives, including potential access and trade restrictions. The Commission's focus is on de-risking critical mineral supply chains and economic relations with countries like China -using tools like export controls and investment screening- aiming to fortify the EU's economic security and strategic autonomy.[171]

3. **Coordinated Export Controls**: The EU is intensifying its efforts to prevent sensitive technologies from falling into the wrong hands by improving the coordination of export controls at the EU level. Preventing technology leakage involves the publication of a consolidated document featuring the national export control lists, enabling member states to require export authorization permits for items listed in the control lists of other member states. This approach is facilitated by the EU's Dual Use Regulation, allowing member states to coordinate and synchronize their export controls on items for which export controls have not yet been subject to multilateral export control agreements.[172]

---

[167] European Commission. (n.d.). Strategic technologies for Europe platform. Retrieved March 16, 2024, from https://commission.europa.eu/strategy-and-policy/eu-budget/strategic-technologies-europe-platform_en
[168] European Commission. (n.d.). Exporting dual-use items. Retrieved March 16, 2024, from https://policy.trade.ec.europa.eu/help-exporters-and-importers/exporting-dual-use-items_en
[169] European Commission. (2023, October 3). Commission recommends carrying out risk assessments on four critical technology areas: advanced semiconductors, artificial intelligence, quantum, biotechnologies. Retrieved from https://defence-industry-space.ec.europa.eu/commission-recommends-carrying-out-risk-assessments-four-critical-technology-areas-advanced-2023-10-03_en
[170] European Commission. (2023). An EU approach to enhance economic security. Retrieved from https://ec.europa.eu/commission/presscorner/detail/en/IP_23_3358
[171] European Commission. (n.d.). Making trade policy. Retrieved March 16, 2024, from https://policy.trade.ec.europa.eu/eu-trade-relationships-country-and-region/making-trade-policy_en

4. **Dual Use Coordination Group (DUCG):** In its January 2025 export control regime report, the European Commission details the initial implementation (covering 2022-2023) of the modernized EU Dual-Use Regulation (EU) 2021/821, which governs the control of dual-use item exports.[173] It marks the first annual report under this regulation, emphasizing increased transparency in export control licensing. The report outlines the evolution of the policy framework amid significant geopolitical changes, covering enhanced transparency guidelines, updates to the EU control list, a specific focus on controlling cyber-surveillance items, and extensive international cooperation, notably with the U.S. through the Trade and Technology Council. It presents aggregated 2022 data showing EUR 57.3 billion in authorized dual-use trade and 813 denials, and details the activities of the Dual-Use Coordination Group (DUCG) and its technical expert groups working on emerging tech, enforcement, and transparency.

5. **Research Security & Outbound Investment Monitoring:** In May 2024, the Council recommended enhancing research security, linking it to export controls for intangible tech transfers. Following this, in early 2025, the Commission recommended Member States monitor outbound investments in sensitive tech like quantum-AI to assess risks of technology leakage, mirroring US concerns.[174]

The EU's approach reflects a multilateral 'Safeguarding' strategy within the bloc, aiming for strategic autonomy and economic security through coordinated risk assessments and harmonized controls. Its effectiveness hinges on consistent implementation across all Member States. These developments echo the European Union's response to shifting geopolitical realities and the imperative of shielding critical technologies like quantum computing, sensing, and communication. The focus lies on finding an equilibrium between maintaining economic openness and the need to safeguard core strategic interests and security concerns.

---

[172] *Dual Use Export Controls. Summary of: Regulation (EU) 2021/821 Setting Up an EU Regime for the Control of Exports, Brokering, Technical Assistance, Transit and Transfer of Dual use Items* (European Commission, 2021); https://eur-lex.europa.eu/EN/legal-content/summary/dual-use-export-controls.html

[173] European Commission, *Report from the Commission to the European Parliament and the Council on the implementation of Regulation (EU) 2021/821 setting up a Union regime for the control of exports, brokering, technical assistance, transit and transfer of dual-use items*, COM(2025) 19 final (Jan. 30, 2025), https://eur-lex.europa.eu/legal-content/EN/TXT/PDF/?uri=CELEX:52025DC0019

[174] Press Release, European Comm'n, *Commission calls on Member States to review outbound investments and assess risks to economic security*, IP/25/261 (Jan. 15, 2025), https://ec.europa.eu/commission/presscorner/detail/en/ip_25_261

**C. China Export Controls**

As of 2025, China's regulations for the import, export, and trade of quantum technology are part of a broader framework that includes controls on various sensitive technologies including semiconductor production minerals and rare earths, laser radar technologies, and drones. The key aspects of these regulations include:

1. **Technology Import and Export Management Regulations**: The cross-border transfer of ownership rights and licenses to use technologies between domestic and international entities in China is governed by both export control law regimes and foreign trade law regimes, which specifically regulate the import and export of civilian and military technologies, and include the Catalogue of Dual-use Items and Technologies Subject to Import and Export Licensing.[175] For the import and export of technologies deemed restricted, authorization from provincial commerce departments is mandatory. This regulatory framework encompasses a routine review process for the import and export of restricted technologies, alongside a specialized approval procedure for the importation of such restricted technology.

2. **Catalogue for Prohibited and Restricted Export Technology**: China utilizes a structured categorization within its export control system, differentiating technologies as prohibited, restricted, or freely exportable. Items deemed prohibited are barred from export under all conditions. In contrast, exporting restricted items necessitates obtaining a specific license. Technologies tagged as 'free for export' are exempt from licensing requirements, albeit they must undergo contract registration management. This classification system is a central element of China's export strategy, aimed at safeguarding its economic interests and national security while promoting international economic and technological collaboration.[176]

3. **Recent Updates to Export Control Regulations**: The 2023 update to China's Catalogue for Prohibited and Restricted Export Technologies streamlined the list of technical items, removed specific technologies from the export ban and restriction lists, and introduced new categories, including those pertinent to AI and quantum technology. The update placed special emphasis on the regulation of technologies related to the extraction and refinement of rare earths.[177] This revision aligns with China's strategic aim to balance export control policies with its technological advancement goals.[178]

4. **Limiting Outbound Transfer of Intellectual Property Rights**: In addition to the standard trade control regulatory frameworks, China has specific rules concerning the limitation of outbound transfer of intellectual property rights in the export of restricted technologies, including copyrights, trade secrets, patents and integrated circuit layout designs.

---

[175] Ministry of Commerce and General Administration of Customs, Joint Announcement No. 66 (29 December 2023),
http://exportcontrol.mofcom.gov.cn/article/zcfg/gnzcfg/zcfggzqd/202312/941.html.
[176] Ministry Of Commerce People's Republic Of China,
Catalogue for Prohibited and Restricted Export Technology, (21 December 2023),
http://www.mofcom.gov.cn/zfxxgk/article/gkml/202312/20231203462079.shtml
[177] Sofia Baruzzi. (2023). China tightens control over management of rare earths. *China Briefing*. Retrieved from https://www.china-briefing.com/news/china-tightens-control-over-management-of-rare-earths/
[178] Lucia Brancaccio (2023). Technologies subject to export control in China: Prohibited & restricted export catalogue. *China Briefing*. Retrieved from https://www.china-briefing.com/news/technologies-subject-to-export-control-in-china-prohibited-restricted-export-catalogue/

China's export control regime, including its Catalogue and controls over critical raw materials, functions as a strategic tool for 'Safeguarding' domestic technological advantages and national interests, while also serving as a lever in response to international pressures and influencing global supply chains. The regimes reflect China's strategic approach to managing its technological advancements and safeguarding its interests in sensitive areas such as quantum technology. The evolving nature of these regulations indicates China's ongoing efforts to align its trade practices with its national security and technology driven economic development objectives.

**D. Rare Minerals & Earths**

As touched upon above, both the United States, the European Union, and China have various trade policies, controls and regulatory strategies that oversee the export of rare critical minerals, metals, and earths such as germanium gallium, graphite, lithium, nickel, cobalt, and helium-3 used in the production of quantum and classical computer chips.[179] Systematically evaluating the supply chain vulnerabilities for these materials is crucial. Analytical methodologies like the 'Quantum Criticality Index (QCI)' utilize data-driven techniques, including machine learning, to assess risks based on indicators such as geopolitical factors, substitutability, and import dependency, offering a quantitative measure of choke points for specific quantum materials like helium-3.[180]

**1. United States**:

According to the 2021 Department of Energy (DOE) Critical Materials Strategy, the guaranteed availability of crucial minerals and materials, along with the resilience of their supply chains, plays a vital role in underpinning the economic well-being and defense capabilities of the U.S..[181] The U.S. is strategically reducing its dependence on critical minerals sourced from countries it considers adversarial, like China. These minerals are crucial for both advanced defense applications, consumer goods, and green technology innovations. In 2022, the Biden-Harris Administration and U.S. industry announced major investments to expand domestic critical minerals supply chains, breaking dependence on China and boosting sustainable practices.[182] By advancing domestic mining and production, the U.S. aims to enhance national security and lessen its reliance on foreign, potentially unfriendly suppliers.[183] Export controls are utilized to regulate the availability of these critical minerals, particularly as a countermeasure to foreign export restrictions that might affect the U.S. 'Made in America supply chains'.[184]

---

[179] For an empirical analysis of rare earths and global supply chains in light of quantum technology, *see* Minha Lee, Quantum Criticality Index: Understanding Critical Raw Materials Supply Chains in Quantum Technologies. (Project). Stanford Center for Responsible Quantum Technology. (2024). Retrieved from https://law.stanford.edu/stanford-center-for-responsible-quantum-technology/projects/quantum-criticality-index/
[180] *Ibid.*
[181] U.S. Department of Energy. (2021). *2021 DOE critical materials strategy*. Office of Energy Efficiency & Renewable Energy. Retrieved from https://www.energy.gov/eere/ammto/2021-doe-critical-materials-strategy
[182] The White House. (2022, February 22). *Fact sheet: Securing a made-in-America supply chain for critical minerals*. Retrieved from https://www.whitehouse.gov/briefing-room/statements-releases/2022/02/22/fact-sheet-securing-a-made-in-america-supply-chain-for-critical-minerals/

**2. European Union**: The European Union has been actively working to secure its supply chains for vital resources, such as rare earth elements, through the establishment of joint ventures and collaborations aimed at the exploration and commercialization of minerals critical for diverse cutting-edge technologies. The 2024 Critical Raw Materials Act is designed to guarantee that the EU's industries can rival those of the United States and China in the production of clean technology products and in securing access to the essential raw materials required for these endeavors.[185] The EU's strategy is focused on diminishing its reliance on imported materials, particularly from regions with geopolitical sensitivities such as China and Russia, with the goal of securing a sustainable and dependable supply of crucial minerals. As the EU isn't a large producer of rare earths itself, it is concerned to become an export controlled region subject to export and trade restrictions on importing strategic metals from China.

**3. China**: From 2016 onwards, China has put in place export restrictions on a range of essential minerals -notably 17 rare earth elements for which mining concentrations are scarce globally- pivotal for numerous advanced technology sectors such as semiconductors.[186] These export limitations form a key part of China's approach to preserving its competitive advantage on the global stage. The measures involve constraints on the export of specific minerals, necessitating that purchasers engage in the procedure to secure export licenses/permits. From 2023 onwards, gallium and germanium require an export license to be outbound traded, too.[187] This continuously updated regime of export control stands as a critical component of China's wider economic and technology driven industrial strategy.

On April 4, 2025, China's Ministry of Commerce and General Administration of Customs imposed new export controls on seven of the seventeen rare earth elements—samarium, gadolinium, terbium, dysprosium, lutetium, scandium, and yttrium—along with their oxides, alloys, compounds, and related magnets.[188] Exporters must now obtain special licenses to ship these materials abroad, citing national security and non-proliferation as justifications for the move. According to the Center for Strategic & International Studies (CSIS), these restrictions, which follow U.S. tariff increases on Chinese products, are expected to disrupt global supply chains, particularly for U.S. defense, energy, and automotive sectors.[189] The licensing regime may cause temporary pauses in exports as it is implemented, and the inclusion of 16 U.S. defense and aerospace entities on China's export control list further limits their access to these critical dual-use materials. Given China's dominance in processing heavy rare earths, the U.S. and other countries will face significant challenges in replacing Chinese critical mineral supply in the near term.

---

[183] King, A. (January 19th, 2024). *King introduces bipartisan bill to reduce dependence on China for key minerals used in military technology, consumer goods*. Senator Angus King. Retrieved from https://www.king.senate.gov/newsroom/press-releases/king-introduces-bipartisan-bill-to-reduce-dependence-on-china-for-key-minerals-used-in-military-technology-consumer-goods

[184] For recommendations to establish trusted and certified, Made in America Quantum Technologies, *see* Kop, Mauritz, *Establishing a Legal-Ethical Framework for Quantum Technology*, Yale Journal of Law & Technology, The Record, (Mar. 30, 2021), https://yjolt.org/blog/establishing-legal-ethical-framework-quantum-technology.

[185] European Commission. (March 16, 2023). Critical Raw Materials Act (Proposal). Retrieved from https://single-market-economy.ec.europa.eu/sectors/raw-materials/areas-specific-interest/critical-raw-materials/critical-raw-materials-act_en and https://single-market-economy.ec.europa.eu/publications/european-critical-raw-materials-act_en

[186] Baruzzi, *supra* note 177. *See also* Ministry of Commerce and General Administration of Customs, Joint Announcement No. 39, 2023 (October 20, 2023), Announcement on Optimizing and Adjusting Temporary Export Control Measures for Graphite Items, http://www.mofcom.gov.cn/article/zcfb/zcdwmy/202310/20231003447368.shtml.

[187] Ministry of Commerce and General Administration of Customs, Joint Announcement No. 23 (July 3, 2023), http://www.mofcom.gov.cn/article/zwgk/gkzcfb/202307/20230703419666.shtml

[188] Ministry of Commerce of the People's Republic of China, Announcement No. 18 (Apr. 4, 2025), http://www.mofcom.gov.cn/article/b/c/202504/20250403467890.shtml

These U.S., EU, and China regulations and controls reflect each region's strategic approach to managing their resources and ensuring security (safeguarding, de-risking) in the supply of critical minerals and rare earth elements essential for high-tech and defense-related technologies such as AI and quantum technologies. However, this strategic competition over critical materials directly contributes to the 'Quantum Divide between Countries'.[190] Unequal access to essential resources, driven by geopolitical considerations and exacerbated by export controls, can create significant path-dependencies. Nations lacking secure supply chains for critical quantum materials may be locked out of certain technological trajectories, deepening global inequalities and hindering their ability to participate in or benefit from the second quantum revolution. Addressing these supply chain vulnerabilities and the resultant quantum divide requires not only national strategies but also robust international dialogue and cooperation to ensure more equitable access and shared prosperity.

**E. Wassenaar Arrangement**

The Wassenaar Arrangement (WA) is a key multilateral export control regime designed to promote transparency, responsibility, and accountability in the transfer of conventional arms and dual-use goods and technologies, with the aim of preventing their proliferation to unauthorized entities.[191] Established in 1996 as the successor to the Coordinating Committee for Multilateral Export Controls (COCOM), the WA has expanded to encompass 42 participating nations.[192] Characterized by its voluntary nature, with member states committing to semiannual reporting on transfers and denials of weapons and dual-use goods, thereby preventing destabilizing accumulations of these items in states of concern.

The arrangement is unique in its approach, not specifically targeting any state or region but maintaining a more geopolitically neutral stance, focusing instead on states of concern like Iran,

---

[189] Emily Benson & Matthew Reynolds, *The Consequences of China's New Rare Earths Export Restrictions*, CSIS (Apr. 14, 2025), https://www.csis.org/analysis/consequences-chinas-new-rare-earths-export-restrictions.

[190] Ayda Gercek & Zeki C. Seskir, Navigating the Quantum Divide(s), March 12, 2024, https://arxiv.org/abs/2403.08033

[191] Wassenaar Arrangement. (n.d.). *The Wassenaar Arrangement on Export Controls for Conventional Arms and Dual-Use Goods and Technologies*. Retrieved from https://www.wassenaar.org/

[192] *See e.g*. Nuclear Threat Initiative. (n.d.). Wassenaar Arrangement. Retrieved from https://www.nti.org/education-center/treaties-and-regimes/wassenaar-arrangement/

Iraq, Libya, and North Korea to avert the accumulation of weaponry and sensitive materials.[193] To be a member state, it is requisite to be a producer or exporter of arms or sensitive industrial equipment, maintain non-proliferation policies, and actively enforce robust export controls.[194]

The WA functions on a consensus basis among its members, representing a diverse array of states from different political and economic backgrounds globally. This framework prompts members to adopt national export control legislation that aligns with the arrangement's requirements. These include the establishment of licensing procedures, customs checks, and maintaining control lists for restricted items, notably the Munitions List and the Dual-Use List. These lists undergo routine revisions and updates by experts from WA member countries to mirror the evolving landscape of global security dynamics and technological progress, also in light of autonomous weapons, AI, nuclear, and quantum technologies.[195]

Despite its comprehensive framework and objectives, the WA has faced challenges, including the controversial membership of Russia, criticized for its partial compliance with the arrangement's objectives, and the absence of key arms exporters like Israel, China, and Belarus. In fact, most Wassenaar countries will align on quantum technology export controls, except for Russia. These issues highlight the complexity of achieving widespread compliance and the challenges of enforcing a regime that relies on the voluntary cooperation of its members. Recognizing the challenge, participating states agreed in December 2023 to establish a dedicated Quantum Technologies Experts Group tasked with developing appropriate control list entries and technical specifications for quantum technologies.

Over the years, the Wassenaar Arrangement's commitment to preventing unauthorized transfers of arms and dual-use items has adapted to the shifting dynamics of global security and the introduction of novel technologies. Notably, the 2013 revisions broadened the scope of technologies restricted for export to encompass Internet-based surveillance systems.[196] This amendment addressed concerns about human rights violations but also encountered opposition from technology corporations over government overreach.

In sum, the Wassenaar Arrangement constitutes a multilateral export control regime that includes 42 states. Its core mission is to enhance transparency and accountability in the transfer of conventional arms and dual-use goods and technologies, with a focus on preventing access by unauthorized parties. Enforcement is primarily the responsibility of each participating state, which is expected to implement national policies aligning with the Arrangement's guidelines. Member states are responsible for maintaining and periodically revising lists of restricted items, ensuring that exporters adhere to these regulations within their jurisdictions. Enforcement mechanisms typically involve national export control laws, licensing requirements, and customs checks.[197]

---

[193] *See e.g.* Bureau of Industry and Security. (n.d.). Multilateral Export Control Regimes. https://www.bis.doc.gov/index.php/policy-guidance/lists-of-parties-of-concern

[194] *See e.g.* Center for Arms Control and Non-Proliferation. (2023, March 8). Fact Sheet: The Wassenaar Arrangement. https://armscontrolcenter.org/fact-sheet-the-wassenaar-arrangement/

[195] For further reading on the strategic risks associated with advanced computing technologies, see: Rand, L. (2023, January). Reducing strategic risks of advanced computing technologies. *Arms Control Today*. Retrieved from https://www.armscontrol.org/act/2023-01/features/reducing-strategic-risks-advanced-computing-technologies

[196] Department of Commerce, Bureau of Industry and Security. (2015, May 20). Wassenaar Arrangement 2013 Plenary Agreements Implementation: Intrusion and Surveillance Items. *Federal Register*, 80(97), 25845-25853. Retrieved from https://www.federalregister.gov/documents/2015/05/20/2015-11642/wassenaar-arrangement-2013-plenary-agreements-implementation-intrusion-and-surveillance-items

While the Wassenaar Arrangement provides a multilateral framework for 'Safeguarding' via transparency and controls, its voluntary nature, consensus requirements, incomplete membership (notably China), and internal disagreements limit its effectiveness, particularly for rapidly evolving technologies like quantum. Adapting such existing regimes may prove more feasible than creating entirely new treaties.

## V. STANDARDIZATION

Moving from restrictive measures like export controls towards mechanisms that facilitate interoperability and safe deployment, technical standardization represents another crucial pillar of technology governance. This section provides a concise overview of emergent U.S., EU, and Chinese quantum technology standards. Standardization serves as a critical mechanism for both 'Advancing' innovation through interoperability and market creation, and 'Safeguarding' via embedded safety, security, and ethical requirements. Establishing key standards *ex ante*, before technologies become entrenched, is vital to avoid fragmentation and ensure alignment with desired values, learning from challenges faced in standardizing the internet.

### A. U.S. Standards

The development of technical interoperability standards for quantum technology in the United States is a collaborative effort that involves several organizations, including the International Organization for Standardization (ISO) and the Institute of Electrical and Electronics Engineers (IEEE). In addition, NIST, DARPA, QED-C, and leading Quantum Technology companies such as IBM, Google and Quantinuum play a role in quantum technology standardization.

**1. IEEE Standards for Quantum Technology**:

The IEEE is at the forefront of formulating standards for different facets of quantum technology.[198] This includes establishing benchmarks for quantum computing performance assessment, quantum communication, and quantum hardware. One notable example is IEEE P7130, designated as the "Standard for Quantum Computing Definitions."[199] This standard introduces a comprehensive glossary of terms and concepts, aiming to foster a unified vocabulary within the quantum computing domain. Furthermore, IEEE spearheads endeavors such as the Quantum Electronics Council and the Quantum Initiative, which are dedicated to

---

[197] *See e.g.* Arms Control Association. (2022, February). The Wassenaar Arrangement at a Glance. https://www.armscontrol.org/factsheets/wassenaar
[198] IEEE Standards Association. (n.d.). Endless possibilities with quantum computing. Retrieved from https://standards.ieee.org/beyond-standards/endless-possibilities-quantum-computing/
[199] IEEE Standards Association. (n.d.). P7130, Standard for Quantum Technologies Definitions. Retrieved from https://standards.ieee.org/ieee/7130/10680/

the development of various standards and the provision of educational materials pertaining to quantum technology.[200] Examples of IEEE standards for quantum technology are:

IEEE P1913 Software-Defined Quantum Communication[201]

IEEE P1943 Standard for Post-Quantum Network Security[202]

IEEE P2995 Trial-Use Standard for a Quantum Algorithm Design and Development[203]

IEEE P3120 Standard for Quantum Computing Architecture[204]

IEEE P3172 Recommended Practice for Post-Quantum Cryptography Migration[205]

IEEE P3185 Standard for Hybrid Quantum-Classical Computing[206]

IEEE P3329 Standard for Quantum Computing and Simulation Energy Efficiency[207]

IEEE P7130 Standard for Quantum Technologies Definitions[208]

IEEE P7131 Standard for Quantum Computing Performance Metrics & Performance Benchmarking[209]

The IEEE Quantum Week 2024 Workshop on operationalizing responsible quantum computing expanded IEEE's quantum standardization efforts beyond technical parameters to include ethical frameworks for quantum technology and examining the societal impacts of quantum technologies, with a focus on quantum computing.[210]

**2. ISO Standards for Quantum Technology**:

The International Organization for Standardization (ISO) is a global standard-setting body that works in concert with its member nations such as the United States, to develop international technical standards. Although ISO's involvement in quantum technology hasn't reached the breadth found in other fields, there's a growing momentum towards standardizing facets of quantum technologies, particularly as they edge closer to practical and commercial viability.

---

[200] IEEE Quantum. (n.d.) Retrieved from https://quantum.ieee.org/
[201] IEEE P1913 Software-Defined Quantum Communication
[202] IEEE P1943 Standard for Post-Quantum Network Security
[203] IEEE P2995 Trial-Use Standard for a Quantum Algorithm Design and Development
[204] IEEE P3120 Standard for Quantum Computing Architecture
[205] IEEE P3172 Recommended Practice for Post-Quantum Cryptography Migration
[206] IEEE P3185 Standard for Hybrid Quantum-Classical Computing
[207] IEEE P3329 Standard for Quantum Computing and Simulation Energy Efficiency
[208] IEEE P7130 Standard for Quantum Technologies Definitions
[209] IEEE P7131 Standard for Quantum Computing Performance Metrics & Performance Benchmarking
[210] National Quantum Computing Centre, IEEE Quantum Week 2024: Workshop on Operationalising Responsible Quantum Computing, NQCC BLOGS (Mar. 25, 2024), https://www.nqcc.ac.uk/blogs/ieee-quantum-week-2024-workshop-on-operationalising-responsible-quantum-computing/.

Given the early stage of second-generation quantum technology's development, ISO's contributions are focused initially on establishing standards for terminology, measurement methods, and interoperability, paving the way for a more unified approach as the technology matures.

As of 2025, the main ISO initiatives, norms and standards for AI & Quantum are:

a. The International Electrotechnical Commission (IEC) and the International Organization for Standardization (ISO) have established a new Joint Technical Committee, JTC 3, focusing on standardization in the field of quantum technologies.[211] This committee aims to develop global standards for quantum technologies, addressing their rapidly evolving nature and anticipated impacts across various sectors. JTC 3 reflects the growing significance of quantum technologies and the need for a unified approach to standardization to ensure their responsible, safe, ethical, and effective use worldwide.

b. ISO 42001, the certifiable framework for an AI Management System (AIMS), is designed to offer organizations a comprehensive guideline for the responsible, ethical, and efficacious management and deployment of AI systems. It provides a framework for organizations to develop and manage AI systems responsibly. The standard outlines a methodical approach to tackle the challenges and risks inherent in AI technologies, emphasizing aspects such as transparency, accountability, and governance. It emphasizes the need for AI governance, impact assessment, and continuous improvement in AI system management across various industries. AIMS intends to ensure that AI systems are aligned with ethical principles and legal requirements, promoting trust and confidence in AI applications. By adhering to ISO 42001, organizations can navigate the intricacies of AI technology, certifying that its advantages are leveraged in a way that upholds human values and rights.[212]

c. ISO 23894, titled Information technology - Artificial intelligence - Guidance on risk management, offers a framework for organizations developing, deploying, or using AI systems to manage AI-specific risks throughout the system's lifecycle. It adopts established risk management principles to address unique AI challenges, such as fairness, bias, transparency, and security. The standard provides guidance on identifying, assessing, evaluating, and treating these risks to promote responsible AI adoption.[213]

d. The ISO/IEC standards for quantum technology, such as JTC 3, FDIS 4879, ISO/IEC 23837-1:2023, ISO 9001, and ISO/IEC JTC 1/WG 14, offer comprehensive guidelines and specifications across various facets of quantum computing and technology. These standards cover critical aspects including terminology, measurement methodologies, and performance metrics, establishing a foundational framework for the advancement, deployment, and assessment of quantum technologies. They play a crucial role in guaranteeing interoperability, reliability, and efficiency within the swiftly progressing domains that constitute the family of quantum technologies.

---

[211] International Organization for Standardization. (January 11, 2024). IEC and ISO launch new joint technical committee on quantum technologies. Retrieved from https://www.iso.org/news/new-joint-committee-quantum-technologies
[212] Koerner, K. (2023, December 19). ISO/IEC 42001: A Leap Forward in Responsible AI Management. *Daiki*. https://dai.ki/blog/iso-iec-42001-a-leap-forward-in-responsible-ai-management/, and ISO/IEC 42001:2023 Information technology — Artificial intelligence — Management system https://www.iso.org/standard/81230.html
[213] ISO/IEC 23894:2023, Information Technology – Artificial Intelligence – Guidance on Risk Management (2023), https://www.iso.org/standard/77304.html

**JTC 3**: This Joint Technical Committee, under the umbrella of ISO/IEC, develops standardization frameworks for the field of quantum technology, including not limited to quantum computing, quantum simulation, quantum sources, quantum metrology, quantum detectors and quantum communications.[214]

**FDIS 4879**: This standard is likely related to specific aspects of quantum computing or technology, focusing on setting guidelines or best practices in this nascent field.[215]

**ISO/IEC 23837-1**: This is a standard related to quantum computing, focusing on specific aspects of quantum technology.[216]

**ISO 9001**: This is a widely recognized standard for quality management systems. It provides guidelines for companies to ensure they meet customer and regulatory requirements and demonstrate continuous improvement.[217]

**ISO/IEC JTC 1/WG 14**: This refers to a working group within the Joint Technical Committee of ISO and IEC, which focuses on specific aspects of information technology, including quantum computing.[218]

These standards collectively contribute to establishing guidelines for best practices, quality management, and technological development in various fields. The initiatives listed above are instrumental in developing standardized approaches and methodologies in the rapidly evolving domain of AI and quantum technology including quantum-classical hybrids, ensuring compatibility, efficiency, and responsible use of these advanced technologies.

**3. NIST's Role in Quantum Standards**:

Though distinct from standard-setting bodies such as ISO or IEEE, the National Institute of Standards and Technology (NIST) in the U.S. plays a crucial role in formulating standards for quantum technology. NIST's contributions encompass research in quantum measurement and quantum information theory, which lay the groundwork for the creation of applicable standards in quantum technology. In addition, NIST is actively working on leveraging nanotechnology to bring the potential of quantum computing into everyday applications, bridging the gap between advanced quantum research and practical, real-world technology.[219] Moreover, NIST is at the forefront of developing post-quantum cryptographic standards (PQC), which are vital for cybersecurity in the quantum computing era.[220]

---

[214] ISO/IEC JTC 3 Quantum technologies, https://www.iso.org/committee/10138914.html
[215] ISO/IEC FDIS 4879 Information technology Quantum computing Vocabulary, https://www.iso.org/standard/80432.html
[216] ISO/IEC 23837-1:2023 Information security Security requirements, test and evaluation methods for quantum key distribution, https://www.iso.org/standard/77097.html
[217] ISO 9001 Quality management systems, https://www.iso.org/standards/popular/iso-9000-family and https://www.iso.org/standard/62085.html
[218] ISO/IEC JTC 1/WG 14 Quantum Information Technology, https://jtc1info.org/sd-2-history/jtc1-working-groups/wg-14/
[219] National Institute of Standards and Technology. (n.d.). Quantum information science. Retrieved from https://www.nist.gov/quantum-information-science

**4. The four standardized NIST quantum-resistant cryptography algorithms**

On July 5, 2022, the National Institute of Standards and Technology (NIST) announced the first standardized set of four quantum-resistant cryptographic algorithms as part of its Post-Quantum Cryptography (PQC) standardization process. This development marked a significant step in the NIST post-quantum cryptography standardization project. These four algorithms - developed through a six-year global effort- are designed to protect sensitive electronic information against future attacks from quantum computers, which could in theory break the encryption systems currently in use. In layman's terms, the chosen algorithms cater to general encryption and digital signatures, ensuring robust security for online transactions and communications in a post-quantum era.[221] Following the 2022 selection of CRYSTALS-Kyber, CRYSTALS-Dilithium, FALCON, and SPHINCS+, NIST concluded its fourth round in late 2024/early 2025, selecting HQC as an additional KEM standard. This nears completion of the initial standards suite, with focus shifting to implementation guidance and migration strategies as organizations were urged in 2025 to prepare actively for a quantum-secure environment.

1. CRYSTALS-Kyber: This algorithm is selected for general encryption, which is typically used to protect information exchanged across public networks, such as accessing secure websites. CRYSTALS-Kyber is known for its comparatively small encryption keys, easy exchange between parties, and speedy operation.

2. CRYSTALS-Dilithium: Chosen for digital signatures, which are commonly used for verifying identities in digital transactions or for remotely signing documents. CRYSTALS-Dilithium is recognized for its efficiency and is recommended as the primary algorithm for digital signatures.

3. FALCON: Also selected for digital signatures, FALCON is an alternative to CRYSTALS-Dilithium for applications requiring smaller signatures. It offers a balance between efficiency and compactness.

4. SPHINCS+: This algorithm is a backup option for digital signatures, notable for its different mathematical approach compared to the other selections. SPHINCS+ is larger and slower but provides a diverse option in the cryptographic arsenal.

---

[220] National Institute of Standards and Technology. (2022, July 5). NIST announces first four quantum-resistant cryptographic algorithms. Retrieved from https://www.nist.gov/news-events/news/2022/07/nist-announces-first-four-quantum-resistant-cryptographic-algorithms
[221] National Institute of Standards and Technology. (2023, August 24). NIST to standardize encryption algorithms that can resist attack by quantum computers. Retrieved from https://www.nist.gov/news-events/news/2023/08/nist-standardize-encryption-algorithms-can-resist-attack-quantum-computers

5. HQC: In March 2025, the U.S. National Institute of Standards and Technology (NIST) announced the selection of the HQC algorithm as its fifth standard for post-quantum cryptography, designated specifically as a backup for key establishment (general encryption). Chosen for its different mathematical foundation (error-correcting codes) compared to the primary standard ML-KEM, HQC provides cryptographic diversity, with a draft standard expected around early 2026 and finalization anticipated in 2027.[222]

These algorithms were chosen after a comprehensive evaluation and are based on different families of mathematical problems, such as structured lattices and hash functions, which are believed to be resilient to attacks from both quantum and conventional computers. This selection methodology incorporated feedback from worldwide cryptography specialists, striving to furnish a solid toolkit adaptable to diverse encryption systems and operational requirements. NIST's selection marks a significant step in advancing cybersecurity in the face of evolving quantum computing capabilities. The focus is not only on maintaining security during the advent of quantum computers but also on ensuring a variety of defense tools to cater to different scenarios and needs.

**5. QED-C's Technical Advisory Committee (TAC) on Standards and Performance Metrics**

The Quantum Economic Development Consortium (QED-C), overseen by SRI International, represents a collaborative effort of stakeholders focused on facilitating the growth and development of the quantum industry. Initiated with backing from the National Institute of Standards and Technology (NIST) as a component of the Federal government's strategy to promote quantum information science, QED-C's foundation was a directive of the National Quantum Initiative Act, which was passed in 2018.[223] The QED-C Technical Advisory Committee (TAC) on Standards and Performance Metrics plays a crucial role in pinpointing standards and metrics poised to expedite the commercialization of quantum-based products and services. This TAC's pioneering work contributes to the development of benchmarks that assess the performance of quantum algorithms across various real-world applications, highlighting improvements in benchmarking methodology and the inclusion of new algorithms for a comprehensive evaluation.[224] The TAC further facilitates connections between its members and pertinent standards development organizations globally, enhancing collaboration and integration within the quantum industry.[225]

**6. DARPA Quantum Benchmarking Initiative (QBI):**

---

[222] Nat'l Inst. of Standards & Tech., *NIST Selects HQC as Fifth Algorithm for Post-Quantum Encryption* (Mar. 11, 2025), https://www.nist.gov/news-events/news/2025/03/nist-selects-hqc-fifth-algorithm-post-quantum-encryption
[223] Quantum Economic Development Consortium. (n.d.). Our mission. Retrieved from https://quantumconsortium.org/#our-mission
[224] Lubinsky, T., Goings, J. J., Mayer, K., Johri, S., Reddy, N., Mehta, A., ... & Mundada, P. S. (2024). Quantum Algorithm Exploration using Application-Oriented Performance Benchmarks. *arXiv preprint arXiv:2402.08985*. Retrieved from https://arxiv.org/abs/2402.08985
[225] Quantum Economic Development Consortium. (n.d.). Technical Advisory Committees (TAC). Retrieved from https://quantumconsortium.org/tac/

The Defense Advanced Research Projects Agency's (DARPA) Quantum Benchmarking Initiative (QBI) aims to establish rigorous performance metrics for quantum computing systems, focusing on industrially relevant applications to accelerate the transition from theoretical research to practical implementations.[226] By developing standardized benchmarks, QBI seeks to address the lack of consensus in evaluating quantum hardware and algorithms, enabling objective comparisons across platforms and fostering interoperability—a critical step toward scalable quantum technologies.[227] This standardization effort aligns with DARPA's broader strategy to partner with private-sector entities advancing quantum computing for defense and commercial applications, as evidenced by its engagement with companies targeting industrially useful systems. This public-private collaboration is structurally enabled by Government Purpose Rights (GPR), which reconcile federal oversight with commercial incentives through tailored IP provisions.

Government Purpose Rights (GPR), defined under 48 C.F.R. § 227.7203-5, govern intellectual property (IP) developed under mixed-funded U.S. government contracts.[228] These rights permit the government to use, modify, and share patented inventions or trade secrets for official purposes while allowing contractors to retain commercial exploitation rights for a negotiable period (typically five years). For quantum technologies developed through initiatives like QBI, GPR ensures DARPA can leverage innovations for national security objectives without undermining private-sector incentives to invest in dual-use applications. Post-GPR expiration, the government gains unlimited rights, creating a balance between public access and proprietary control.

In the AI domain, NIST initiated the AI Standards "Zero Drafts" pilot project in early 2025. This aims to accelerate standards development by generating preliminary stakeholder drafts on key topics for submission to formal SDOs, seeking broader input and faster response to AI's rapid evolution.[229]

The U.S. standardization model, combining industry initiatives (IEEE, QED-C) with government R&D and guidance (NIST, DARPA), fosters innovation and aims for performance/security benchmarks. However, this reliance on market dynamics and consortia risks fragmentation and slower consensus compared to more centralized approaches. Concludingly, the constructive efforts by IEEE, ISO, NIST, DARPA, and QED-C described in this section represent a growing recognition of the need for standardized definitions, measurements, and practices in quantum technology to facilitate its development, commercialization, and integration into existing technological systems. As the field of quantum technology rapidly evolves, we can expect these and other standard-setting bodies to expand their scope and depth of quantum technology standards.

---

[226] Defense Advanced Research Projects Agency, *QBI: Quantum Benchmarking Initiative*, https://www.darpa.mil/research/programs/quantum-benchmarking-initiative

[227] Defense Advanced Research Projects Agency, *DARPA Eyes Companies Targeting Industrially Useful Quantum Computers*, https://www.darpa.mil/news/2025/companies-targeting-quantum-computers

[228] 48 C.F.R. § 227.7203-5 (2025), https://www.ecfr.gov/current/title-48/chapter-2/subchapter-E/part-227/subpart-227.72/section-227.7203-5 and https://www.acquisition.gov/dfars/227.7203-5-government-rights.

[229] Nat'l Inst. of Standards & Tech., *NIST's AI Standards Zero Drafts Pilot Project to Accelerate AI Standards*, https://www.nist.gov/artificial-intelligence/ai-research/nists-ai-standards-zero-drafts-pilot-project-accelerate

**5. A comparison between QKD and PQC**

QKD and PQC are terms related to the field of cryptography, particularly in the context of securing communication against the theoretical threat posed by quantum computing.[230] Here's a breakdown of what each term means and how they differ:

1. QKD (Quantum Key Distribution): QKD is a method of secure communication that uses quantum mechanics principles to ensure the security of cryptographic keys. It enables two parties to generate a shared, secret random key known only to them, which can then be used to encrypt and decrypt messages. The security of QKD comes from the fundamental principles of quantum mechanics – any attempt at eavesdropping on the key can be detected because it inevitably alters the quantum states involved in the key distribution. QKD is currently seen as a practical method for achieving secure communication that is theoretically immune to the capabilities of quantum computers. However, the NSA recently outlined technical limitations and practical challenges of these technologies, ultimately recommending against their use in NSS until certain limitations are overcome, favoring quantum-resistant cryptographic -or Post-Quantum Cryptography (PQC)- solutions as more practical and cost-effective.[231]

2. PQC (Post-Quantum Cryptography): PQC refers to cryptographic algorithms that are believed to be secure against an attack by a quantum computer.[232] Unlike QKD, which is a specific technology, PQC is a broader category that includes various cryptographic algorithms designed to secure digital communications against quantum computers. These algorithms are based on mathematical problems that are currently believed to be hard for quantum computers to solve, such as lattice-based cryptography, hash-based cryptography, code-based cryptography, and multivariate polynomial cryptography.[233]

In sum, QKD is a specific quantum-based method for secure key distribution, and PQC encompasses a class of cryptographic algorithms designed to be secure in the age of quantum computing. Each approach has its own advantages and potential applications, and they could be used in complementary ways to ensure the security of communications in a future where quantum computing is prevalent.[234] Here, combination is the key. A holistic approach to ensuring ubiquitous, reliable, and secure encryption across various platforms and technologies, would ideally integrate different cryptographic techniques and strategies, including elements of QKD and PQC, to provide robust quantum safe security in a variety of scenarios.

---

[230] Quantropi. (n.d.). TrUE vs. QKD vs. PQC – Know the Difference. Retrieved from https://www.quantropi.com/true-vs-qkd-vs-pqc-know-the-difference/
[231] National Security Agency. (n.d.). Quantum Key Distribution (QKD) and Quantum Cryptography (QC). Retrieved from https://www.nsa.gov/Cybersecurity/Quantum-Key-Distribution-QKD-and-Quantum-Cryptography-QC/
[232] *See e.g.* Li S, Chen Y, Chen L, Liao J, Kuang C, Li K, Liang W, Xiong N. Post-Quantum Security: Opportunities and Challenges. Sensors (Basel). 2023 Oct 26;23(21):8744. doi: 10.3390/s23218744. PMID: 37960442; PMCID: PMC10648643.
[233] National Institute of Standards and Technology. (2024, February 26). Post-Quantum Cryptography. Retrieved from https://csrc.nist.gov/projects/post-quantum-cryptography
[234] For a Public Discussion on QKD and PQC at QCrypt 2023, see: https://www.youtube.com/watch?v=yXj_g3ywyh0

**B. EU Standards**

Within the European Union, the formulation of standards for quantum technology is a collaborative endeavor that encompasses a multitude of organizations and standard-setting bodies. This cooperative framework is designed to streamline and harmonize the development of quantum technology standards across the region and beyond. Here's a brief summary of the initiatives underway in this domain:

**1. European Telecommunications Standards Institute (ETSI)**:

ETSI is one of the prominent bodies in the EU working on quantum technology standards, particularly in quantum cryptography and quantum key distribution (QKD).[235] ETSI's Industry Specification Group on Quantum Key Distribution (ISG QKD) works on standardizing aspects of QKD, which is essential for secure communications in the quantum computing era. These standards cover components like security specifications, hardware, and network aspects of QKD systems. ETSI also collaborates with the ITU (International Telecommunications Union). For example, the ITU-T Focus Group on Quantum Information Technology for Networks (FG-QIT4N) and ETSI ISG QKD held a joint e-meeting on June 10, 2020, to discuss the development of quantum information technology, focusing particularly on Quantum Key Distribution (QKD). This collaboration aimed at exploring areas of mutual interest and cooperation between the two groups.[236] The ETSI Cyber Quantum Safe Cryptography (QSC) Working Group focuses on practical implementation, including performance and security assessments of quantum-safe primitives for industrial use. Emphasizing the documented threats posed by quantum computing to current public-key cryptography systems like RSA and ECC, ETSI is actively working on quantum safe migration strategies, to ensure secure, futureproof digital systems.[237]

**2. CEN and CENELEC**:

The European Committee for Standardization (CEN) and the European Committee for Electrotechnical Standardization (CENELEC) are important EU standardization organizations. The CEN and CENELEC Joint Technical Committee 22 (CEN/CLC/JTC 22) is currently developing standards for the various quantum technology pillars, addressing European market needs and supporting EU policies and values.[238] To further guide standardization efforts in this swiftly progressing domain, CEN and the CENELEC Focus Group on Quantum Technologies (FGQT) have recently unveiled two significant publications: a Standardization Roadmap and a report on Quantum Technologies Use Cases. Together, these documents offer a comprehensive overview of Europe's standardization requirements for quantum computing, quantum communication, and quantum metrology, charting a clear path for the future of quantum technology development and implementation.[239]

---

[235] European Telecommunications Standards Institute. (n.d.). Quantum-safe cryptography. Retrieved from https://www.etsi.org/technologies/quantum-safe-cryptography

[236] International Telecommunication Union. (2020, June 10). Joint ITU-T FG-QIT4N/ETSI ISG QKD meeting on "Quantum Information Technology". Retrieved from https://www.itu.int/en/ITU-T/focusgroups/qit4n/Pages/202006.aspx

[237] European Telecommunications Standards Institute. (2020, August 11). ETSI releases migration strategies and recommendations for Quantum-Safe schemes. Retrieved from https://www.etsi.org/newsroom/press-releases/1805-2020-08-etsi-releases-migration-strategies-and-recommendations-for-quantum-safe-schemes

[238] CEN-CENELEC. (n.d.). Quantum Technologies. Retrieved from https://www.cencenelec.eu/areas-of-work/cen-cenelec-topics/quantum-technologies/

**3. EU Quantum Flagship**:

The Quantum Flagship, a large-scale and long-term research initiative by the European Union, focuses on driving the second quantum revolution through significant advancements in detecting and manipulating single quantum objects. Launched in 2018, with a budget of at least €1 billion over 10 years, it aims to unite research institutions, academia, industry, enterprises, and policymakers in collaborative quantum research and development efforts.[240] Its Strategic Research and Industry Agenda (SRIA) acknowledges the importance of promotion of coordinated, tailored standardization and certification efforts. This document outlines the EU's strategy for quantum technology development, covering quantum computing, simulation, communication, sensing, and metrology.[241] The SRIA aims to align existing and new initiatives within the EU's quantum ecosystem, offering a roadmap for 2030 and recommendations for programs like the Chips Act and EuroHPC Joint Undertaking. The EU Quantum Flagship's mission contributes indirectly to standardization and best practices by advancing the state of quantum technologies.

**4**. **Key Performance Indicators (KPIs) for Quantum Technologies in Europe 2024:**

The KPIs 2024 outlined by the EU Quantum Flagship are essential for aligning EU quantum standardization efforts with strategic priorities such as interoperability, ethical governance, and market scalability.[242] These KPIs track progress in areas like quantum hardware supply chain resilience, cross-border research collaboration, and adoption of EU-led quantum communication protocols (e.g., the European Quantum Communication Infrastructure (EuroQCI)).[243] By quantifying advancements in quantum readiness levels and certification frameworks, the KPIs ensure EU standards reflect shared values like privacy, security, and equitable access while fostering global competitiveness.

**5. European Commission's Rolling Plan for ICT Standardization 2025:**

The European Commission's Rolling Plans 2020 to 2025 for ICT Standardization inter alia address the standardization needs for Quantum Technologies (QT), highlighting QT's potential in various domains and its importance for Europe's safety and independence.[244] The Rolling Plan 2025 is important for EU quantum standards as it strategically guides the development of harmonized, interoperable frameworks across quantum computing, communication, and sensing.[245] By identifying critical standardization needs and policy objectives, the plan ensures alignment between industry, academia, and regulators, fostering a cohesive quantum ecosystem. It prioritizes security, market scalability, and technological sovereignty, addressing fragmentation risks while supporting initiatives like the EU Quantum Declaration and the Digital Decade goals. The Rolling Plan's focus on collaborative benchmarks and ethical governance reinforces Europe's competitiveness in the global quantum race.

---

[239] CEN-CENELEC. (2023, March 22). Standardization for Quantum Technologies: Read our two newly published documents. Retrieved from https://www.cencenelec.eu/news-and-events/news/2023/brief-news/2023-03-22-standardization-for-quantum-technologies/
[240] Quantum Flagship. (n.d.). Homepage. Retrieved from https://qt.eu/
[241] Quantum Flagship. (2022, November 21). Quantum Flagship publishes preliminary Strategic Research and Industry Agenda. Retrieved from https://qt.eu/news/2022/quantum-flagship-publishes-preliminary-strategic-research-and-industry-agenda
[242] *See* KPIs for Quantum Technologies in Europe 2024 Values, QT.EU (Apr. 10, 2025), https://qt.eu/news/2025/2025-04-10_kpis_for-quantum-technologies-in-europe-2024-values.
[243] European Commission, European Quantum Communication Infrastructure (EuroQCI), DIGITAL STRATEGY, https://digital-strategy.ec.europa.eu/en/policies/european-quantum-communication-infrastructure-euroqci

**6. Coordination with Global Standards – ISO and IEEE**:

The EU actively participates in global standardization efforts through organizations like ISO and IEEE.[246] ISO -through its member bodies from EU countries- contributes to global standardization efforts in quantum technologies. ISO also collaborates with the International Electrotechnical Commission (IEC) on standards that may impact quantum technologies. This coordination ensures that European developments in quantum technology align with international standards, facilitating global interoperability and collaboration.

7. **JRC Quantum Standards Landscape Analysis:** This European Commission report published in March 2024 mapped existing quantum standardization activities, identified critical gaps (esp. in metrology, PQC implementation), and recommended priority areas for European leadership to ensure technological sovereignty and alignment with EU values.[247]

8. **Ongoing Standards Development Organizations (SDO) Activities:** Standardization continues via CEN/CENELEC JTC 21 (AI Act support focusing on trustworthiness/risk management)[248], ETSI (QKD protocols, QSC migration strategies), guided by the EU's Rolling Plan for ICT Standardisation which designates quantum as a crucial area requiring coordinated standards for market development and interoperability.

The EU's approach strategically coordinates standardization through formal bodies (ETSI, CEN/CENELEC), aligning technical work with broader policy goals like the AI Act, Quantum Flagship, and EU values. This promotes coherence within the bloc but can sometimes involve more complex bureaucratic processes than purely industry-driven efforts. As the field of quantum technology progresses, we can expect more focused efforts from these and other organizations within the EU to develop comprehensive standards and performance metrics.

---

[244] *See e.g.* European Commission. (2023). Quantum Technologies (RP2023). Retrieved from https://joinup.ec.europa.eu/collection/rolling-plan-ict-standardisation/quantum-technologies-rp2023
[245] European Commission, Rolling Plan for ICT Standardisation 2025, INTEROPERABLE EUROPE, https://interoperable-europe.ec.europa.eu/collection/rolling-plan-ict-standardisation/rolling-plan-2025
[246] *See e.g.* van Deventer, O., Spethmann, N., Loeffler, M. *et al.* Towards European standards for quantum technologies. *EPJ Quantum Technol.* **9**, 33 (2022). https://doi.org/10.1140/epjqt/s40507-022-00150-1
[247] *See* European Comm'n, Joint Rsch. Ctr., *Quantum Technologies*, https://joint-research-centre.ec.europa.eu/projects-and-activities/quantum-technologies_en
[248] Eur. Comm. for Standardization [CEN] & Eur. Comm. for Electrotechnical Standardization [CENELEC], *Artificial Intelligence*, https://www.cencenelec.eu/areas-of-work/cen-cenelec-topics/artificial-intelligence/

These standards will be crucial for the commercialization, safeguarding, advancing, and adoption of responsible quantum technologies across various sectors and domains.

### C. China Standards

In China, the development of standards for quantum technology reflects the country's significant investment and interest in this field. Several organizations and governmental bodies are engaged in these efforts:

**1. National Standardization Development Program (NSDP):**

In July 2022, China announced its intention to become a leading force in setting international technological standards, particularly in quantum computing, among other sectors. This ambition is outlined in the National Standardization Development Program (NSDP), marking a strategic push since 2017 to influence global standards. The NSDP, resulting from the China Standards 2035 project, aims to streamline and enhance China's standards-setting processes and encourage domestic and international collaboration. This move aligns with global trends, highlighting the geopolitical significance of technological standards and China's commitment to assert its influence in the evolving technological landscape.[249] 'China's approach can be contrasted with a 'Standards-first Approach', which prioritizes early international consensus on foundational standards before national strategies diverge significantly.

**2. Chinese National Standardization Development Outline:**

Issued in October 2021, the Chinese National Standardization Development Outline represents China's inaugural long-term strategic plan focusing on standardization. One of its seven key missions is to promote the simultaneous development of standardization along with scientific and technological innovation. This encompasses studies on standardization in areas such as AI, quantum information, and biotechnologies.[250]

**3. National Technical Committee 578 on Quantum Computing and Metrology of Standardization Administration of China (SAC/TC578):**

In January 2019, the Standardization Administration of China established the National Technical Committee on Quantum Computing and Metrology (SAC/TC578). This committee, led by the Standardization Administration with its secretariat at the Jinan Institute of Quantum Technology, oversees national standards in quantum computing and metrology. Its work spans the standardization of terminology, classification, hardware, software, architecture, and application platforms within these fields. SAC/TC578 oversees standardization in quantum computing and cryptography, including efforts to develop post-quantum cryptographic algorithms aligned with national security goals. It includes members from government, research institutes, industry associations, educational institutions, and enterprises.[251] In May 2023, SAC/TC578 published China's first national standard in the quantum computing field. This QC standard was implemented on December 1, 2023. Over 2024, China published several voluntary quantum metrology standards.

---

[249] Quantum Economic Development Consortium. (2022, July). China raising the ante on standards setting. Retrieved from https://quantumconsortium.org/blog/china-raising-the-ante-on-standards-setting/

[250] Wei, K. (2022). China's National Standardization Development Outline: Policy Implications and Future Directions, The University of Tokyo, https://ifi.u-tokyo.ac.jp/en/wp-content/uploads/2022/03/SSUessay_5_Wei20220209_EN.pdf.

**4. Chinese National Standardization Management Committee (SAC)**:

SAC, responsible for overseeing standardization efforts in China, plays a key role in the development of quantum technology standards. The committee works on formulating and disseminating national standards, including those for emerging quantum technologies as the sector evolves.[252]

**5. China Electronics Standardization Institute (CESI)**:

CESI, under the Ministry of Industry and Information Technology (MIIT), contributes to the development of standards in the electronics sector, which encompasses quantum technology. Their work may include standardization in quantum communication and quantum computing hardware.

**6. Chinese Association for Cryptologic Research (CACR)**

The CACR announced a national post quantum cryptographic (PQC) algorithm design competition on June 11, 2018.[253] This initiative aims to advance quantum safe cryptographic research and application, foster talent, and contribute to the technological progression of cryptographic algorithms, aligning with national strategic goals. The question remains whether non-Western countries like China and Russia will adopt the NIST, ISO, and IETF (Internet Engineering Task Force) PQC standards, or that they will adopt their own PQC standards.[254]

**7. Research Institutions and Universities**:

Major Chinese research institutions and universities, such as the University of Science and Technology of China (USTC) and the Chinese Academy of Sciences, are at the forefront of quantum technology research. These institutions not only contribute to technological advancements but also influence the development of domestic and international industry standards through their research findings.[255]

---

[251] National Technical Committee 578 on Quantum Computing and Metrology of Standardization Administration of China. (n.d.). About Us. Retrieved from https://en.tc578.org.cn/portal/index/about

[252] National Certification and Standardization Electronic Service Platform. (n.d.). Retrieved from http://ncse.sac.gov.cn/sacen/

[253] Chinese Cryptography Society. (2018, June 11). 关于举办全国密码算法设计竞赛的通知 [Notice on holding a national cryptographic algorithm design competition]. Retrieved from https://www.cacrnet.org.cn/site/content/259.html

[254] Liu, N. (2022, October). China, Russia to adopt slightly different PQC standards from U.S.. *SDxCentral*. Retrieved from https://www.sdxcentral.com/articles/analysis/china-russia-to-adopt-slightly-different-pqc-standards-from-us/2022/10/

[255] *See e.g.* University of Science and Technology of China. (n.d.). Division of Quantum Physics and Quantum Information. Retrieved from https://quantum.ustc.edu.cn/web/en

## 8. China's state-centric model aimed at bolstering national industrial and strategic policies while facing decoupling pressures:

China's approach to technical standardization, particularly evident in its advancements in emerging technologies like quantum technologies, 6G and AI, reflects a state-centric model aimed at bolstering national industrial and strategic policies. This has led to gains in some sectors without displacing Western dominance in standard development organizations (SDOs). However, as China pioneers in certain technologies and faces 'decoupling' pressures, its "multi-bilateralist" approach emphasizes both international collaboration and promoting domestic advancements. The evolution of China's standardization efforts, emphasizing both collaboration and national interests, has significant implications for global technology development and policy and should be carefully monitored.[256]

## 9. Government-Led Strategic Geopolitical Initiatives:

The Chinese government's emphasis on quantum technology in its strategic planning indicates a top-down approach to standardization. Government-led initiatives and funding often come with guidelines and standards for research and development in quantum technology. This can be exemplified by China's strategic initiative to become a global leader in setting international technology standards, specifically focusing on quantum computing, sensing, and communication. Through the National Standardization Development Program (NSDP) highlighted above, China aims to streamline its influence in global standards, highlighting the geopolitical importance of technology standards in the modern era.[257]

## 10. International Collaboration with ISO and IEEE:

While China develops its own standards following a state-centric approach, it also participates in international standardization efforts by NIST, ISO and IEEE.[258] This participation ensures some level of global interoperability and alignment with international standards, which is crucial for technologies with worldwide applications like quantum computing. Moreover, adoption of international standards will increase, which comprises conversion of ISO and IEEE standards into domestic Chinese standards, including translating these into Mandarin.[259]

## 11. Quantum Communication and Cryptography:

China has made significant strides in quantum communication – and is arguably leading the global quantum communication race-, particularly in quantum key distribution (QKD). The

---

development of QKD networks, like the Beijing-Shanghai quantum communication link, is accompanied by efforts to standardize quantum communication protocols and hardware.[260]

**12. Embedding Autocratic Values into Quantum Technology Standards:**

As China continues to be a key player in the quantum technology arena, its approach to standardization is expected to be both comprehensive and aligned with the autocratic Chinese Communist Party (CCP) norms, values and strategic objectives in science and technology. These values are potentially embedded into the design, architecture and infrastructure of Chinese technology, and are often in superposition with the fundamental rights and freedoms, universal democratic norms and principles, and rule of law thinking of democratic countries.[261] There is a risk of China -in its new role as technological trendsetter and pioneer- setting the rules of the road for quantum standards via its Digital Silk Road, by exporting its autocratic norms and values to democratic countries and to states participating in the Belt & Road Initiative, through its technology – such as SAC/TC578. The world has witnessed this before with AI based surveillance technologies and 5G. It is therefore important that China adopts the Responsible Quantum Technology paradigm, actively steering the field towards beneficial outcomes for all groups of our societies. As a leader in several quantum domains, the standards developed by China will be critical in guiding regional and global advancements and commercialization efforts in quantum technology.

China views standardization (NSDP, SAC/TC578) as a core strategic instrument to promote domestic industry, achieve technological leadership, and project influence globally ('Beijing effect'). Its dual strategy involves active participation in international SDOs while simultaneously developing national standards aligned with state objectives and potentially embedding autocratic norms, particularly through initiatives like the Digital Silk Road. The development of distinct national PQC standards exemplifies this approach.

**D. South Korea, Japan, and Taiwan Standards**

South Korea, Japan, and Taiwan are actively advancing quantum technology standardization as integral components of their national strategies to secure leadership positions in quantum computing and related high-technology sectors, reflecting developments current as of 2025.

In South Korea, the Korea Advanced Institute of Science and Technology (KAIST) spearheads research in quantum algorithms and error correction, actively collaborating with major industry partners like Samsung and LG to translate academic breakthroughs into commercial applications.[262] Concurrently, the Korean Agency for Technology and Standards (KATS) works in coordination with the nation's 2025 National Quantum Computing Development Plan to ensure domestic standards align with international frameworks, particularly those of ISO/IEC

---

JTC 3.[263] Underscoring this commitment, South Korea's Ministry of Science and ICT has allocated a significant budget of ₩300 billion for the 2024–2028 period towards quantum R&D, explicitly emphasizing interoperability and the crucial standardization of post-quantum cryptography (PQC).[264]

Japan, meanwhile, leverages the Japanese Standards Association (JSA) in collaboration with the ambitious Moonshot Research and Development Program to establish critical benchmarks for quantum hardware reliability and quantum communication protocols.[265] The National Institute of Information and Communications Technology (NICT), through its Quantum ICT Consortium, has developed influential guidelines for quantum key distribution (QKD) interoperability, which have gained traction internationally, being adopted by bodies such as ETSI and ITU.[266] Japan's robust industry involvement is pivotal, with contributions like Toshiba's advancements in quantum annealing and NTT's research in quantum-photonic networking significantly shaping Japan's input into global standards development.[267]

In Taiwan, the Bureau of Standards, Metrology and Inspection (BSMI) launched a dedicated Quantum Technology Standards Working Group in 2023, focusing efforts on quantum-safe encryption standards and the integration of quantum devices with semiconductor platforms.[268] The Industrial Technology Research Institute (ITRI) collaborates closely with global semiconductor leader TSMC to develop vital cryogenic control standards specifically for silicon-based qubits, essential for building scalable quantum computers.[269] Furthermore, TSMC's own 2024 technology roadmap includes quantum-inspired CMOS architectures, directly influencing standardization efforts within IEEE and ISO concerning hybrid quantum-classical computing systems.[270]

A common thread across all three jurisdictions is the high priority placed on ensuring interoperability with established global standards, such as NIST's PQC algorithms and ISO/IEC JTC 3 terminology, while simultaneously nurturing vibrant domestic quantum ecosystems. Their strategic focuses exhibit nuances, however: South Korea and Taiwan strategically integrate quantum standards development with broader goals of semiconductor supply chain resilience, whereas Japan places a notable emphasis on QKD standardization, particularly for defense and secure communication applications.[271] Each nation clearly recognizes the profound strategic importance of quantum technology and is vigorously cultivating an ecosystem supportive of its research, development, and eventual commercialization.

---

[263] Korean Agency for Technology and Standards, *Quantum Technology Standardization Roadmap* 7–9 (2024), https://www.kats.go.kr
[264] Ministry of Science and ICT (South Korea), *2025 National Quantum Strategy* 4.2 (2024), https://www.msit.go.kr
[265] Council for Science, Technology and Innovation (Japan), *Moonshot R&D Program: Quantum Standards Initiative* 22–25 (2023), https://www8.cao.go.jp/cstp
[266] National Institute of Information and Communications Technology, *Quantum ICT White Paper* 45–48 (2024), https://www.nict.go.jp
[267] Toshiba Corporation, *Advancing Quantum Annealing: Technical Brief* 3 (2025), https://www.toshiba.com (last visited June 10, 2025).
[268] Bureau of Standards, Metrology and Inspection (Taiwan), *Quantum Standards Working Group Charter* 1 (2023), https://www.bsmi.gov.tw
[269] Industrial Technology Research Institute, *Cryogenic Control Standards for Silicon Qubits* 10–12 (2024), https://www.itri.org.tw
[270] TSMC, *Quantum-Inspired CMOS Roadmap* 5–7 (2024), https://www.tsmc.com
[271] Council of Canadian Academies, *Quantum Potential* 118–122 (2024), https://cca-reports.ca

# VI. COMPARATIVE ANALYSIS OF U.S., EU AND CHINA EMERGING TECH REGULATION

Having reviewed the distinct regulatory, export control, and standardization landscapes in the US, EU, and China, this section offers a direct comparative analysis of their approaches to governing emerging quantum and AI technologies. A comparative analysis of the approaches taken by the U.S., the EU, and China towards regulating emerging technologies like quantum and artificial intelligence (AI), including associated export controls and standardization efforts, reveals distinct philosophies rooted in their respective innovation systems and geopolitical priorities.

## A. Regulation and Export Controls

Regarding regulation and export controls, the U.S. generally prioritizes market-driven innovation, relying heavily on private sector investment and striving for flexible regulatory environments, particularly in nascent technological fields, to avoid stifling progress. However, driven by escalating national security concerns, particularly in dual-use areas like advanced AI and quantum computing, the U.S. employs stringent export controls and investment screening mechanisms aimed at protecting its technological edge and preventing misuse by strategic rivals. This security focus, while intended to safeguard national interests, carries acknowledged risks of hindering international collaboration and provoking retaliatory measures, impacting global innovation dynamics.[272] The EU seeks to balance innovation with social welfare, environmental sustainability, and fundamental rights. Its approach emphasizes collaborative research funded through programs like Horizon Europe and relies on comprehensive, harmonized regulatory standards (e.g., GDPR, AI Act) that often embed ethical considerations and aim to build public trust. EU export controls are typically integrated within this broader regulatory framework, reflecting a values-based approach to technology governance. China executes a state-led innovation strategy characterized by significant government investment, centralized planning, and a focus on achieving national economic goals and technological self-sufficiency. Regulation, export controls, and industrial policies are tightly integrated to serve these strategic objectives, enabling rapid scaling and deployment while maintaining state control over critical technology sectors. While the U.S. and EU share concerns regarding security and ethics, their primary mechanisms differ – market adaptation and security controls in the U.S. versus proactive, comprehensive regulation in the EU. China's approach prioritizes state control and strategic advantage above all, utilizing regulation and controls instrumentally towards those ends.

## B. Standards

---

[272] *See e.g.* Gregory C. Allen, Ctr. for Strategic & Int'l Stud., Choking Off China's Access to the Future of AI (Oct. 2022), https://www.csis.org/analysis/choking-chinas-access-future-ai.

These differing philosophies extend to standardization. The U.S. typically favors industry-led standardization, where standards often emerge *de facto* from market competition, consortia agreements (like those involving NIST participation), or private sector initiatives, valuing flexibility and responsiveness to technological change. In contrast, the EU champions more formal, inclusive standardization processes often managed through bodies like CEN-CENELEC, deliberately integrating ethical, safety, and environmental requirements to create *de jure* standards that reflect European values and support its regulatory objectives. China employs state-driven standardization, viewing standards as critical tools of industrial policy and international influence. It works actively to align domestic standards with national strategic goals and promotes their adoption internationally, sometimes leveraging initiatives like the Belt and Road.[273]

**C. SEA-Framework for Responsible Quantum Innovation**

Applying the SEA framework (Safeguarding, Engaging, and Advancing) for Responsible Quantum Innovation illuminates these differences further.[274] All three regions engage in Advancing quantum and AI technologies, albeit through distinct means: market dynamism and targeted funding in the U.S., collaborative projects within a regulated ecosystem in the EU, and massive state investment in China. The approaches to Engaging stakeholders also vary: the EU emphasizes broad, multi-stakeholder consultation and public participation; the U.S. relies more on public-private partnerships and market feedback; China's engagement is primarily state-directed. The concept of Safeguarding is perhaps where interpretations diverge most significantly. The EU focuses on safeguarding fundamental rights, privacy, and ethical principles through its regulatory posture. The U.S. prioritizes safeguarding national security and technological leadership, primarily through mechanisms like intellectual property rights and export controls. However, it is crucial to recognize, as analysts note, that such controls, while intended to safeguard national interests, are not synonymous with safeguarding global scientific progress or societal well-being; they can fragment the research landscape and have counterproductive economic and diplomatic effects, such as forcing the export-controlled country to be more innovative. China's safeguarding efforts are focused on protecting state interests, maintaining social stability according to its political model, and securing strategic autonomy. Therefore, while elements of each approach towards technology driven innovation touch upon the SEA principles, the relative emphasis, underlying motivations, and chosen tools differ profoundly, reflecting the deep-seated variations in their emerging tech governance models and worldviews.

---

[273] *See e.g.* Kenneth Propp, Atl. Council, Standardizing the Future: How Can the United States Navigate the Geopolitics of International Technology Standards? (Aug. 15, 2023), https://www.atlanticcouncil.org/in-depth-research-reports/report/standardizing-the-future-how-can-the-united-states-navigate-the-geopolitics-of-international-technology-standards/.
[274] *See* Kop *et al.*, *supra* note 166.

## VII. COMPARING U.S., EU AND CHINA INNOVATION SYSTEMS

Building on the distinct regulatory and standardization approaches of the US, EU, and China outlined above, this section comparatively analyzes their underlying national innovation systems. This comparative analysis of innovation systems directly informs the pathways and challenges for global quantum and AI governance. It provides a comparative analysis of the U.S., EU, and China innovation systems, and discusses the key elements needed to create a common strategy for global innovation. From a bird's eye view, the U.S. system's strength in rapid, disruptive innovation contrasts with the EU's focus on responsible, sustainable development, and China's state-directed implementation. Ideally, a harmonized global innovation strategy could strive to leverage complementary strengths observed across the three leading ecosystems.

### 1. United States Innovation System

The United States innovation system is predominantly characterized by a market-driven approach, where private sector dynamics and competitive pressures strongly influence technological advancement. This environment cultivates flexibility and adaptability among firms, often leading to the rapid commercialization of breakthroughs. Complementing this is a renowned entrepreneurial culture, supported by a robust ecosystem comprising venture capital, incubators, and accelerators, which actively encourages risk-taking and fosters the growth of startups developing disruptive technologies.

Generally, the regulatory environment in the U.S. maintains flexibility, especially concerning emerging technologies, intending to avoid stifling innovation. While this facilitates faster adoption and diffusion, it periodically gives rise to calls for more stringent oversight, particularly regarding issues such as data privacy and ethical AI. [275] This overall framework enables American firms to strive for global market competitiveness, frequently leveraging technological innovation as a primary competitive advantage and positioning the U.S. as a leader across various high-tech sectors, including information technology, biotechnology, and aerospace.

More recently, confronting contemporary economic and geopolitical challenges, the Biden administration has advanced a twenty-first century pro-American industrial policy.[276] This comprehensive strategy is designed to revitalize and strategically position the U.S. industrial base. It is founded upon five key pillars: enhancing supply chain resilience, directing targeted public investments, utilizing public procurement strategically, promoting climate resilience, and ensuring equity.[277] Formulated partly in response to economic vulnerabilities exposed during the COVID-19 pandemic and the competitive pressures posed by other nations, notably China, this industrial policy acknowledges the transformed global economic landscape and the imperatives of technological change and the climate crisis. Through this concerted approach, the policy endeavors to secure a renewed era of industrial and economic leadership for the United States.

---

[275] Moutii, M. *The Future of Innovation in the United States: Permissionless or Regulated?* Econlib. (Oct. 2023), Retrieved April 7 2024, from https://www.econlib.org/the-future-of-innovation-in-the-united-states-permissionless-or-regulated/

[276] Ang, Y. Y. (2021, April 19). *The Limits of State-Led Innovation: Evidence from Chinese Patents*. Center for the Study of Contemporary China, University of Pennsylvania. https://cscc.sas.upenn.edu/node/3744

**2. European Union Innovation System**

The European Union innovation system distinctly emphasizes collaborative research and development, significantly facilitated through public funding initiatives like Horizon Europe.[278] This approach actively encourages cross-border partnerships among businesses, universities, and research institutions, thereby fostering innovation ecosystems that effectively leverage diverse strengths and expertise from across the member states. A defining characteristic of the EU's strategy is the strong integration of innovation with sustainable development goals, prioritizing green technologies and solutions aimed at addressing key societal challenges. Furthermore, the EU upholds high regulatory standards, particularly evident in data protection through the GDPR and stringent environmental protection measures. This reflects a concerted effort to balance technological advancement with ethical considerations and social welfare. Consequently, the EU's approach to innovation is often characterized by its commitment to ensuring that technological progress aligns with principles of social equity and environmental sustainability, frequently involving a more precautionary stance and careful assessment of the societal and environmental impacts of new technologies.

**3. China Innovation System**

China's innovation system is distinctly characterized by state-led strategies, featuring strong government direction and substantial public investment in R&D, particularly targeting strategic sectors such as artificial intelligence, renewable energy, and telecommunications. The state plays a pivotal role in defining national priorities and mobilizing resources toward major national technology projects.[279] A key operational strength of this model is its capacity for rapid scaling and implementation, enabling innovations to be quickly applied across China's vast domestic market, supported by policies conducive to rapid industrial growth and technological adoption.

---

[277] Malloy, D. (2021, June 23). The White House's case for crafting a new industrial strategy. Atlantic Council. https://www.atlanticcouncil.org/blogs/new-atlanticist/the-white-houses-case-for-crafting-a-new-industrial-strategy/

[278] European Commission. (n.d.). *Horizon Europe*. https://research-and-innovation.ec.europa.eu/funding/funding-opportunities/funding-programmes-and-open-calls/horizon-europe_en

[279] Zhang, L., & Lan, T. (2023). The new whole state system: Reinventing the Chinese state to promote innovation. Environment and Planning A: Economy and Space, 55(1), 201-221. https://doi.org/10.1177/0308518X221088294

Underpinning these efforts are clear national goals centered on achieving technological independence and securing leadership positions in critical technological areas, largely motivated by concerns related to economic security and global competitiveness.[280] This drive often manifests as significant state support for domestic firms and a concerted push to develop indigenous capabilities. The entire system operates within a unique governance framework that blends authoritarianism with surveillance capitalism.[281] This approach utilizes advanced digital monitoring and data analysis not only to manage the economy but also to surveil the populace and suppress dissent, effectively merging economic objectives with mechanisms for political control.[282] This fusion of authoritarian governance with the economic logic of surveillance capitalism facilitates unparalleled levels of social, political, and economic oversight and intervention by the state, a model that China is actively exporting to other regions globally.[283]

**Comparative Analysis**

Each innovation system – those of the United States, the European Union, and China – exhibits distinct strengths and inherent challenges, largely shaped by their differing prioritizations of market freedom, social welfare, and state intervention. The U.S. system traditionally excels in fostering a dynamic, competitive environment conducive to risk-taking, entrepreneurial ventures, and the generation of groundbreaking, often disruptive, innovations, although it is increasingly incorporating elements of strategic industrial policy. Conversely, the EU framework emphasizes ensuring that innovation aligns with broader societal well-being and environmental sustainability goals. This is pursued through substantial collaborative research initiatives, often publicly funded like Horizon Europe, and the establishment of high regulatory standards intended to embed ethical considerations and protect citizens, reflecting a more precautionary approach to technological deployment. China's system demonstrates the potent effectiveness of state-led strategies and substantial government investment in accelerating technological advancement, achieving rapid scaling and implementation across its vast domestic market to meet defined national goals, particularly technological independence. However, this model faces international scrutiny regarding market access, intellectual property practices, and the implications of its fusion of authoritarianism with surveillance capitalism.[284]

---

[280] Reuters. (2022, September 6). China to strengthen state-led system in core tech breakthroughs, Xi says. *Reuters*. https://www.reuters.com/world/china/china-improve-mechanism-core-tech-innovations-state-media-2022-09-06/

[281] Polyakova, A & Meserole, A., 2019. *Exporting digital authoritarianism: The Russian and Chinese models*, Brookings Institution. United States of America. Retrieved from https://policycommons.net/artifacts/3527460/exporting-digital-authoritarianism/4328250/ on 07 Apr 2024. CID: 20.500.12592/rwbp8q.

[282] Wang, M. (2021, April 8). *China's Techno-Authoritarianism Has Gone Global*. Human Rights Watch. https://www.hrw.org/news/2021/04/08/chinas-techno-authoritarianism-has-gone-global

[283] Cave, D., Hoffman, S., Joske, A., Ryan, F., & Thomas, E. (2019). Enabling & exporting digital authoritarianism. In *Mapping China's technology giants* (pp. 08–15). Australian Strategic Policy Institute. http://www.jstor.org/stable/resrep23072.8

[284] James A. Lewis, *Rethinking Technology Transfer Policy Toward China*, Center for Strategic and International Studies (Aug. 8, 2023), https://www.csis.org/analysis/rethinking-technology-transfer-policy-toward-china.

The divergence extends to the fundamental role envisaged for the state and the philosophical underpinnings of regulation. The U.S. model leverages market forces but is adapting to strategic competition with targeted interventions and investments. The EU acts as a normative regulator, seeking to shape global standards based on its values through comprehensive legal frameworks like the GDPR and the AI Act, prioritizing fundamental rights alongside economic progress. China utilizes the state apparatus directly to orchestrate innovation ecosystems, tightly integrating technological development with national security objectives and political control, often blurring the lines between civilian and military applications.[285] Governing quantum requires learning from prior experiences with AI, nano, biotech and nuclear. The economic stakes, reflected in patent trends and market assessments[286], further underscore the competitive implications of these divergent systems.

These contrasting paradigms create significant implications for global governance, particularly in sensitive, data-intensive fields like AI and quantum technologies. Differing approaches to data governance – market-oriented in the U.S. (though evolving), rights-centric in the EU, and state-controlled in China – present substantial barriers to interoperability and trust. Furthermore, the competition extends to the setting of international technical standards, where divergence risks fragmenting global markets and creating incompatible technological ecosystems, thereby hindering scientific collaboration and increasing security risks.[287]

Achieving greater global coherence might involve hybridizing strengths: blending U.S. dynamism with EU value-based guardrails and acceptable aspects of Chinese long-term strategic planning, provided it aligns with universal human rights and democratic principles ('ons verbindt' - what connects us). This necessitates robust international cooperation mechanisms, shared R&D platforms, harmonized standards where possible, mutual recognition agreements, and pooled funding for global challenges. While these diverse innovation ecosystems contribute unique strengths and perspectives to the global technological landscape – driving advancements with broad implications for economic development, competitiveness, and societal progress worldwide – their foundational differences also fuel geopolitical tension. Effectively managing the global commons in areas like cybersecurity, AI ethics, and quantum non-proliferation requires navigating these disparate models, seeking pathways for cooperation and coordination despite systemic rivalry and contrasting values.

---

[285] Australian Strategic Policy Institute, *Persuasive Technologies in China: Implications for the Future of National Security* (Nov. 2024), https://www.aspi.org.au/report/persuasive-technologies-china-implications-future-national-security

[286] Kop, Aboy & Minssen, Intellectual property in quantum computing and market power: a theoretical discussion and empirical analysis, *Journal of Intellectual Property Law & Practice*, Volume 17, Issue 8, August 2022, Pages 613–628, https://doi.org/10.1093/jiplp/jpac060

[287] Bacchus, J., Borchert, I., Marita-Jaeger, M., & Ruiz Diaz, J., *Interoperability of Data Governance Regimes: Challenges for Digital Trade Policy*, CITP Briefing Paper No. 12 (Apr. 8, 2024), https://citp.ac.uk/publications/interoperability-of-data-governance-regimes-challenges-for-digital-trade-policy

**The Downsides of Having 3 Distinct Innovation Systems**

The ongoing competition among the distinct innovation systems of the United States, the European Union, and China, while a potent driver of technological advancement and economic growth, concurrently generates significant downsides. These drawbacks arise from fundamental differences in regulatory philosophies, approaches to market access, strategic priorities, and underlying values. The resulting complexities increasingly impact the global innovation landscape, international trade dynamics, supply chain stability, and geopolitical relations.

1. **Fragmentation and Incompatibility**

A primary challenge relates to fragmentation and incompatibility. Persistent and arguably widening regulatory divergence creates significant operational hurdles and market fragmentation. Beyond the established contrast between the EU's GDPR and approaches elsewhere, the divergence is intensifying in critical areas like Artificial Intelligence. The EU's comprehensive, risk-based AI Act, now entering implementation, contrasts with the U.S.'s sector-specific guidance, state-level initiatives, and framework-based approach (e.g., NIST AI Risk Management Framework following the 2023 AI Executive Order), and China's evolving state-centric AI regulations focused on content control and algorithmic oversight. This divergence complicates cross-border data flows, AI model training and deployment, and compliance for international technology firms.[288]

Furthermore, competing efforts to shape international technology standards continue, moving beyond 5G into arenas like 6G, the Internet of Things (IoT), quantum technologies, and AI standardization. Diverging protocols and technical specifications risk creating incompatible ecosystems, hindering global interoperability, increasing costs for businesses, resulting in a 'splinternet' or bifurcated technology stacks aligned with geopolitical blocs.

2. **Trade Tensions and Protectionism**

These distinct approaches also fuel trade tensions and protectionism. Geopolitical rivalry, particularly between the U.S. and China, has escalated such tensions, leading to more explicit forms of protectionism often framed as national security measures. This impacts market access and involves significant trade barriers, including stringent U.S. export controls on advanced semiconductors, AI chips, and manufacturing equipment aimed at China, alongside inbound and outbound investment screening mechanisms (e.g., related to EO 14105 on quantum-AI investments). Reciprocal actions, concerns about economic coercion (e.g., restrictions on critical mineral exports from China), and the EU's "de-risking" strategy further restrict the global flow of technology, capital, and talent, fragmenting critical supply chains and dampening innovation.[289] Persistent intellectual property disputes remain a significant source of friction, encompassing concerns over IP theft, knowledge siphoning, and forced technology transfers, particularly involving China, thereby undermining trust and complicating R&D collaboration.[290]

---

[288] IFAC. (2018, April 11). *Regulatory Divergence: Costs, Risks and Impacts*.
https://www.ifac.org/knowledge-gateway/contributing-global-economy/publications/regulatory-divergence-costs-risks-and-impacts

3. **Security Concerns and Technological Sovereignty**

Moreover, these dynamics heighten security concerns and intensify the drive for technological sovereignty. Increased cybersecurity and surveillance risks arise from the proliferation of technologies embedded with differing security standards and vulnerabilities, amplified by state-sponsored cyber espionage and critical infrastructure threats. The export of surveillance technologies from authoritarian regimes exacerbates concerns about human rights and data privacy globally.[291] Consequently, reducing technological dependence has become a primary national security objective, evident in major industrial policies (e.g., U.S. and EU CHIPS Acts) targeting critical sectors like advanced semiconductors, AI foundation models, quantum computing, biotechnology, and essential raw materials, driving significant efforts towards supply chain restructuring and resilience.

4. **Ethical and Social Impact Concerns**

There are also substantial ethical and social impact concerns stemming from this competitive environment. The relentless pursuit of technological supremacy can sideline crucial social and environmental considerations. While the EU attempts to proactively integrate these through regulation like the AI Act, concerns persist that rapid, market-driven or state-directed development in the U.S. and China may inadequately address complex issues like algorithmic bias, AI-driven misinformation, labor market disruptions, the significant energy consumption of large AI models, and broader environmental sustainability impacts. This intense competition also risks exacerbating global inequality, widening the digital divide as developing nations face increasing difficulty in accessing or participating in frontier technological advancements, leading to greater economic dependency.

5. **Innovation Echo Chambers**

---

[289] *See e.g.* European Think-tank Network on China (ETNC), *National Perspectives on Europe's De-risking from China* (Patrik Andersson & Frida Lindberg eds., with Bernhard Bartsch, Una Aleksandra Bērziņa-Čerenkova, Andreea Brinza, Lucas Erlbacher, Miguel Otero-Iglesias, John Seaman, Plamen Tonchev & Mariana Trifonova, June 2024), available at https://www.merics.org/en/report/national-perspectives-europes-de-risking-china.
[290] *See e.g.* Shivakumar, S. (2022, March 3). Securing Intellectual Property for Innovation and National Security. Center for Strategic and International Studies (CSIS). https://www.csis.org/analysis/securing-intellectual-property-innovation-and-national-security
[291] *See e.g.* Kop, M. (2021, March 14). Democratic Countries Should Form a Strategic Tech Alliance. *Transatlantic Antitrust and IPR Developments*. Stanford Law School. https://law.stanford.edu/publications/democratic-countries-should-form-a-strategic-tech-alliance/, discussing the imperative for democratic countries to establish a Strategic Tech Alliance to counteract the advances and influences of authoritarian regimes in the global technology arena, emphasizing the importance of shared economic interests and common values.

Finally, the concentration of technological development within these three dominant systems can foster innovation echo chambers. This potential homogenization of innovation priorities, such as the current intense focus on specific architectures for large language models in AI, might inadvertently neglect diverse societal needs or alternative, arguably valuable technological pathways, thereby stifling global creativity and limiting the variety of solutions developed for complex global challenges.[292]

While competition remains a powerful engine for innovation, these escalating downsides underscore the urgent need for enhanced international dialogue and cooperation. Finding common ground on regulatory approaches (including IP protection and antitrust policies sensitive to innovation dynamics)[293], technical standards harmonization where feasible, and establishing norms for responsible technology development and deployment is critical. Balancing intensifying competition with necessary collaboration, particularly on managing shared risks like AI safety or quantum threats to cryptography, is essential for fostering a more inclusive, secure, and sustainable global innovation ecosystem. This fragmentation, driven by differing innovation models and priorities, represents a significant barrier to achieving the harmonized 'Quantum Acquis Planétaire' and hinders collaborative progress on global challenges requiring coordinated action.

## VIII. A 'WASHINGTON EFFECT' FOR QUANTUM LAW

The U.S.'s proactive implementation of quantum policies focused on economic safety and national security (through several Trump and Biden quantum related Executive Orders and Memoranda[294]) suggests the emergence of a 'Washington effect' for quantum law, potentially

---

[292] Carr, A.. (2023, August 3). Breaking out of the innovation echo chamber. *Forbes Tech Council*. https://www.forbes.com/sites/forbestechcouncil/2023/08/03/breaking-out-of-the-innovation-echo-chamber/?sh=1bcf454820fc

[293] *See e.g.* Lemley, M. A. (2012). Industry-Specific Antitrust Policy for Innovation. *Columbia Business Law Review*, 2011(3), 537–653. https://doi.org/10.7916/cblr.v2011i3.2911, discusses the relationship between intellectual property law and antitrust law, challenging the assumption that competition impedes innovation while monopoly promotes it. The author argues for a nuanced approach recognizing that different industries require different balances between competition and protection to foster innovation. *See also* Lemley, M. A. (2009). A Cautious Defense of Intellectual Oligopoly with Fringe Competition. *Review of Law and Economics*, 5(3), 1025-1035. Retrieved from Stanford Law School, https://law.stanford.edu/publications/a-cautious-defense-of-intellectual-oligopoly-with-fringe-competition/ discussing discusses the implications of intellectual property rights, suggesting that while they deviate from perfect competition, they do not necessarily lead to monopoly pricing and may not impose the significant social harms often associated with monopolies.

[294] *See e.g.* Presidential Memorandum on United States Government-Supported Research and Development National Security Policy, Trump White House (Jan. 14, 2021), https://trumpwhitehouse.archives.gov/presidential-actions/presidential-memorandum-united-states-government-supported-research-development-national-security-policy/;
National Security Memorandum on Promoting United States Leadership in Quantum Computing

setting global rules for quantum technologies analogous to the 'Brussels effect' in data regulation (i.e. the GDPR, MDR and EU AI Act). This influence manifests through the extraterritorial reach of U.S. export controls and investment screening (E.O. 14105, BIS controls), compelling global actors seeking U.S. market access or collaboration to align with American standards, particularly regarding security and Post-Quantum Cryptography (PQC) by NIST.

The 'Brussels Effect' describes the European Union's unilateral ability to project its regulatory standards globally by leveraging the size and attractiveness of its internal market; multinational companies often adopt EU rules not only for their European operations but across their worldwide activities, leading to a *de facto* Europeanization of global commerce in areas such as data protection, environmental standards, and digital markets.[295] This phenomenon occurs because compliance with EU regulations is often less costly than maintaining divergent standards for different markets, and because the EU possesses both the regulatory capacity and political will to enforce high standards, distinguishing it from other large economies such as the U.S. or China.

The United States is taking a proactive lead in quantum technology regulation, and the emergence of a "Washington effect" could reshape global governance in this domain – though driven by different motivations. If the U.S. sets comprehensive, security- and ethics-focused standards for quantum technologies, these rules, enforced via mechanisms like export controls and investment screening, are likely to have extraterritorial reach: multinational companies and allied governments seeking access to the lucrative U.S. market or collaboration would be compelled to adopt American requirements, thereby creating a global baseline for responsible quantum development and deployment. This first-mover advantage could cement U.S. leadership in quantum innovation and standard-setting, but it also carries significant risks. In a rapidly evolving field with limited empirical data, early regulatory interventions based on incomplete information may inadvertently stifle innovation, lock in suboptimal approaches, or trigger counterproductive market fragmentation if they are not evidence-based or sufficiently

---

While Mitigating Risks to Vulnerable Cryptographic Systems, Biden White House (May 4, 2022), https://bidenwhitehouse.archives.gov/briefing-room/statements-releases/2022/05/04/national-security-memorandum-on-promoting-united-states-leadership-in-quantum-computing-while-mitigating-risks-to-vulnerable-cryptographic-systems/; Exec. Order No. 14,073, Enhancing the National Quantum Initiative Advisory Committee, 87 Fed. Reg. 28,899 (May 4, 2022), https://www.presidency.ucsb.edu/documents/executive-order-14073-enhancing-the-national-quantum-initiative-advisory-committee, "Executive Order 14105 on Addressing United States Investments in Certain National Security Technologies," Federal Register, vol. 88, no. 170, August 2023, https://www.federalregister.gov/documents/2023/08/11/2023-17449/addressing-united-states-investments-in-certain-national-security-technologies-and-products-in, and Executive Order on Strengthening and Promoting Innovation in the Nation's Cybersecurity, Biden White House (Jan. 16, 2025), https://bidenwhitehouse.archives.gov/briefing-room/presidential-actions/2025/01/16/executive-order-on-strengthening-and-promoting-innovation-in-the-nations-cybersecurity/.
[295] Anu Bradford, *The Brussels Effect: How the European Union Rules the World* (Oxford Univ. Press 2020), https://academic.oup.com/book/36491

adaptable[296]. Unlike the market-access driven Brussels Effect, the quantum Washington Effect appears primarily security-motivated, highlighting the need for balance and international coordination to ensure its legitimacy and effectiveness and avoid detrimental fragmentation. Strategic 'recoupling' efforts with key players like China may be necessary to mitigate these risks. Thus, while the Washington effect could drive global convergence around U.S. values and priorities in quantum technology, it also underscores the need for regulatory humility, flexible policy design, and international coordination to avoid repeating the pitfalls of unilateral tech governance.

**IX. A 'BEIJING EFFECT' FOR QUANTUM STANDARDS**

Contrasting with the potential security-driven 'Washington Effect', a different dynamic—a potential 'Beijing Effect'—may arise from China's distinct approach to standardization and technological diffusion. A "Beijing Effect" for quantum standards could emerge if China leverages its manufacturing scale and strategic investments to flood global markets with cost-competitive quantum systems and services, embedding authoritarian governance norms into technical standards. By prioritizing affordability over transparency-such as quantum communication networks with state-mandated backdoors or quantum sensors optimized for biometric surveillance-China could export standards that normalize quantum-AI driven mass monitoring, central bank digital currencies with embedded social credit systems, and metaverse infrastructures enabling virtual autocracy. This approach, exemplified by its Digital Silk Road initiatives and state-driven ISO/IEC participation, risks fragmenting global interoperability while advancing technical benchmarks that prioritize control over privacy or ethical safeguards. For instance, China's National Standardization Development Program (NSDP) explicitly ties standards to national security objectives, raising concerns that quantum-enabled encryption protocols or quantum-AI hybrid architectures could institutionalize digital authoritarianism in critical infrastructure.[297]

Such a scenario underscores the urgency of counterbalancing China's first-mover standardization push with inclusive, rights-respecting frameworks to prevent quantum technologies from becoming vectors of techno-authoritarian diffusion. Concretely, a "Beijing Effect" in quantum standards-where China exports authoritarian-aligned technical norms via affordable quantum systems and infrastructure-can be mitigated through coordinated efforts by international standards bodies such as ISO/IEC, CEN-CENELECT and strategic initiatives like DARPA's Quantum Benchmarking Initiative (QBI).

---

[296] Gary Marchant et al., Learning From Emerging Technology Governance for Guiding Quantum Technology (August 09, 2024). Arizona State University Sandra Day O'Connor College of Law Paper No. 4923230, UNSW Law Research Paper No. 24-33, Available at SSRN: https://ssrn.com/abstract=4923230 or http://dx.doi.org/10.2139/ssrn.4923230 .
[297] Nat'l Standardization Dev. Program (China), *Standardization Development Plan* 4.2 (2021), https://www.sac.gov.cn

# X. QRITE: A QUANTUM-RELATIVISTIC INNOVATION THEORY OF EVERYTHING (A PHILOSOPHICAL THOUGHT EXPERIMENT)

To delve deeper into the fundamental dynamics shaping these complex global governance challenges, this section explores a speculative philosophical thought experiment, blending fundamental physics with social science. There is currently no established "quantum-relativistic innovation theory of everything." What follows is a hypothetical outline for such a theory, drawing conceptual parallels and inspirations from physics to frame a unified understanding of innovation dynamics. The outline should be understood as a theoretical exploration, using physics concepts primarily as powerful metaphors and perhaps as sources for mathematical or conceptual tools, rather than claiming direct physical causation of innovation phenomena in most instances. It explores how a synthesis analogous to physics' quest for unification might offer novel perspectives on governing transformative technologies.

**1. Premise and Ambition**

The Quantum-Relativistic Innovation Theory of Everything (QRITE) is envisioned as a unified conceptual framework aiming to describe the fundamental processes governing innovation across all scales – from individual cognition and small-team creativity to sectoral transformations and global technological diffusion. Its core premise is that innovation dynamics exhibit characteristics analogous to, or perhaps operate within constraints illuminated by, principles derived from quantum mechanics (describing uncertainty, discreteness, and interconnectedness at the micro-level) and general relativity (describing the influence of context, structure, and propagation limits at the macro-level). Reflecting the inherent quantum-classical nature of socio-technical systems, this theoretical lens seeks to move beyond disparate models by identifying universal principles governing the emergence, selection, and impact of novelty within complex socio-economic systems. It interprets the "theory of everything" not as unifying physical forces, but as providing a holistic, fundamental perspective on innovation itself. This remains a conceptual and largely analogical framework requiring rigorous development and empirical exploration.

**2. Core Principles and Components**

A. Quantum Dynamics of Novelty Generation (Micro-Level):
   1. *Innovation Quanta:* Novel ideas and technological breakthroughs often emerge as discrete, non-incremental advancements, akin to quantized energy levels. Incremental improvements exist, but radical innovation represents distinct "jumps."
   2. *Superposition of Potential Innovations:* In the early stages, multiple potential solutions, designs, or research pathways coexist simultaneously within minds, teams, or nascent markets. This represents a state of superposition, where diverse possibilities are explored before resources, decisions, or market feedback "collapse" the state onto a dominant trajectory.
   3. *Inherent Uncertainty and Complementarity:* Analogous to the Heisenberg Uncertainty Principle and Bohr's Complementarity, there exists a fundamental limit to simultaneously knowing or precisely defining the precise final form and the eventual market/societal impact of a radically new idea at its inception. Certain key aspects (e.g.

technical refinement vs. market applicability) might exist as complementary pairs, where optimizing for one obscures the full potential or nature of the other during the development process. Focusing resources sharply on one aspect may obscure understanding of the other. This uncertainty necessitates adaptive governance approaches.
   4. *Entanglement of Actors and Ideas:* Innovations are rarely isolated phenomena. Key concepts, researchers, funding sources, technological components, and market needs become deeply interconnected ("entangled"). The success or failure of one element can have instantaneous correlated effects on others within the innovation network, irrespective of conventional "distance" (e.g., geographical separation). Knowledge spillovers can appear to operate non-locally.
   5. *Source of Quantum Analogies in Cognition:* While largely metaphorical, QRITE acknowledges speculative theories like 'Quantum Brain' hypotheses. These suggest macroscopic quantum coherence might physically underpin cognitive processes like creativity and consciousness, in theory providing a deeper, albeit highly contested, physical layer to these quantum-like dynamics of idea generation and superposition observed at the micro-level of innovation.

B. Relativistic Dynamics of Innovation Diffusion and Context (Macro-Level):
   1. *Innovation Spacetime:* The socio-economic and institutional landscape within which innovation occurs can be conceptualized as a dynamic "spacetime." Existing structures (dominant firms, established regulations, technological paradigms, cultural norms) act like "mass," warping this landscape and influencing the trajectories ("geodesics") that new innovations follow. Path dependency is a manifestation of this curvature.
   2. *Frame-Dependent Perception and Value:* The significance, viability, and perceived value of an innovation are relative and depend heavily on the observer's frame of reference – their market position, regulatory perspective, cultural background, or existing technological infrastructure. There is no absolute, universal valuation independent of context.
   3. *Finite Speed of Diffusion:* The propagation of innovative ideas, practices, and technologies through the global system is not instantaneous. It is limited by the "speed of light" equivalents in socio-economic systems: communication network bandwidth, institutional absorption rates, cognitive limits, regulatory hurdles, and cultural adaptation speeds.
   4. *Time Dilation in Innovation Cycles:* Different sectors or regions experience innovation cycles at different rates. Highly dynamic environments ("high velocity" sectors) experience a faster "proper time" – more innovation cycles occur within a given external observer's time frame compared to slower, more static environments. Similarly, the rapid pace of technological progress relative to regulatory adaptation can be seen as a form of relativistic tension, mirroring time dilation effects and necessitating agile governance.
   5. *Equitable Access and the Equivalence Principle*: Analogous to the equivalence principle suggesting gravity affects all masses equally, QRITE implies that the 'gravity' of societal structures should not disproportionately hinder access to innovation's benefits. This motivates mechanisms for democratizing access, particularly for transformative technologies like quantum computing, ensuring benefits are shared equitably across nations and socio-economic groups.

C. Unification via Information, Complexity, and Fields:
   1. *Innovation Potential Field:* Borrowing from field theory, innovation might arise from fluctuations or excitations within an underlying "potential field" generated by the distribution of knowledge, capital, human talent, and unmet needs. Interactions within this field generate novel configurations.
   2. *Information as the Fundamental Substance:* QRITE posits that innovation is fundamentally about the creation, processing, and propagation of information. Quantum principles may illuminate the generation of novel information bits (perhaps conceptual "qubits of innovation"?) and their entanglement, while relativistic principles govern the constraints on information flow and its contextual interpretation within the global system.
   3. *Scale Transition and Emergence:* A key challenge for this quantum-classical framework is bridging the scales: How do the micro-level quantum-like phenomena (uncertainty, entanglement, complementarity) aggregate and interact to produce the emergent macro-level relativistic effects (landscape curvature, diffusion patterns)? Concepts from complexity science and statistical mechanics are needed to model this emergence.

**3. Potential Implications and Research Directions**

1. *Predictive Modeling:* Develop new classes of models for innovation forecasting and diffusion, incorporating quantum uncertainty and relativistic context-dependency. A promising application area is Responsible Quantum Simulation for Financial Forecasting: leveraging quantum algorithms (possibly within a QRITE framework acknowledging inherent uncertainties and systemic contexts) to enhance macroeconomic analysis, stress testing, and the modeling of complex financial instruments, while embedding ethical guardrails from the outset to manage the dual-use risks of powerful predictive capabilities.
2. *Policy Design:* Frame policy interventions (funding, regulation, infrastructure) in terms of their effects on the "innovation spacetime geometry" or their ability to foster "quantum jumps." This perspective supports the use of agile regulatory sandboxes for high-risk technologies and dynamic sunset clauses for rules governing rapidly evolving fields like quantum computing and AI, ensuring governance adapts at relevant timescales.
3. *Ecosystem Analysis:* Analyze the health and resilience of innovation ecosystems through the lens of entanglement and network interconnectedness.
4. *Strategic Management:* Guide R&D portfolio management by explicitly managing the "superposition" of exploratory projects and understanding "collapse" triggers and complementarity trade-offs.
5. *Ethical and Legal Frameworks:* Inform the development of adaptive ethical guidelines and legal theories for emerging technologies by acknowledging inherent uncertainty, complementarity, frame dependence, and the complex system dynamics highlighted by QRITE. This includes embedding human rights-by-design principles and exploring novel questions posed by quantum phenomena for legal concepts:
- Can micro-level (quantum-like) decision properties affect macro-level legal outcomes, modeled using quantum probability or cognition theories?
- Can quantum mechanical effects (superposition, indeterminism, complementarity) offer insights into understanding conflicting legal interpretations or judicial reasoning?
- Could quantum theory enrich legal principles like causation, responsibility, and justice?
- Explore Quantum Computational Law and Antitrust: Investigate the viability of using quantum computing (possibly guided by QRITE principles recognizing complexity and entanglement in markets) for computational antitrust analysis. This involves examining its relationship to existing computational antitrust doctrines and assessing how quantum simulation could enable ex-ante testing of antitrust policy interventions within complex macroeconomic digital twins, running on quantum-classical hybrid systems.

6. *Explore Micro-foundations:* Investigate links between QRITE's micro-level dynamics, cognitive science, and speculative quantum biology/brain theories (e.g., macroscopic coherence) to understand the fundamental origins of innovation.
7. *Mathematical Formalism:* Move beyond analogy to develop rigorous mathematical tools, potentially drawing from quantum information theory, network science, chaos theory, entropy concepts, and differential geometry adapted to socio-economic contexts.
8. *Empirical Validation:* Design methodologies (complex simulations, large-scale data analysis, comparative case studies) to test QRITE's conceptual propositions.

**4. A Quantum-Classical Perspective on Innovation**

The QRITE framework, as a philosophical thought experiment, represents an attempt to synthesize insights from fundamental physics to achieve a deeper, more unified understanding of innovation – a process arguably as fundamental to societal evolution as physical laws are to the cosmos. Its value lies not in asserting direct physical causation across disparate domains, (though it remains open to exploring theoretical quantum biological underpinnings of cognition) but in leveraging powerful conceptual tools and a holistic, quantum-classical perspective to ask new questions about how novelty arises and transforms our world. Pursuing this direction, particularly in exploring outcomes like quantum-enhanced finance or computational law, requires extensive interdisciplinary collaboration, a tolerance for exploring unconventional theoretical frontiers, and a commitment to ensuring innovation's profound influence bends towards global equity, justice, and equal abundance.

**XI. TOWARDS A UNIFIED GLOBAL INNOVATION SYSTEM**

While theoretical frameworks like QRITE offer novel perspectives on global governance challenges and divergent national effects, the practical challenge remains: forging a more unified global innovation system amidst the current geopolitical realities. Forging a unified global innovation system faces significant headwinds in the current era of heightened geopolitical competition, techno-nationalist impulses, and diverging governance models. The trend towards strategic decoupling or "de-risking" among major powers complicates prospects for deep integration. Nevertheless, the escalating downsides of fragmentation and the sheer scale of shared global challenges—from climate change and pandemics to ensuring the safe development of artificial intelligence—arguably make pursuing greater coherence and cooperation more critical than ever.

Ideally, such a system could strive to leverage complementary strengths observed across major ecosystems: harnessing the dynamism of the U.S.'s market-driven and entrepreneurial approach, integrating the EU's normative leadership and focus on sustainability and regulation anchored in societal values, and potentially adapting lessons from China's demonstrated strategic investment and scaling capabilities including long-term planning in combination with decentralized experimentation, *provided* these could operate within a framework of transparency and internationally accepted norms. Realizing even aspects of this vision demands extraordinary international cooperation, the articulation of shared goals for sustainable and ethical innovation (perhaps building on frameworks like the UN Sustainable Development Goals), and painstaking efforts towards the harmonization of regulatory frameworks, especially concerning data governance, AI ethics, and competition policy, to foster a diverse yet functionally cohesive global innovation landscape.

Within any effort towards a more unified or cooperative global innovation framework, universal human rights, fundamental freedoms, and democratic values must play a crucial, non-negotiable role in shaping ethical guardrails and operational practices. These principles are paramount for ensuring that innovations are conceptualized, developed, and deployed in ways that respect individual dignity and autonomy, promote equitable access and opportunity, counter bias and discrimination, and foster open and transparent ecosystems accountable to the public. Adherence to these values is fundamental to addressing global challenges like inequality and ensuring the benefits of technological progress contribute to broad societal well-being, distributed according to good governance principles of distributive justice.[298]

Upholding and actively promoting these values internationally also serves as a vital counterpoint to the rise of digital authoritarianism and ensures technology serves humanity.
Given current realities, progress may be more pragmatic through enhanced collaboration within alliances of like-minded nations (e.g., via platforms like the EU-U.S. Trade and Technology Council) aiming to align standards and approaches. However, tackling existential risks and truly global challenges necessitates broader multi-stakeholder dialogues and mechanisms capable of facilitating cooperation across geopolitical divides where vital common interests exist. Furthermore, a truly global system must actively empower and integrate innovators from the Global South, ensuring diverse perspectives shape the future of technology. Ultimately, embedding a strong commitment to human rights and democratic governance within innovation processes offers the most robust foundation for cultivating a more inclusive, responsible, and sustainable global innovation landscape that justly balances rapid technological progress with the welfare and flourishing of all people and groups in society.

---

[298] Kop, M., *Abundance & Equality*, in Scarcity, Regulation, and the Abundance Society, Mark A. Lemley ed., Lausanne, Switzerland: Frontiers in Science, 2022,
https://www.frontiersin.org/articles/10.3389/frma.2022.977684/full

## XII. THREE HYPOTHETICAL VISIONS ON A QUANTUM GOVERNANCE ACT

Given the distinct innovation ecosystems and strategic priorities of the United States, the European Union, and China, it is instructive to envision how each might hypothetically structure a dedicated legislative framework for quantum technologies. Such an exercise highlights the differing national or regional approaches to balancing innovation promotion, national security imperatives, economic competitiveness, and ethical considerations within the rapidly evolving quantum domain. The following outlines hypothetical "Quantum Governance Acts" for each, reflecting their respective governance philosophies and innovation models, while also considering pathways towards greater international alignment based on shared values.

### A. United States: Removing Barriers for U.S. Quantum Technology Act

A hypothetical U.S. legislative initiative, perhaps titled the "Removing Barriers for U.S. Quantum Technology Act," would likely be designed to operate within its predominantly market-driven innovation system. Its core focus would be on accelerating innovation while safeguarding national security and embedding ethical considerations. Key features would likely involve robust promotion of quantum research and development through incentives for both private and public sector investment. Concurrently, it would establish rigorous security standards and protocols, particularly addressing quantum threats to cryptography, to protect critical infrastructure and national interests. Rather than prescriptive rules, the Act might favor a flexible regulatory framework for safe use, employing soft law guidelines to steer the ethical application of quantum technologies without unduly hindering innovation. Emphasis would be placed on fostering partnerships and collaboration among government agencies, industry players, academic institutions, and key international allies. Crucially, given the pace of technological change, the Act would incorporate adaptability and review mechanisms, ensuring policies can be regularly updated. Furthermore, the Act could integrate principles of Responsible Quantum Innovation, mandating ethical impact considerations for funded research. This approach aims to leverage inherent U.S. strengths in technological innovation and commercialization, seeking a dynamic equilibrium between rapid development, robust security, and responsible governance.

This hypothetical U.S. Act strongly reflects the 'Advancing' (leadership) and 'Safeguarding' (national security) dimensions of RQI. Its market-oriented, flexible approach aims to accelerate innovation but may require vigilance to ensure ethical guardrails keep pace.

**B. European Union: The EU Quantum Act**

Reflecting the EU's emphasis on harmonization, fundamental rights, and collaborative research, a hypothetical "EU Quantum Act" would aim to establish a comprehensive and harmonized regulatory framework across its member states.[299] Central to this Act would be the embodiment of Responsible Quantum Innovation principles. Operating within the EU's collaborative research innovation system, it would intrinsically seek to balance technological advancement with social welfare and ethical principles. Central elements would include clearly defined ethical and safety standards to ensure quantum technologies are developed and deployed responsibly. Building on precedents like the GDPR and the AI Act, it would incorporate stringent data privacy and security measures, acknowledging quantum computing's potential impact on current encryption standards. The Act would actively support collaborative research and innovation through EU-wide funding mechanisms and initiatives, ensuring alignment with societal needs. Following the recent emphasis on anticipatory governance and responsible innovation tools in the EU, it would introduce Quantum Technology Impact Assessments (QIA). To let innovation breathe, the Act would institute agile regulatory sandboxes for high-risk quantum-AI hybrids. Recognizing the global nature of quantum science, it would foster international cooperation and advocate for global standards alignment. To avoid stifling progress, the framework would need built-in regulatory flexibility to adapt to rapid technological developments. Furthermore, consistent with EU practice, it would mandate public engagement and transparency, involving diverse stakeholders to enhance legitimacy and public trust. This envisioned Act leverages the EU's strengths in regulatory governance, prioritizing the ethical, secure, and socially beneficial trajectory of quantum technologies.

The envisioned EU Act emphasizes 'Safeguarding' through its values-based, rights-centric approach (building on GDPR/AI Act precedents) and 'Engaging' via collaboration. While conceivably slower on market 'Advancing', it aims for socially robust and ethical outcomes, incorporating QIAs and sandboxes.

**C. China: Comprehensive Quantum Law**

Envisioning a hypothetical Chinese "Comprehensive Quantum Law" requires considering its state-driven innovation system, which blends elements of authoritarian governance with surveillance capitalism, alongside its strategic technological ambitions. Initially, such a law might primarily focus on accelerating state-led development and application, prioritizing national security and economic competitiveness. It would likely detail guidelines for R&D, institute strict export controls, define parameters for international collaboration acceptable to the state, and potentially integrate quantum technologies into key industries under significant government oversight, reflecting China's established approach to technology governance.

However, contemplating a pathway towards greater global harmonization necessitates imagining a different structure for such a law – one potentially more aligned with the values underpinning the US and EU systems, fostering responsible innovation and enabling international cooperation. In this alternative vision, a Chinese Comprehensive Quantum Law could be drafted to include several key components, including explicitly incorporating Responsible Quantum Innovation (RQT) principles, potentially operationalized through frameworks like the 10 Principles for Responsible Quantum Innovation that operationalize the SEA framework.[300] It would define its purpose around promoting responsible development and establish its scope over all relevant entities. A regulatory framework could feature a National Quantum Office for oversight, clear licensing, certification, and compliance requirements based on legal norms and technical standards, and robust auditing, benchmarking, and verification schemes tailored to various industrial sectors (Healthcare, Energy, Transportation, Agrifood, Defense).

---

[299] For in depth reading on a potential EU Quantum Act, *see* Mauritz Kop, Constanze Albrecht & Urs Gasser, Towards an EU Quantum Act, forthcoming 2025.

Regarding research and development, it would outline funding mechanisms and encourage collaboration networks. Crucially, it would address ethical and legal considerations by implementing ethical standards addressing privacy, security, and dual-use concerns, alongside stringent data protection measures. In terms of international cooperation, it would aim to promote global partnerships for standards alignment and establish harmonized export controls consistent with international security agreements. For industrial and commercial application, it would develop sectorial guidelines and support innovation hubs. Significant investment in workforce development through education, training programs, and exchange initiatives would be crucial. Finally, robust monitoring and evaluation mechanisms, including periodic assessments and requirements for reporting and transparency, would ensure accountability and public trust.

Adopting such a structured, responsible framework would signal China's willingness to address the complex ethical, legal, and security challenges inherent in quantum technologies proactively. Furthermore, by genuinely aiming for international cooperation on global challenges and fostering fair competition, it could make a potential "recoupling" or at least a more stable and cooperative relationship with the U.S. and EU innovation ecosystems more feasible.[301]

A Chinese Act following current trends would prioritize 'Advancing' state goals and 'Safeguarding' state control. The alternative, globally aligned vision presented would represent a significant shift towards genuine international 'Engagement' and responsible 'Safeguarding' beyond purely national interests, a prerequisite for Sino-American recoupling.

These hypothetical legislative visions underscore how distinct national and regional contexts shape approaches to governing emerging technologies like quantum. The U.S. model prioritizes market dynamism and security, the EU emphasizes harmonized regulation rooted in fundamental rights and societal benefit, while China's approach is heavily influenced by state strategic objectives, though pathways for alignment based on responsible innovation principles can be conceived. While a single, unified global quantum governance act remains unlikely in the near term due to these divergences, understanding these national frameworks is crucial for navigating the complex geopolitical landscape, identifying areas for potential cooperation, managing risks, and working towards greater international coherence in ensuring quantum technologies benefit humanity safely and equitably.

---

[300] Kop *et al.*, *supra* note 166.
[301] *See* in this context: Jon Bateman, *U.S.-China Technological "Decoupling": A Strategy and Policy Framework*, Carnegie Endowment for International Peace (Apr. 2022), https://carnegieendowment.org/research/2022/04/us-china-technological-decoupling-a-strategy-and-policy-framework?lang=en.

**XIII. BRIDGING CULTURAL DIFFERENCES BETWEEN THE TECH BLOCKS**

The distinct innovation pathways pursued by the United States, the European Union, and China are deeply interwoven with their unique cultural tapestries, historical trajectories, institutional arrangements, political systems, and societal habits. The U.S. system, often characterized by individualism, market fundamentalism, and a tolerance for disruption, fosters rapid, often market-driven innovation cycles and a vibrant entrepreneurial ecosystem. The European Union, emphasizing communitarian values, social welfare, and regulatory prudence, cultivates a more collaborative research environment focused on aligning technological progress with sustainability, ethical considerations, and fundamental rights, often acting as a global standard-setter through comprehensive legislation. China's approach reflects its millennia-old civilization that is rich in cultural depth, emphasis on collective goals, and modern state-led development model, enabling immense resource mobilization, rapid scaling, and strategic long-term technological planning within a framework that integrates political control with economic ambition. Recognizing and appreciating this cultural diversity is essential; each system possesses unique strengths contributing varied perspectives and capabilities to the global innovation landscape. However, we must simultaneously confront the stark political realities of 2025, marked by intensifying geopolitical competition, diverging values regarding governance and human rights, and widespread techno-nationalism.[302] Despite these profound differences and tensions, the imperative to align on addressing shared global challenges – such as combating climate change, reducing systemic inequality, solving the depth crisis, and responding to pandemics – remains undeniable and increasingly urgent for collective survival and prosperity.

Framing the emerging great quantum power rivalry as an inherent zero-sum game requires critical evaluation; indeed, analyses suggest this perspective can be dangerously misleading. Viewing contemporary geopolitical and technological competition through a simplistic "Cold War redux" narrative risks repeating historical errors, often ignoring the instability and violence associated with past superpower confrontations while fueling domestic nativism and xenophobia.[303] This adversarial framing, often offering limited and unequal economic benefits domestically, significantly hinders the international cooperation urgently needed to address transnational threats like climate change, pandemics, and nuclear proliferation, often alienating global actors and nations in the Global South who may not perceive the rivalry in the same existential terms and seek partnership rather than alignment in a bipolar contest.[304] Applied specifically to the quantum domain, perceiving advancements solely through this zero-sum lens risks stimulating a costly and destabilizing quantum arms race, obscuring the complex, interconnected nature of scientific progress. Similar to findings in the AI sector, the quantum field involves significant global interdependencies, including the international circulation of talent, foundational research collaborations, and cross-border materials and devices supply chains, suggesting that breakthroughs are not purely national gains or losses.

---

[302] For further reading on techno-nationalism, *see* Luo Y. Illusions of techno-nationalism. J Int Bus Stud. 2022;53(3):550-567. doi: 10.1057/s41267-021-00468-5. Epub 2021 Sep 7. PMID: 34511654; PMCID: PMC8421717, https://pmc.ncbi.nlm.nih.gov/articles/PMC8421717/

[303] Van Jackson & Michael Brenes, *The Rivalry Peril: How Great-Power Competition Threatens Peace and Weakens Democracy* (Yale Univ. Press 2025), available at https://yalebooks.co.uk/book/9780300272895/the-rivalry-peril/.

Policies derived from a nativist, zero-sum viewpoint – such as broad restrictions on international scientific exchange, talent mobility, or access to foundational research tools – can directly undermine the trust and cooperation essential for establishing global norms and verification regimes. Such approaches actively hinder vital efforts towards global quantum non-proliferation by reducing transparency and disincentivizing adversaries from participating constructively in safety and security frameworks. Policies should instead foster a multidimensional relationship shaped by quantum-AI governance, differing technological architectures, talent mobility, and deeply intertwined innovation ecosystems. Evidence supporting this includes the significant two-way migration of top AI talent, the demonstrably high impact of U.S.-China collaborative research (reportedly exceeding U.S.-U.K. academic collaborations in impact), and existing corporate synergies like those between Apple and Alibaba or Ford and DeepSeek, which highlight practical interdependencies.[305] Consequently, policies derived from a zero-sum assumption, such as broad restrictions on semiconductor access through the 2025 Framework for Artificial Intelligence Diffusion or proposed limitations on academic visas like the "Stop CCP VISAs Act,"[306] risk disrupting these productive linkages and hindering global innovation progress more than securing a decisive national advantage.[307] Therefore, portraying the quantum challenge as a purely zero-sum contest ignores complex realities, potentially accelerates dangerous quantum arms race dynamics fueled by suspicion and techno-McCarthyism, and obstructs the crucial international cooperation required to navigate the profound implications of quantum technologies safely and equitably.

Beneath the surface of political and cultural divergence lies an ethical baseline of values that connects all humans: the inherent dignity of people, the desire for security and well-being, the aspiration for future generations to flourish, and the shared vulnerability of our species on a single planet. Appealing to these fundamental, universally resonant values may offer a starting point for building bridges, even between systemic rivals. This foundation is crucial when considering governance for transformative technologies like quantum computing. To achieve broad participation, including China's, in vital initiatives like quantum non-proliferation and the development of a global quantum acquis (a common body of principles and rules), carefully calibrated incentives are paramount. These might include mutual security assurances against quantum-enabled threats (like decryption of critical secrets), guaranteed participation in shaping global technical standards essential for economic competitiveness, access to collaborative international scientific research on fundamental quantum science, and mechanisms ensuring equitable access to beneficial applications, perhaps echoing the NPT's framework differentiating peaceful uses from proliferation.[308]

---

[304] Mike Cummings, *Cold War Redux? Weighing the Risks of a 'Great-Power' Contest with China*, Yale News (Feb. 24, 2025), https://news.yale.edu/2025/02/24/cold-war-redux-weighing-risks-great-power-contest-china.
[305] *The AI Superpower Rivalry: A Zero-sum Game Between China and the United States?*, The Diplomat (Mar. 21, 2025), https://thediplomat.com/2025/03/the-ai-superpower-rivalry-a-zero-sum-game-between-china-and-the-united-states/.
[306] S. 1086, 119th Cong. (2025), available at https://www.congress.gov/bill/119th-congress/senate-bill/1086/text.
[307] *ibid.*

History suggests that forging global consensus often requires significant moments or shared perceived threats. The horrors of World War II catalyzed the formation of the United Nations; the existential fear of nuclear annihilation during the Cold War provided the grim impetus for the NPT and the establishment of the IAEA to oversee its verification.[309] Is the dawning quantum-AI age itself such a moment, akin to discovering complex life elsewhere or facing an imminent global catastrophe? Or will it require a more acute shock to compel unified action? Would Q-Day, the day that quantum computers break classical encryption, be devastating enough? Regardless, we must learn from the negotiation processes that led to the NPT and IAEA.[310] Understanding the delicate balance of national interests, security concerns, verification challenges, and negotiated incentives (like access to peaceful nuclear technology under safeguards) that brought rival nations to the table is critical for designing analogous frameworks for quantum governance.[311]

Furthermore, the quantum-AI era demands evolving moral values and ethical frameworks.[312] Ethics are not static; they are inherently contextual, culturally sensitive, and dynamic. What constitutes responsible innovation, fair data use, or acceptable risk in the quantum age requires continuous global dialogue and deliberation, sensitive to diverse cultural perspectives yet striving for broadly applicable principles. These evolving quantum age ethics must form the normative organ point of future global quantum institutions, laws, and treaties. Without such a foundation, attempts at global governance risk being perceived as illegitimate or failing to address the unique challenges posed by these powerful new technologies.[313] Ultimately, bridging cultural divides in the face of political headwinds requires acknowledging anthropological differences, leveraging shared values, learning from history, and committing to an ongoing, inclusive ethical dialogue to navigate our shared technological future responsibly.

---

[308] *See, e.g.*, Treaty on the Non-Proliferation of Nuclear Weapons art. IV, July 1, 1968, 21 U.S.T. 483, 729 U.N.T.S. 161, https://disarmament.unoda.org/wmd/nuclear/npt/text/.

[309] *See e.g.* Lalla, Vijay, "*The Effectiveness of the Comprehensive Test Ban Treaty on Nuclear Weapons Proliferation: A Review of Nuclear Non-Proliferation Treaties and the Impact of the Indian and Pakistani Nuclear Tests on the Non-Proliferation Regime*" 8 Cardozo Journal of International and Comparative Law (2000).

[310] Bertrand Goldschmidt, *The Negotiation of the Non-Proliferation Treaty (NPT)*, 27 IAEA Bulletin 2 (1985), available at https://www.iaea.org/sites/default/files/publications/magazines/bulletin/bull27-2/27204782024.pdf.

[311] *See e.g.* Dyral Kimball, *Timeline of the Nuclear Nonproliferation Treaty (NPT)*, Arms Control Association, https://www.armscontrol.org/factsheets/timeline-nuclear-nonproliferation-treaty-npt and Ionut Suseanu, *The NPT and IAEA Safeguards*, IAEA Bulletin, https://www.iaea.org/bulletin/the-npt-and-iaea-safeguards.

[312] *See e.g.* Kop, *supra* note 298.

**XIV. TOWARD A HARMONIZED ACQUIS PLANETAIRE FOR QUANTUM**

The concept of an *acquis communautaire* within the European Union denotes a foundational body of common rights, obligations, laws, and norms binding upon all member states, creating a zone of legal predictability and shared understanding.[314] Extrapolating this idea to the global stage suggests the possibility of an *acquis planétaire* – a worldwide framework encompassing shared principles, rules, standards, and best practices. Applying this concept to the rapidly advancing field of quantum technologies yields the notion of a quantum acquis: a globally recognized body of knowledge, norms, and potentially harmonized regulations designed to guide the development, deployment, and governance of this transformative technological suite. Establishing such an acquis represents a profound challenge, yet possibly an essential step for navigating the complex opportunities and risks presented by quantum science and engineering on a planetary scale.

Recent developments reflect this tension. Calls for international dialogue on quantum governance, such as by the UN Secretary-General following the 2024 AI Resolution, and bottom-up efforts like the joint statement by global quantum research institutions advocating for open science principles, show momentum towards harmonization. However, persistent geopolitical rivalries, fragmented supply chains, and competing national security interests remain significant barriers, necessitating identification of common ground (e.g., preventing catastrophic misuse, enabling quantum for global challenges) that transcends competition.

Building upon the preceding comparative analysis, the distinct approaches of the US (market-driven, security-focused), the EU (values-based, regulatory), and China (state-led, strategic) highlight significant tensions rooted in differing political systems, economic philosophies, and societal priorities. Harmonizing these perspectives into a unified framework requires reconciling fundamental divergences. This trend towards legislative frameworks is seen in both the EU's comprehensive, risk-based AI Act and (proposed) U.S. efforts like the Algorithmic Accountability Act, which, despite differing scopes—targeting 'AI systems' versus 'automated decision systems' respectively—both signal a move beyond self-regulation toward balancing innovation with accountability.³¹⁵ While common ground exists – including a shared acknowledgment of quantum's strategic importance and the necessity of addressing risks responsibly – deep divisions persist regarding the preferred governance models (market vs. values vs. state-centric) and the balance between openness and control. Nonetheless, common grounds also exist: all major players acknowledge the profound strategic importance of quantum technologies; there is a shared, albeit differently interpreted, recognition of the need for responsible development encompassing security, safety, and ethical considerations; and all value technological advancement as critical for future prosperity and influence. These shared interests, coupled with the unique global risks posed by quantum technologies (e.g., the potential to break current encryption standards universally, fueling a cryptographic transition crisis, or the unforeseen capabilities of quantum-AI hybrids), provide a compelling rationale for pursuing greater international coherence.³¹⁶

---

[313] *See generally* Mireille Hildebrandt, *Law for Computer Scientists and Other Folk* 187-218 (Oxford Univ. Press 2020), available at https://global.oup.com/academic/product/law-for-computer-scientists-and-other-folk-9780198860884. (discussing the challenges of regulating disruptive technologies and the need for adaptable legal-ethical frameworks).
[314] *See* European Parliament, *The Enlargement of the Union*, Fact Sheets on the European Union, https://www.europarl.europa.eu/factsheets/en/sheet/167/the-enlargement-of-the-union.

A globally harmonized quantum acquis could be instrumental in fostering international collaboration needed to tackle planetary-scale challenges. Universally accepted technical standards (for interoperability, security, and safety), responsible technology metrics and benchmarks (for performance and ethical impact assessment), and trust enhancing certifications could significantly lower friction for joint research and deployment efforts targeting climate change modeling, drug discovery, materials science for sustainable energy, or optimizing resource allocation to combat inequality. By creating a common operational language (building upon efforts by ISO/IEC JTC 3) and predictable regulatory environment, such tools facilitate trust and streamline cooperation, allowing resources to be pooled more effectively towards shared global goals. Harmonization could apply across the full suite of quantum technologies – including computing, sensing, metrology, and networking/communication – as well as integrating with related fields. This would involve international agreement on ethical use guidelines, robust safety protocols, common security standards (crucially, for post-quantum cryptography), frameworks for handling quantum data, and ensuring the reliability of increasingly complex systems, such as those involving quantum effects in advanced semiconductors and quantum-AI hybrids. Inspiration can be drawn from ongoing efforts towards globally harmonized AI regulation, such as the recent UN Resolution on AI, which seek consensus on principles like safety, fairness, transparency, and accountability.³¹⁷

However, the pursuit of a quantum acquis planétaire must traverse the complex realities of competition versus cooperation. It is likely, as observed in AI, that a mixed scenario will prevail, with intense competition in areas deemed critical for national security or economic advantage, alongside cooperation in fundamental research, global risk mitigation (like non-proliferation), or addressing shared humanitarian challenges. National and economic safety and security, alongside the protection of intellectual property, will inevitably act as boundaries limiting openness.[318] The principle of innovation being "as open as possible, and as closed as necessary" becomes central, yet defining this balance is inherently disputatious. Decisions on control levels are made within national contexts by government bodies (like the U.S. Dept. of Commerce and China's MIIT) based on strategic assumptions about security and competitiveness, heavily influenced by the existing geopolitical climate. A functional global quantum acquis would require mechanisms for ongoing dialogue and dispute resolution regarding these sensitive boundaries. An example of an international dispute resolution mechanism for emerging technologies is the International Atomic Energy Agency (IAEA), which provides a framework for nuclear technology governance. Its safeguarding principles - including verification, compliance, technical expertise, neutrality, shared standards, and arbitration panels- offer a template for globally harmonized quantum governance.[319]

---

³¹⁵ Mökander, Jakob and Juneja, Prathm and Watson, David and Floridi, Luciano, The US Algorithmic Accountability Act of 2022 vs. The EU Artificial Intelligence Act: What can they learn from each other? (August 18, 2022), Minds & Machines (2022). https://doi.org/10.1007/s11023-022-09612-y, Available at SSRN: https://ssrn.com/abstract=4268345
³¹⁶ See Kop, Mauritz, *Establishing a Legal-Ethical Framework for Quantum Technology*, Yale Journal of Law & Technology, The Record, (Mar. 30, 2021), https://yjolt.org/blog/establishing-legal-ethical-framework-quantum-technology.
³¹⁷ *See* UN News, *General Assembly adopts landmark resolution on artificial intelligence* (Mar.21, 2024), https://news.un.org/en/story/2024/03/1147831. The text of the AI Resolution can be found here: https://docs.un.org/en/A/78/L.49.

Achieving such harmonization faces obstacles noted in AI governance comparisons, including differing legal focal points (e.g., EU fundamental rights vs. U.S. consumer protection emphasis), varied enforcement structures, and complex interoperability challenges across technical, legal, and jurisdictional norms.[320] The rationale for establishing a harmonized acquis rests on promoting stability, facilitating beneficial cooperation, managing shared existential risks, fostering interoperability and fair trade, and embedding ethical principles globally. Arguments against center on the immense difficulty of achieving consensus among diverse political systems, the risk of regulations becoming outdated quickly or stifling innovation through premature standardization, concerns over national sovereignty, and the potential for frameworks to be dominated by powerful actors or reflect only a lowest-common-denominator compromise. Indeed, the inherent uncertainty, complexity, rapid pace, and pervasive dual-use nature of quantum technologies challenge traditional regulatory models, reinforcing the necessity for incorporating agile and adaptive governance tools like guardrails, sandboxes, and impact assessments into any effective global acquis.

A pragmatic pathway towards a global quantum acquis might involve several steps: leveraging existing international bodies (UN, ISO/IEC, adapted Wassenaar) rather than immediately seeking new treaties; achieving initial consensus on foundational principles via a non-binding UN Declaration on Quantum; developing modular technical standards for interoperability and safety; promoting strategic U.S.-China recoupling in specific non-sensitive areas; and designing smart regulations that incentivize responsible behavior (RQT by design, ethical procurement) rather than relying solely on restrictions.

---

[318] Kop *et al., supra* note 165
[319] International Atomic Energy Agency, *Statute of the IAEA* art. XVII (Settlement of Disputes), https://www.iaea.org/about/statute.
[320] Halim, Noha Lea and Gasser, Urs, Vectors of AI Governance - Juxtaposing the U.S. Algorithmic Accountability Act of 2022 with The EU Artificial Intelligence Act (May 9, 2023). Available at SSRN: https://ssrn.com/abstract=4476167 or http://dx.doi.org/10.2139/ssrn.4476167

Despite the challenges, outlining the constituent elements of a quantum acquis is key to guiding future efforts. The acquis should encompass:

1. Foundational Ethical Principles: Rooted in universal human rights, promoting fairness, accountability, transparency, safety, and responsible stewardship.
2. Common Lexicon: Standardized terminology and definitions for quantum concepts and technologies (building on work by ISO/IEC JTC 3).
3. Technical Standards: Interoperability protocols (e.g., for quantum networks), security standards (including PQC algorithms and QKD benchmarks), safety guidelines, and metrology standards.
4. Regulatory Best Practices: Frameworks for risk assessment, ethical impact evaluation, data governance for quantum systems, and responsible development lifecycle management. Furthermore, regulatory best practices within the acquis should incorporate 'guardrail' principles, focusing on guiding human decision-making processes that utilize quantum-AI systems rather than solely regulating the technology itself, ensuring human agency and oversight are maintained in critical contexts.[321]
5. Non-Proliferation Norms: Mechanisms for transparency, verification (inspired by IAEA safeguards for nuclear technology), and coordinated export controls for sensitive dual-use quantum applications.
6. Adaptive Governance Mechanisms: Multi-stakeholder platforms for continuous dialogue, review, and updating of the acquis to keep pace with rapid technological evolution.

Striving towards a harmonized *acquis planétaire* for quantum technologies is not merely an idealistic pursuit but a pragmatic necessity for ensuring this powerful technological revolution unfolds in a way that is safe, equitable, and beneficial for humanity as a whole, demanding a delicate balance between national interests and collective responsibility.

**XV. A NON-PROLIFIRATION TREATY & IAEA FOR QUANTUM**

As we contemplate regulatory interventions for quantum technologies, historical precedents from adjacent fields offer invaluable lessons. Experience with nuclear physics (both fission and fusion), biosciences, nanotechnology, spatial computing, and agentic artificial intelligence underscores the importance of foresight, adaptability, and international cooperation in governing transformative and dual-use capabilities. Building upon these insights, recent discourse increasingly advocates for a dedicated international framework for quantum technologies, specifically drawing inspiration from the landmark Treaty on the Non-Proliferation of Nuclear Weapons (NPT) and its verification body, the International Atomic Energy Agency (IAEA), to address the unique strategic risks and opportunities quantum

---

[321] *See generally* Urs Gasser & Viktor Mayer-Schönberger, Guardrails: Guiding Human Decisions in the Age of AI (2024), https://press.princeton.edu/books/hardcover/9780691150680/guardrails

presents.³²² Operationalizing key components of a Quantum Acquis, particularly concerning non-proliferation and verification, necessitates examining established international institutional models, such as the NPT and IAEA.

The NPT, established in 1968, serves as a cornerstone of global security, resting on three pillars: preventing the spread of nuclear weapons and related technology (non-proliferation); committing signatory states, particularly declared nuclear-weapon states, to pursue negotiations towards eventual disarmament; and guaranteeing the inalienable right of states to develop and use nuclear energy for peaceful purposes under international safeguards.³²³ The IAEA provides the critical verification mechanism, conducting inspections and monitoring activities in non-nuclear-weapon states to ensure nuclear materials and technology are not diverted towards weaponization, thereby building international confidence in treaty compliance. Recent discussions, such as the March 2024 United Nations Institute for Disarmament Research (UNIDIR) expert roundtable on quantum governance models, explored lessons from existing regimes (IAEA, Chemical Weapons Convention (CWC)) and highlighted the value of a dedicated international body for quantum monitoring, verification, capacity building, and confidence-building measures. Enforcement relies on this verification system, backed by international diplomacy and the potential for collective action, including sanctions, in response to non-compliance.³²⁴

Adapting this proven, albeit imperfect, framework to quantum technologies, related AI applications (recognizing that quantum needs AI and AI increasingly needs quantum for advanced capabilities) and Lethal Autonomous Weapons (LAWS) offers a pathway to manage emerging unacceptable risks.³²⁵ Similar to nuclear technology, many quantum applications possess inherent dual-use characteristics; advancements in quantum sensing, computing, and communication could revolutionize civilian industries but also enable novel military capabilities or undermine global cybersecurity (e.g., via cryptographically relevant quantum computers). An NPT-like structure – realized through a dedicated UN Quantum Treaty modeled on precedents like the 2024 UN AI Resolution, the Council of Europe Framework Convention on AI, and the NPT itself – could aim to: (i) control the proliferation (Non-Proliferation) of specific high-risk quantum technologies; (ii) establish future limitations (Disarmament/Limitation) on certain quantum-enabled weapons, especially LAWS, ensuring meaningful human control (which could involve implementing adaptable 'guardrail' principles focused on governing the human decision-making processes augmented by these systems, rather than solely regulating the underlying technology itself; (iii) guarantee the right to peaceful quantum R&D under safeguards (Peaceful Uses), designed explicitly to align quantum advancements with global imperatives such as the UN SDGs; and (iv) mandate transparency and establish Verification mechanisms to ensure development remains oriented towards peaceful uses and adheres to international humanitarian law and ethical principles.

---

³²² *See e.g.* Kop, Albrecht & Gasser, *supra* note 299; Mauritz Kop, Regulating Transformative Technology in The Quantum Age: Intellectual Property, Standardization & Sustainable Innovation (October 7, 2020). Stanford - Vienna Transatlantic Technology Law Forum, Transatlantic Antitrust and IPR Developments, Stanford University, Issue No. 2/2020, https://law.stanford.edu/publications/regulating-transformative-technology-in-the-quantum-age-intellectual-property-standardization-sustainable-innovation/ or https://ssrn.com/abstract=3653544; and Gary Marchant et al., Learning From Emerging Technology Governance for Guiding Quantum Technology (August 09, 2024). Arizona State University Sandra Day O'Connor College of Law Paper No. 4923230, UNSW Law Research Paper No. 24-33, Available at SSRN: https://ssrn.com/abstract=4923230 or http://dx.doi.org/10.2139/ssrn.4923230 .
³²³ Treaty on the Non-Proliferation of Nuclear Weapons, July 1, 1968, 21 U.S.T. 483, 729 U.N.T.S. 161, https://disarmament.unoda.org/wmd/nuclear/npt/text/.
³²⁴ Ionut Suseanu, *The NPT and IAEA Safeguards*, IAEA Bulletin (Sept. 10, 2020), https://www.iaea.org/bulletin/the-npt-and-iaea-safeguards.
³²⁵ United Nations Office for Disarmament Affairs, *Lethal Autonomous Weapon Systems (LAWS) – Background on LAWS in the CCW*, https://disarmament.unoda.org/the-convention-on-certain-conventional-weapons/background-on-laws-in-the-ccw/.

The rationale for establishing such a dedicated Quantum Non-Proliferation Treaty (QNPT) and an associated oversight body (an Atomic Agency for Quantum-AI, or "IAEA-Q") is compelling, despite its obvious challenges. Quantum technologies harbor the potential for strategic disruption on a scale perhaps unseen since the advent of nuclear weapons. A QNPT could proactively mitigate the risk of a destabilizing quantum arms race by establishing norms against weaponization and fostering transparency. It could provide the security assurances necessary to enable robust international cooperation on peaceful applications, preventing the entire field from becoming securitized and fragmented. A verification regime, overseen by an IAEA-Q inspired by the IAEA's expertise in safeguards implementation, would be essential for building trust and predictability among nations regarding quantum activities. Such a body would focus specifically on proliferation risks, complementing other regulatory layers like national market authorization agencies or broader UN resolutions on responsible quantum technology.

However, significant challenges and arguments against must be recognized. For example, defining what constitutes "weaponizable" quantum technology is far more complex than for nuclear materials, given the intangible nature of algorithms, software, and data, and the pervasive dual-use ambiguity. Verification will be immensely difficult, requiring unprecedented technical expertise and at times intrusive access to research labs and data centers, likely facing strong resistance based on national security (including state secrets) and intellectual property (including trade secrets) concerns. Concerns about national sovereignty (states wanting to be 'boss in their own backyard') may also lead to reluctance regarding new binding international treaties, suggesting pathways built upon adapting existing frameworks might be more politically feasible. Achieving political consensus among major powers with deviating interests and values represents a formidable diplomatic hurdle. Furthermore, achieving the necessary political consensus for a QNPT/IAEA-Q must navigate fundamentally different regulatory philosophies, such as the contrast between more focused, impact-based approaches seen in US AI proposals versus the EU's broader, rights-centric, system-level AI regulation. Moreover, any global quantum regime must effectively navigate the complex 'maze of norms'—interfacing with diverse national laws, technical standards, ethical guidelines, and differing enforcement models—presenting significant interoperability challenges.[326] Experiences comparing AI governance approaches, such as the EU AI Act's rights-based focus versus proposed US legislation emphasizing consumer impacts and FTC enforcement, illustrate the deep-seated differences in legal priorities and structures that would need to be reconciled within a global quantum framework. Furthermore, overly broad or premature restrictions risk stifling legitimate scientific research and beneficial commercial innovation.

---

[326] Halim & Gasser, *supra* note 320

Specifically, verification challenges for quantum are immense due to intangible software/algorithms, the rapid pace of change, and pervasive dual-use ambiguity. Defining thresholds for control (e.g., cryptographically relevant capability) is difficult. Achieving political consensus and ensuring compliance among rival powers, while respecting sovereignty and avoiding stifling innovation, requires overcoming significant hurdles and necessitates adaptive approaches like focusing on specific high-risk capabilities, using Confidence Building Measures (CBMs), or starting with non-binding codes.

Critically, learning from the NPT's history involves proactively addressing its known imperfections when designing a QNPT and its oversight body. This includes striving for greater universality from the outset and structuring commitments related to limiting weaponization ('disarmament') in parallel with non-proliferation efforts to ensure a more balanced and widely accepted 'grand bargain'. Furthermore, acknowledging the profound verification challenges posed by quantum's often intangible nature necessitates embedding more robust, technologically assisted monitoring mechanisms and stronger transparency norms than were feasible for the NPT regime. Finally, a quantum framework should perhaps incorporate more dynamic review and adaptation processes to better cope with the anticipated rapid evolution of quantum and AI technologies, ensuring the regime remains relevant and effective over time.

To break the ice, envisioning key elements of a hypothetical QNPT is crucial. This framework might include:

- **Pillars:** Explicit commitments to non-proliferation of designated high-risk quantum capabilities, flanked by future commitments towards limitations or disarmament concerning specific quantum-enabled weapons, and affirmation of the right to peaceful quantum R&D under agreed safeguards.
- **Definitions:** Attempting clear distinctions between prohibited activities/technologies and permitted peaceful applications (a core challenge).
- **Obligations:** Tailored commitments for states based on their quantum capabilities (e.g., advanced research states vs. others).
- **Verification & Transparency:** Mandates for declarations, data sharing, and inspections managed by the oversight body.
- **Peaceful Use Incentives:** Mechanisms for facilitating access to non-sensitive quantum knowledge or technology for compliant states.

A corresponding Atomic Agency for Quantum, or "IAEA-Q" would require:

- **Mandate:** Verifying QNPT compliance, implementing international safeguards (drawing on IAEA expertise) including principles of responsible quantum technology, promoting quantum safety and security culture globally, possibly developing technical standards for peaceful uses. Beyond verification, the IAEA-Q's mandate could incorporate developing 'guardrails' – flexible norms and best practices focused on guiding the responsible use of quantum systems in critical decision-making processes, ensuring human oversight and mitigating risks arising from the technology's inherent limitations, rather than solely focusing on bans.
- **Structure:** An international governance board, a technical secretariat with deep quantum and AI expertise, and a specialized inspectorate, flanked by interdisciplinary subcommittees specializing in application areas like energy, health, climate change, environment, finance, industry, and fundamental quantum and AI science.
- **Functions:** Analyzing declarations, conducting technical monitoring (remotely and on-site where agreed), facilitating information exchange, reporting compliance findings to treaty members and relevant international bodies (e.g., UN); integrating or harmonizing Quantum Technology Impact Assessments (QIA) internationally.

Furthermore, alongside the non-proliferation oversight functions of an IAEA-Q, collaborative research platforms akin to CERN (emulating its model for international resource pooling and research coordination) or refined international nuclear fusion megaprojects (like ITER) could be established specifically to boost coordinated responsible innovation in quantum science and technology.[327] Achieving ambitious goals like universal fault tolerant quantum computers or quantum-centric supercomputing along with algorithmic development and use case discovery toward quantum benefit requires such collective global expertise, making protectionist measures counterproductive.

While establishing a global quantum non-proliferation framework mirrored after the NPT/IAEA structure presents political and technical challenges in our current geopolitical climate, these difficulties should galvanize, rather than deter, concerted international effort. Beneath the surface of political and cultural divergence lies an ethical baseline of values that connect all humans: the inherent dignity of people, the desire for security and well-being, the aspiration for future generations to flourish, and the shared vulnerability of our species on a single planet. Appealing to these fundamental, universally resonant values offers a powerful starting point for building bridges, even between systemic rivals, providing a crucial foundation for governing transformative technologies like quantum computing.

Securing China's participation in both a global quantum acquis and a potential Quantum Non-Proliferation Treaty (QNPT) is challenging but crucial for the framework's effectiveness. A strategy focused solely on pressure or exclusion is unlikely to succeed; instead, incentives must appeal to China's own perceived strategic interests. Offering China a significant, genuinely influential role in *co-designing* these international frameworks – rather than presenting it with predetermined rules – could address concerns about fairness and allow it to help shape norms governing a technology central to its future ambitions. Emphasizing the economic benefits of participation, such as guaranteed access to global markets through interoperability standards and reduced trade friction (export controls, tariffs), aligns with China's economic priorities. Furthermore, the mutual security assurances offered by a QNPT – reducing the shared risks of a destabilizing quantum arms race, mitigating the universal threat of quantum decryption ("Q-Day"), and preventing accidental escalation – directly serve China's interest in stability. Carefully structured access to international scientific collaboration in fundamental, non-sensitive quantum domains could also be appealing, complementing China's drive for scientific leadership while being managed under agreed principles and safeguards. Framing participation not as a concession, but as integral to securing China's own long-term security, economic prosperity, and international standing as a responsible leader in a pivotal technological era, likely offers the most promising pathway to garnering its essential cooperation.

---

[327] Alberto Di Meglio et al., CERN Quantum Tech. Initiative, *CERN Quantum Technology Initiative: Strategy and Roadmap* (Oct. 7, 2021), https://cds.cern.ch/record/2789149?ln=en.

The success of any comparable international regime, much like the NPT's enduring relevance, will heavily depend on the cooperation and commitment of signatory states to genuinely uphold its principles. Achieving the necessary broad participation, including that of major quantum powers like China, hinges on designing carefully calibrated incentives. Such incentives could include mutual security assurances against quantum-enabled threats (such as widespread cryptographic failure), guaranteed roles in shaping essential global technical standards, continued access to collaborative international scientific research in fundamental quantum science, and equitable mechanisms for sharing the benefits of peaceful applications, perhaps echoing the principles enshrined in NPT Article IV regarding peaceful uses.

History teaches that forging global consensus on critical security matters *is* possible, often catalyzed by significant moments or shared perceived threats – the aftermath of WWII spurred the UN's creation, and the existential fear of nuclear annihilation drove the formation of the NPT/IAEA regime. Whether the dawn of the quantum-AI age itself, or a future disruptive event like "Q-Day" when quantum computers demonstrably break classical encryption, will serve as a sufficient catalyst remains uncertain. Regardless, we must learn from the negotiation processes[328] that successfully established the NPT and IAEA, understanding the delicate balance achieved between national security interests, credible verification challenges, and negotiated incentives that brought rival nations to agreement.[329] This historical experience provides invaluable lessons for designing analogous, effective frameworks for quantum governance. Ultimately, the strategic stakes and proliferation risks associated with advanced quantum capabilities highlight the imperative for a dedicated international framework, flanked by sector-specific best practices and modeled on precedents like the NPT, UN AI Resolution, and CoE AI Convention, overseen by an IAEA-like agency for quantum and quantum-AI hybrids, to manage global security and ensure responsible development. These institutional steps represent crucial measures to help steward quantum advancements for humanity's shared benefit, demanding our urgent global collaborative focus starting today.

---

[328] Bertrand Goldschmidt, *The Negotiation of the Non-Proliferation Treaty (NPT)*, 27 IAEA Bulletin 2 (1985), *available at* https://www.iaea.org/sites/default/files/publications/magazines/bulletin/bull27-2/27204782024.pdf.

[329] Daryl G. Kimball, *Timeline of the Nuclear Nonproliferation Treaty (NPT)*, Arms Control Ass'n, https://www.armscontrol.org/factsheets/timeline-nuclear-nonproliferation-treaty-npt

## XVI. CONCLUSION AND FURTHER RESEARCH

The intensifying quantum competition necessitates robust global governance structures, yet the landscape in 2025 is defined by diverging approaches among the major technological blocs: the United States, the European Union, and China. While China leverages massive state investment to achieve dominance in quantum communications and advantages in sensing, the US maintains an edge in computing hardware and algorithms, reflecting different innovation models, ethics, and strategic priorities.[330] For example, China's emphasis on quantum infrastructure like Hefei's Quantum Avenue and applications like the Beijing-Shanghai QKD network underscores its ambition to set global technical standards embedded with autocratic norms, potentially entrenching authoritarian surveillance capitalism. Amidst this complex techno-systemic rivalry, the U.S., through early legislative actions and security-focused directives, is exhibiting characteristics of a *de facto* 'Washington effect' for quantum law, setting global precedents, particularly concerning national security safeguards and investment screening, though risking premature regulatory lock-in before the full quantum prospective is clear. Concurrently, the EU's influence, often termed the 'Brussels effect' in data and AI regulation, may shape quantum governance through its emphasis on fundamental rights, ethical frameworks, and comprehensive market standards, albeit at a slower pace. China's state-centric strategy prioritizes national goals and technological sovereignty, utilizing standardization and industrial policy as strategic tools, which risks creating market fragmentation and propagating state-aligned values globally – a potential 'Beijing effect'. Monitoring this evolving, competitive landscape requires systematic tracking to understand the interplay of these influences.

An analysis through the lens of the SEA framework (Safeguarding, Engaging, Advancing) for Responsible Quantum Innovation reveals critical imbalances in current national strategies. The U.S. and China demonstrate strong focuses on 'Advancing' technological capabilities and 'Safeguarding' national or state interests, respectively. However, the U.S. approach, heavily reliant on export controls and security measures, may inadvertently hinder global scientific 'Engagement' and even prove counterproductive to its own long-term 'Advancing' goals if it provokes rivals towards self-sufficiency or disrupts critical supply chains. The EU excels at 'Engaging' stakeholders and 'Safeguarding' through its values-based regulatory model but needs to ensure this does not unduly impede timely 'Advancing' in the global race. This focus on national tech supremacy and the prevalent zero-sum rivalry narrative incurs significant opportunity costs, hindering the international collaboration essential for leveraging quantum (and AI) technologies (which are increasingly complementary and interdependent) to address pressing planetary challenges like climate change, pandemics, and inequality. The costs of non-cooperation may sooner or later outweigh the perceived benefits of 'winning' a fragmented technological competition.

Despite these challenges, the imperative and opportunity for global harmonization remain. Striving towards a Quantum Acquis Planétaire – a shared body of principles, norms, and rules – necessitates finding common ground rooted in universal values like human dignity, security, and shared planetary vulnerability ('ons verbindt' – what connects us). This requires moving beyond purely competitive dynamics towards pragmatic cooperation, likely involving strategic 'recoupling' between the US and China in specific scientific or pre-competitive areas. An effective global innovation model might hybridize the strengths of the major systems: combining US market dynamism and rapid innovation, the EU's emphasis on ethical guardrails and human-centric regulation, and leveraging acceptable aspects of long-term strategic planning seen in China, all within a framework grounded in democratic principles and human rights.

---

[330] Hmaidi & Groenewegen-Lau, *supra* note 2.

Achieving such harmonization requires more than just ethical guidelines. It demands the development and implementation of smart regulation – agile, pro-innovation, problem-based frameworks that utilize incentives effectively alongside necessary restrictions ('hard law'). This involves looking beyond traditional regulation towards a broader suite of responsible innovation tools with greater plasticity, such as Quantum Technology Impact Assessments (QIA), legal sandboxes, and modular approaches. Given the unique governance challenges posed by quantum's uncertainty, complexity, dual-use potential, and rapid evolution, reliance on traditional regulatory approaches alone is likely insufficient, underscoring the critical need for these more flexible and adaptive instruments. Embedding values directly into technology development through Responsible Quantum Technology (RQT) by design, supported by robust technical standards (ideally developed *ex ante* through international bodies like ISO/IEC and IEEE to ensure interoperability), flexible 'guardrails' focused on guiding human decision-making, and RQT metrics and benchmarks, is essential. Harmonization with existing sectoral quality management systems will also be needed. Institutionally, progress towards a global acquis could be advanced through a UN Quantum Treaty modeled on precedents like the UN AI Resolution and NPT to align with global goals like the SDGs, complemented by an adapted IAEA-like agency ('Atomic Agency for Quantum-AI') focused on non-proliferation and implementing international safeguards, learning from nuclear governance by distinguishing beneficial applications from military uses and building upon existing treaty frameworks rather than creating entirely new ones to ease state adoption concerns, and collaborative research platforms akin to CERN or ITER to boost coordinated responsible innovation. Realizing ambitious goals like fault tolerant quantum-centric supercomputing alongside algorithmic development and use case discovery toward quantum benefit requires such collective global expertise, challenging protectionist measures that stifle necessary collaboration and supply chains. These institutional measures should be flanked by effective national regulatory oversight bodies (perhaps akin to an 'FDA for Quantum' for market authorization).

Further research is critically needed across several domains. Empirical analysis is required to compare the outcomes of the different national innovation systems and regulatory approaches. Robust RQT metrics, benchmarks, and Quantum Technology Impact Assessment (QIA) methodologies must be developed and validated. Effective and politically feasible verification mechanisms for quantum controls, particularly for software and dual-use applications, need exploration. Economic modeling comparing the long-term costs and benefits of global cooperation versus fragmentation in quantum is essential. Furthermore, research into effective global governance models that respect national sovereignty while ensuring responsible innovation and addressing the unique challenges of regulating baseline technologies like quantum is paramount. The path forward demands immediate global actions focused on leveraging quantum for the UN Sustainable Development Goals (SDG), ensuring equitable access to mitigate a 'Quantum Divide,' and promoting broad quantum literacy. Concerted international effort, grounded in shared principles and informed by continuous research, is essential to navigate the complexities of the quantum age for the benefit of all humanity.

**END**

**EXECUTIVE SUMMARY**

This essay provides an overview of emerging AI and quantum technology regulation, export controls, and technical standards in the three main tech blocks: the U.S., EU, and China. Through the lens of systemic similarities and differences between the tech blocks' innovation mechanisms and vision for the future, the paper compares and contrasts legislative efforts seeking to strategically balance benefits and risks of the extravagant suite of quantum technologies including quantum-AI hybrids, anno 2025.

Analytically, the paper finds convergence across transatlantic and transpacific lines on the significance of adopting a principled approach to responsible technology governance, despite conflicting views on market driven, values-based, and state driven innovation.

The paper posits that the U.S. taking the lead in implementing quantum economic safety and national security centric policies results in the contours becoming visible of a *de facto* 'Washington effect' for quantum law, setting out the rules of the road along with associated extraterritorial impact and first mover advantage vis-à-vis Brussels and Beijing. However, regulating under profound uncertainty about quantum's future trajectory presents a core challenge. This U.S. first-mover position carries the inherent risk of premature regulation based on incomplete future insights, creating suboptimal technological lock-in or stifling innovation before quantum's full potential is understood.

China has positioned itself as a global pioneer in developing cutting edge technology software and hardware solutions. This marks a 'Beijing effect' for standardization and certification of AI, 6G, and quantum networking & communication. The consequence of China setting national and extra-domestic technical interoperability standards (such as SAC/TC578 to access Belt & Road initiative funding) with embedded autocratic norms and values, is that these values are being exported to democratic countries though their products, services, platforms and devices, propelled by China's Belt and Road and Digital Silk Road initiatives. Broad, unilateral export controls -often with counterproductive results- and decoupling pressure from recent import and trade tariffs exacerbate these trends, and are expected to disrupt global interoperability. The above potentially results in irresponsible quantum technologies by design and default, and an expected 'Beijing effect' for technical standards that lack universally accepted values in quantum technologies.

Faced with planetary scale challenges pertaining to inequality, climate change, and eroded digital trust, the world has a once in a lifetime opportunity to align on Responsible Quantum Technology norms, standards, certification, and verification. The time to prepare these regulatory interventions is now. In an era of great power competition, the paper investigates pathways toward a globally harmonized Quantum Acquis Planétaire, whilst pursuing a pro-innovation regulatory strategy that supports global market access. This strategy should be anchored in a global search for 'what connects us'—universal values and common denominators—translated into foundational standards and agile legal guardrails. Critically, smart regulation must leverage flexible instruments like sandboxes and impact assessments alongside foundational standards, moving beyond mere restrictions to actively incentivize responsible behaviors from nation-states and industry actors alike, for instance through frameworks promoting 'Responsible Quantum Technology (RQT) by design'. Such an approach necessitates frameworks for the equitable redistribution of quantum's benefits and risks, even requiring a fundamental redesign of societal institutions to align with principles of post-Rawlsian distributive justice in a quantum-influenced age of widespread relative abundance. Ultimately, it calls for nations to work together, efficiently stewarding the world's quantum resources for the shared benefit of humanity and the planet.

The paper conceptualizes the Quantum Acquis Planétaire as a collection of common rights and obligations constituting the body of Quantum Law, that is binding and incorporated into the legal systems of all countries of our world. It should be flanked by sector-specific guardrails and best practices, building upon existing industry best practices and quality management systems (QMS). A globally harmonized body of Quantum Law requires inter-continental quantum policy making. Instead of promoting division by restricting a bilateral relationship between China and the U.S., to be effective, this coordinated international approach requires a Sino-American "recoupling" in terms of academic, industrial-economic and policy cooperation and quantum community building. The U.S. should take the initiative for a recoupling, from a position of strength, quantum competency, and Responsible Quantum Technology thought leadership. A culturally sensitive, restorative approach towards aligned tech governance based on universally accepted human rights, values, and guardrails is envisioned to enable the three tech blocks to compete and innovate responsibly while working together on big picture trends.

Codification of a unified body of Quantum Law can be achieved by implementing a designated United Nations Declaration on Quantum -mirrored after the 2024 UN Declaration on AI and the 1968 Nuclear Non-Proliferation Treaty (NPT). A United Nations Declaration on Quantum should endorse a principled approach to responsible quantum technology governance, applied to all countries of our world and in space. To enforce such a harmonized global acquis, disincentivize a quantum arms race, and safeguard quantum non-proliferation, the paper suggests to further examine the possibilities of establishing an 'Atomic Agency for quantum-AI' modelled after the International Atomic Energy Agency (IAEA) for dual-use second generation quantum technologies including quantum-classical hybrid approaches. It also provides the rationale and model for a global quantum-AI non-proliferation treaty, and outlines responsibilities of a potential Atomic Agency for quantum-AI, alongside establishing collaborative research platforms for quantum and AI (which increasingly rely on, and complement each other) like CERN or nuclear fusion (ITER) projects to coordinate responsible innovation. Achieving breakthroughs like quantum-centric supercomputing requires global collaboration, making protectionist measures counterproductive.

∗ ∗ ∗